\documentclass[aps,showpacs]{revtex4}
\usepackage{graphics}
\usepackage{latexsym}
\usepackage{graphics}
\usepackage{hyperref}
\usepackage{indentfirst}
\usepackage{bm}
 \hypersetup{colorlinks=true, urlcolor=blue, 
linkcolor=blue, citecolor=blue} 
\usepackage{lineno,hyperref}
\usepackage{graphicx}
\usepackage{amsmath}
\usepackage{amssymb}
\usepackage{mathrsfs}
\usepackage{filecontents}
\usepackage{booktabs}
\usepackage{siunitx}
\usepackage{blindtext}

\DeclareMathAlphabet\mathbfcal{OMS}{cmsy}{b}{n}
\usepackage{mathtools}

\begin{document}


\title{Enhanced magnetocaloric effect in a mixed spin-(1/2, 1)  Ising-Heisenberg two-leg ladder with strong-rung interaction}
 
\author{Hamid Arian Zad$^{1,2}$, Nerses Ananikian$^{1,3}$}
\email{arianzad.hamid@yerphi.am}
\affiliation{$^{1}$A.I. Alikhanyan National Science Laboratory, 0036, Yerevan, Armenia}
\affiliation{$^{2}$ICTP, Strada Costiera 11, I-34151 Trieste, Italy}
\affiliation{$^{3}$ CANDLE Synchrotron Research Institute, Acharyan 31, 0040 Yerevan, Armenia}




\begin{abstract}
The magnetic and magnetocaloric properties of the mixed spin-(1/2,1)  Ising-Heisenberg model on a two-leg  ladder with dimer-rung
alternation are exactly examined under an adiabatic demagnetization process using the transfer-matrix formalism. 
We notify that the magnetization curve of the model exhibits plateaux as a function of the applied magnetic field and cyclic four-spin Ising interaction at certain rational fractions of the saturation value. 
We precisely investigate the ability of cooling/heating of the model nearby the  critical points at which discontinuous ground-state phase transition occurs.
It is evidenced that the model manifests an enhanced magnetocaloric effect in a proximity of the magnetization steps and jumps, accompanying with the plateaux and jumps of correlation function of the dimer spins.
We conclude that not only the cooling/heating capability of the model could be pleasantly demonstrated by the applied magnetic field variations, but also a typical cyclic four-spin Ising interaction plays essential role to determine an efficiency of the magnetocaloric effect of the model.
\end{abstract}

\maketitle


 \section{Introduction} \label{sec:level1}
Various Heisenberg spin models defined on the two-leg ladders have attracted a great deal of attention in theoretical condensed matter due to reveal extremely rich behaviors, dominated by quantum effects \cite{Cabra1,Cabra2,Langari2000,Hida2004,Vekua2004}. Two-leg ladders with antiferromagnetic exchange along their rungs \cite{Koga1998,Avalishvili2019}, as well as, both of antiferromagnetic and ferromagnetic exchanges along legs have been investigated in previous studies \cite{Japaridze2006,Amiri2015,Amiri2017,Eggert2018}. 
 From this perspective, M. T. Batchelor {\it et al.} \cite{Batchelor2007} have comprehensively investigated the magnetic properties, ground-state phase transition and thermodynamics of various versions of exactly solvable two-leg ladders, both pure spin-1/2 models and mixed spin (1/2, 1) ones, and discussed their implementation in the physics of strong-rung interaction ladder compounds.
  
Magnetocaloric effect (MCE) can be defined as the temperature variation of magnetic materials upon changing the external magnetic field. 
In many-body problem, MCE has attracted renewed attentions because of having a strong potential of cooling applications in science and technology \cite{oja97,Tishin2003,gsch05,Cho2014}. 
One another important application of the MCE is the study of phase transitions by using the magnetocaloric anomalies at the magnetic phase transitions \cite{Bez2016,Law2018}.  
The standard quantity to characterize the MCE is the so-called Gr{\" u}neisen parameter $\Gamma_B$ which can be counted as one of applicable tools for detecting and investigating quantum critical points  \cite{Garst2003}. The Gr{\" u}neisen parameter for magnetic systems under adiabatic conditions can be defined by
\begin{equation}\label{SLhamiltonian}
\begin{array}{lcl}
\Gamma_B=\frac{1}{T}\big(\frac{\partial T}{\partial B}\big)_S=-\frac{T}{{C}_B}\big(\frac{\partial S}{\partial B}\big)_T=-\frac{1}{{C}_B}\big(\frac{\partial {M}}{\partial T}\big)_B,
\end{array}
\end{equation}
where ${C}_B$ is the heat capacity at the constant magnetic field, $T$ is the temperature (for simplicity we consider $k_B=1$) and $B$ is the applied magnetic field. In the recent decades a series of  theoretical and computational researches have been conducted on the magnetic properties and MCE in low-dimensional quantum and Ising spin models \cite{zhit04,hon09,Honecker2009,Vadim2010,Vadim2012EPJB,gal14,str15,strPhysE2018,strPhysB2018,str17a,kar17,zuk18,Beckmanna2018, Vadim2012, str14,gal16,tor16,ale18,gal18,Hamedoun2018,Masrour2018,Masrour2010,Masrour2015,MasrourPLA2008}. The exact results obtained within the low-dimensional quantum and mixed classical-quantum interacting spin models figure out many important features of the MCE nearby the quantum critical points, particularly the enhancing role of frustration and residual entropy, the possibility of magnetic cooling and magnetic heating during the adiabatic demagnetization, deep connection of the MCE and the ground-state phase transitions, etc. Besides, by examining the behavior of adiabatic cooling rate, important information about the MCE can be obtained from the plots of the isentropes in the temperature-magnetic field plane.

 During the current decade, a number of exact results on the MCE in the so-called Ising-Heisenberg one-dimensional spin models have been obtained \cite{Vadim2012,str14,gal16,tor16,ale18,gal18}. The main feature of these models is the special alternation of the small clusters of quantum spins and Ising interaction bond in such a way that the local Hamiltonians for the blocks commute with each other. This allows one to obtain an exact solution in terms of the generalized classical transfer-matrix method.  
 
 For one-dimensional  Ising-Heisenberg models,  the average spin value of the $j-$th  block when the system reaches thermodynamic equilibrium  is the same for all values of $j$. Since the Hamiltonian is translationally invariant, all the unit-blocks are identical, and the average spin will be the same no matter which block we look at. Useful details about the solution of the spin-1/2 Ising-Heisenberg diamond chain within the transfer matrix technique can be found, for example, in Ref. \cite{Moises2019}. Analogously,  O. Rojas {\it et al.} solved a mixed spin-(1/2, 1) Ising-Heisenberg double-tetrahedral chain in an external magnetic field using the classical transfer-matrix formalism and reported comprehensive results on the ground-state phase transition and the thermodynamics of such a model in their recent work \cite{Onofre2020}. 
 
 The four-spin Ising interaction can be identified as the Ising limit of the cyclic permutation of the quantum spins localized on the vertices of a plaquette. It was demonstrated that this particular term is important to realize the magnetic properties of the solid He$^3$ \cite{roger}, as well as in some cuprates  Ref. \cite{Muller}. The  Ising-Heisenberg spin models with additional four-spin Ising interaction have been also examined in several papers as an Ising limit of the four-spin cyclic permutation \cite{gal14,ara03, oha05, ana07, hov09,ana12,gal13,Arian1,Arian2,Arian3}.

Although, there is a great interest on the magnetic and thermodynamic properties of the Spin-1/2 two-leg ladder systems, mixed spin-(1/2, S)  Ising-Heisenberg ladders have been much less studied. They exhibit many interesting aspects that would definitely attract a numerous attentions in theoretical condensed matter and magnetic material science.
 Undoubtedly, the mixed spin-(1/2, S)  Ising-Heisenberg  two-leg ladders can be viewed as decorated Heisenberg ladders \cite{Masrour2020}. 
From the experimental point of view, various magnetic materials with obvious quantum nature have been detected whose structures can be 
characterized in terms of the mixed spin  ladders. With this regard, a widespread of organic mixed spin ladders with strong rung coupling have been synthesised \cite{Arnaudon2004,Hida2010,Batchelor2007}.  
 In a resent report \cite{Lu2017}, H. Lu {\it et al.} experimentally studied the synthesis, structure, and magnetic properties of a novel diamond chain  $\text{Cu}_2\text{FePO}_4\text{F}_4(\text{H}_2\text{O})_4$ composed of  mixed spins
   $S_{\text{Cu}^{2+}} = 1/2$ and $S_{\text{Fe}^{3+}} = 5/2$. The model possesses  a noncollinear spin order with successive ground-state phase transitions.
    In the current paper, we introduce a mixed spin-(1/2,1)  Ising-XYZ  two-leg ladder whose nodal Ising spins play the decoration role. 
    One of applications of this model is that, it can effectively 
  reproduce the mixed spin diamond chain compound $\text{Cu}_2\text{FePO}_4\text{F}_4(\text{H}_2\text{O})_4$ by rigorously mapping  to equivalence with the introduced mixed-(1/2,1) Ising-Heisenberg two-leg ladder. We prove that the generalized mixed spin-(1/2, 1)  Ising-Heisenberg model on a decorated two-leg ladder can be exactly solved through the classical transfer matrix technique and discloses many interesting magnetic properties.

The mixed spin-(1/2,1)  Ising-XYZ model on a  two-leg ladder can be constructed from unit blocks with the square spin configuration which linked together through their rungs possessing various kinds of Heisenberg exchange interactions. 
Indeed, the introduced model  is characterized by full anisotropic  XYZ interaction between the interstitial Heisenberg dimers, the Ising coupling between spins localized on the  legs and rungs, furthermore an additional cyclic four-spin Ising interaction in the square plaquettes of each sub-unit block.
The motivation of considering this unique model  is to schematically represent an exactly solvable two-leg ladder whose nodal sites effectively involve with  higher spins ($S>1/2$). 
In similar fashion, Ising-Heisenberg variant of the saw-tooth chain has been considered earlier in Refs. \cite{Ohanyan2009,Bellucci,bel13}, however, the quantum cluster was there considered as a two-spin bond along each second tooth of the chain. Moreover, pure spin-1/2 and mixed spin-(1,1/2) Ising-Heisenberg double saw-tooth ladders have been intensively investigated in previous works  \cite{Arian2,Arian3}.
  
  In the present paper, we will focus on the MCE, particularly, on the effects of a typical cyclic four-spin Ising term, as well as, on how the Heisenberg exchange interaction influences the MCE of the model.  Our motivation to this model comes from its superb magnetic phase diagram that indicates complex quantum nature, especially in the case of existing anisotropy, cyclic four-spin Ising term, and external magnetic field. The model describes a new scenario for magnetization process, MCE as well as the thermodynamic medium. Due to quantum correlations are very important in quantum information processing, quantum computing, and spintronics  \cite{Arian1,Paulinelli2013,Rojas2016,Masrour2014QC,Masrour2016QC,Masrour2017QC,Masrour2014QC1}, we also test the correlation function of the Heisenberg dimer-rungs, compare its behavior with other thermodynamic parameters under the same conditions, and eventually quote obtained results for interaction dependencies of such a function. 
  
  The current progress contributes to manufacture low-dimensional magnetic materials with desired parameters, particularly, when the system involves with the alternating dimer-rung exchange. Moreover, the selected model properly creates the possibility of understanding an 
  Ising-Heisenberg ladder system with the four-spin Ising interaction and leg-rung exchange modulations.

The organization of this paper is as follows. In the next section we describe in detail the mixed spin-(1/2,1) Ising-Heisenberg two-leg  ladder and remind the reader the main points of the solution within the transfer-matrix technique. In Sec. \ref{cooling}, magnetic phase transition, the behavior of entropy and the cooling rate versus the magnetic field and temperature for several fixed values of the model parameters are particularly discussed. The effects of the both parameters, four-spin Ising interaction and the temperature, on the entropy, cooling rate and correlation function of the interstitial Heisenberg dimers are examined as well. Finally, several concluding remarks are mentioned in Sec. \ref{conclusion}.

\begin{figure}
  \centering
  \resizebox{1\textwidth}{!}{
 \includegraphics[trim=10 130 60 130, clip]{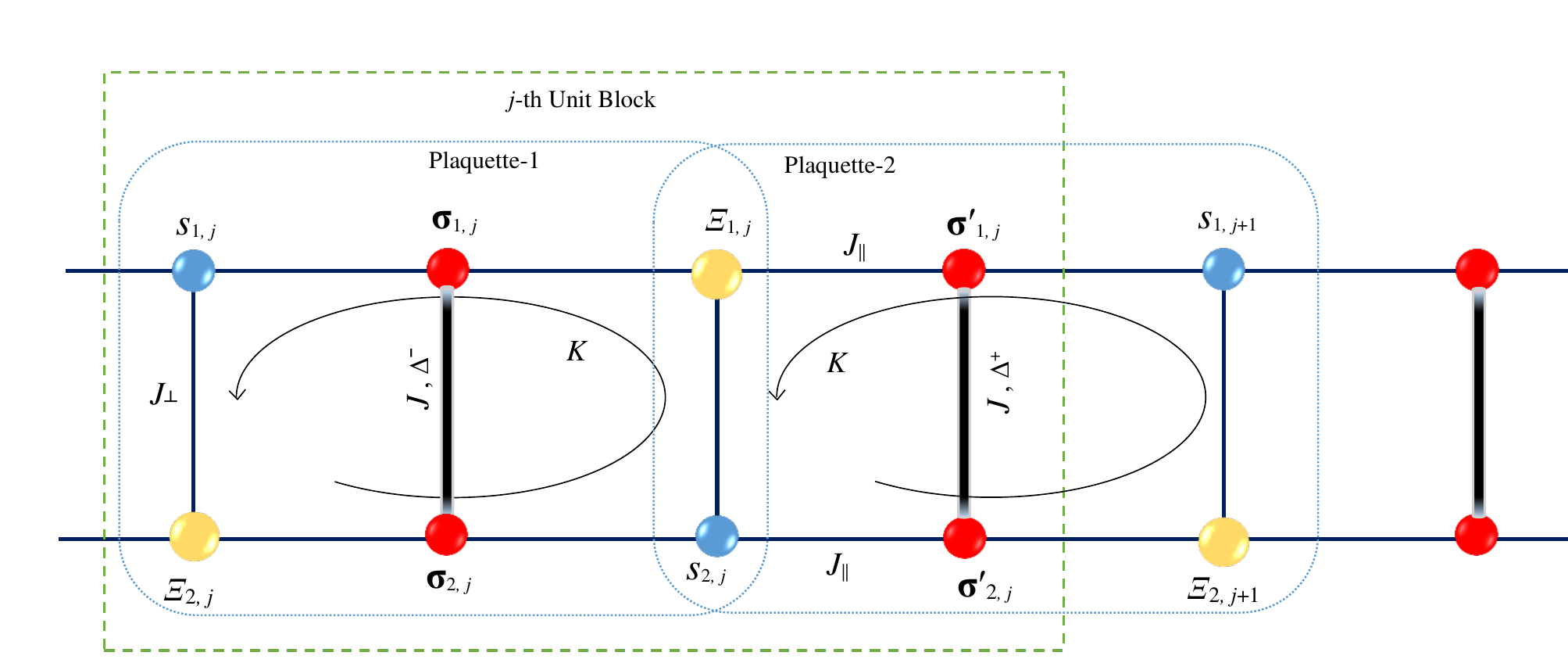} 
}
\caption{Sketch  of the mixed spin-(1/2,1) Ising-Heisenberg two-leg model on a decorated ladder. Gold balls present effective spin-1 particles, while blue balls denote effective spin-1/2 particles. They interact with their nearest-neighbor sites through Ising-type coupling constant $J_{\parallel}$ along the legs and through $J_{\perp}$ along the rungs.  Red balls linked together with tick lines show the spin-1/2 interstitial Heisenberg dimers. The region indicated by dashed rectangle displays a unit block (including two plaquettes) that repeats throughout the ladder.} 
\label{figure1}
\end{figure}
\section{Model and  its exact solution}\label{Model}

The most general Hamiltonian of the two-leg ladder could be then obtained by the sum of all unit blocks represented by dashed rectangle in Fig. \ref{figure1}. Due to better characterizing the Hamiltonian, we split all unit blocks to two sub-unit blocks as
\begin{equation}
\label{SLHamiltonian}
\begin{array}{lcl}
H = \sum\limits_{j=1}^N \big(\mathcal{H}_{\square}^{j,1}+\mathcal{H}_{\square}^{j,2}\big),
\end{array}
\end{equation}
 The operators $\mathcal{H}_{\square}^{j,1}$ and $\mathcal{H}_{\square}^{j,2}$ define Hamiltonians of two sub-unit blocks so-called square Plaquette-1 ($\mathcal{P}_1$) and Plaquette-2  ($\mathcal{P}_2$), respectively. Hamiltonians of the two square palquettes  can be written as below abbreviated forms
\begin{equation}\label{HLHamiltonian}
\begin{array}{lcl}
\mathcal{H}_{\square}^{j,1}= \big\{-J\left({\boldsymbol \sigma}_{1,j}\cdot{\boldsymbol \sigma}_{2,j}\right)_{\Delta^-}-
\sum\limits_{\kappa=1,2}{\sigma}_{\kappa,j}^z \big(g_1B+ \alpha J_{\parallel} \big[{s}_{\kappa,j}+{\Xi}_{\kappa,j}\big]\big)\big\}_q+ \\
\quad\quad\quad \big\{-\frac{J_{\perp}}{2}\big[{s}_{1,j}\;{\Xi}_{2,j}+{\Xi}_{1,j}\;{s}_{2,j}\big]+ K \left(s_{1,j}\;\Xi_{2,j}\;s_{2,j}\;\Xi_{1,j}\right)
-\frac{B}{2}\big[g_2(s_{1,j}+s_{2,j})+g_3(\Xi_{1,j}+\Xi_{2,j})\big]\big\}_c,\\
\\
\mathcal{H}_{\square}^{j,2}= \big\{-J\left({\boldsymbol \sigma}_{1,j}^{\prime}\cdot{\boldsymbol \sigma}_{2,j}^{\prime}\right)_{\Delta^{+}}-
\sum\limits_{\kappa=1,2}{\sigma}_{\kappa,j}^{\prime z} \big(g_1B+ \alpha J_{\parallel} \big[{s}_{\kappa,j+j^{'}}+{\Xi}_{\kappa,j+j^{''}}\big]\big)\big\}_q+\\
\quad\quad\quad \big\{-\frac{J_{\perp}}{2}\big[\Xi_{1,j}\;{s}_{2,j}+{s}_{1,j+1}\;{\Xi}_{2,j+1}\big]+ K \left(\Xi_{1,j}\;s_{1,j+1}\;\Xi_{2,j+1}\;s_{2,j}\right)
-\frac{B}{2}\big[g_2(s_{2,j}+s_{1,j+1})+g_3(\Xi_{1,j}+\Xi_{2,j+1})\big]\big\}_c,
\end{array}
\end{equation}
where ${\boldsymbol \sigma}^{a}_{\kappa, j} ({\boldsymbol \sigma}^{\prime a}_{\kappa, j})$ indicates spatial components of being the Pauli operators to the leg $\kappa$ and $j$-th ($j+1$-th) rung under periodic boundary conditions. $a= \{x, y, z\}$ denotes the direction of the Pauli matrices in the $x-$, $y-$ and $z-$axis, respectively. Subscripts $q$ and $c$ represent, respectively, the quantum and classical parts of the Hamiltonians. 
Here, $N$ is the number of unit blocks which is supposed to be even, and we assume that each pair of quantum spins ${\boldsymbol \sigma}_{1,j}$ (${\boldsymbol \sigma}^{\prime}_{1,j}$) and ${\boldsymbol \sigma}_{2,j}$ (${\boldsymbol \sigma}^{\prime}_{2,j}$), interact through the fully anisotropic XYZ exchange coupling
\begin{equation}\label{FourSpin}
\centering
\begin{array}{lcl}
J \left({\boldsymbol \sigma}_{1,j}\cdot{\boldsymbol \sigma}_{2,j}\right)_{\Delta^-}=
J \big[\big((1+\gamma){\sigma}^{x}_{1,j}{\sigma}^{x}_{2,j}+(1-\gamma){\sigma}^{y}_{1,j}{\sigma}^{y}_{2,j}\big)+\Delta^-\;{\sigma}^{z}_{1,j}{\sigma}^{z}_{2,j}\big],\\

J\left({\boldsymbol \sigma}_{1,j}^{\prime z}\cdot{\boldsymbol \sigma}_{2,j}^{\prime z}\right)_{\Delta^{+}}=
J \big[\big((1+\gamma){\sigma}^{x}_{1,j}{\sigma}^{x}_{2,j}+(1-\gamma){\sigma}^{y}_{1,j}{\sigma}^{y}_{2,j}\big)+\Delta^+\;{\sigma}^{z}_{1,j}{\sigma}^{z}_{2,j}\big],
\end{array}
\end{equation}
 where, $J$ represents the ferromagnetic exchange interaction between the interstitial Heisenberg dimers, while $\gamma$ is the XY-anisotropy. 
 Subscripts  $\Delta^-=(1-\Delta)/2$ and  $\Delta^+=(1+\Delta)/2$ denote alternative rung-exchange anisotropy in the interstitial dimers respectively, $\mathcal{P}_1$-dimer and $\mathcal{P}_2$-dimer, where 
 $\Delta$ ($-1\leq\Delta\leq 1$) stands for the exchange anisotropy. 
 In addition we supposed $j^{'} =-\frac{(-1)^{\kappa}-1}{2}$ and $j^{''} =\frac{(-1)^{\kappa}+1}{2}$.
${J}_{\perp}$ and ${J}_{\parallel}$ are the Ising-type couplings on the rungs and along the legs, respectively. 
Tunable coefficient $\alpha$ typically represents the strength of $J_{\parallel}$  with respect to other parameters. To invoke the strong ferromagnetic Ising-rung interaction condition $J_{\perp}\gg 0$, in the current research we consider $0<\alpha<1$.
$K$ manifests four-spin Ising interaction between four nodal sites of each plaquette. These couplings are supposed to include only $z$-component of the quantum spins. 
$s_{\kappa,j}$ and $\Xi_{\kappa,j}$ are the $S=1/2$ and $S=1$ Ising spin variables, taking values $(1,-1)$ and $(1,0,-1)$, respectively.  $B$ is the external homogeneous magnetic field  applied  in the $z$-direction. Motivated by assuming mixed spin parties in the ladder, we here optionally consider three different static Land{\'e} g-factors $g_1$, $g_2$ and $g_3$, denoting three different particles in the spin model. Multiplicity of the parameters in the Hamiltonian makes enable us to introduce a more flexible and eligible model for both of the theoretical investigations and experimental analysis specially for mixed-spin two-leg ladders and metal ions doped spin ladders \cite{Mizuno1997,Wang2009,Exius2010}.
\subsection{The exact solution within the classical transfer-matrix formalism}
We perform the generalized classical transfer-matrix technique to obtain the partition function of the model. To study the thermodynamics of the mixed spin-(1/2,1) Ising-Heisenberg two-leg  ladder, we realize that Hamiltonians of each pair of unit block commute with each other. Consequently, the partition function of the model could be expressed as the product of Boltzmann factors corresponding to the unit blocks possessing the same transfer matrix $\mathbf{T}$, namely,
\begin{equation}\label{BZ}
\begin{array}{lcl}
\mathcal{Z}_N=\mbox{tr}\;[\mathbf{T}^N]=\mbox{tr}\left( \prod\limits_{j=1}^N \mathrm{e}^{\big[ -\beta\big( \mathcal{H}_{\square}^{j,1} + \mathcal{H}_{\square}^{j,2}\big)\big]}\right).
\end{array}
\end{equation} 
To obtain above partition function, one can apply the transfer-matrix approach using the product of  Boltzmann factors for the sub-unit blocks
 with Hamiltonians $\mathcal{H}_{\square}^{j,1} $ and $\mathcal{H}_{\square}^{j,2} $. Hence, the $6\times 6$ transfer-matrix of a unit block can be written as
\begin{equation}\label{TMww}
\begin{array}{lcl}
\mathbf{T}={\scriptstyle\mathbfcal{W}}(s_{1,j},\;\Xi_{2,j}|\Xi_{1,j},\;s_{2,j}){\scriptstyle \overline{\mathbfcal{W}}}(\Xi_{1,j},\;s_{2,j}|s_{1,j+1},\;\Xi_{2,j+1}).  
\end{array}
\end{equation}
$6\times 6$ fully symmetric matrices ${\scriptstyle \mathbfcal{W}}$ and ${\scriptstyle \overline{\mathbfcal{W}}}$ are Boltzmann factors for the sub-unit blocks with Hamiltonians $\mathcal{H}_{\square}^{j,1} $ and $\mathcal{H}_{\square}^{j,2}$, respectively.
The procedure of deducing eigenvalues of the Hamiltonians of  4-sites plaquettes and their corresponding transfer-matrix coefficients is given  below.   

Consider the mixed spin-(1/2,1) Ising-Heisenberg  two-leg model on a decorated ladder. The corresponding $6\times 6$ transfer-matrix per block is given by Eq. (\ref{TMww}). For simplicity, we divide the Boltzmann factors  ${\scriptstyle \mathbfcal{W}}$ and ${\scriptstyle \overline{\mathbfcal{W}}}$ into classical and quantum parts such that
\begin{equation}\label{W}
\begin{array}{lcl}
{\scriptstyle \mathbfcal{W}}\left(s_{1,j},\;\Xi_{2,j}|\Xi_{1,j},\;s_{2,j}\right)=
\mathrm{e}^{-\beta \mathcal{H}_{c}^{1,j}\left(s_{1,j},\;\Xi_{2,j}|\Xi_{1,j},\; s_{2,j}\right)}\times
{\scriptstyle \mathbfcal{W}}_q\left(s_{1,j},\;\Xi_{2,j}|\Xi_{1,j},\; s_{2,j}\right),
\\
\\
{\scriptstyle \overline{\mathbfcal{W}}}\left(\Xi_{1,j},\;s_{2,j}|s_{1,j+1},\;\Xi_{2,j+1}\right)=
\mathrm{e}^{-\beta \mathcal{H}_{c}^{2,j}\left(\Xi_{1,j},\;s_{2,j}|s_{1,j+1},\;\Xi_{2,j+1}\right)}\times
 {\scriptstyle \overline{\mathbfcal{W}}}_q\left(\Xi_{1,j},\;s_{2,j}|s_{1,j+1},\;\Xi_{2,j+1}\right).
\end{array}
\end{equation}
On above, we have the Boltzmann weight for the quantum parts of the Hamiltonians (\ref{HLHamiltonian})
\begin{equation}\label{W}
\begin{array}{lcl}
{\scriptstyle \mathbfcal{W}}_q\left(s_{1,j},\;\Xi_{2,j}|\Xi_{1,j},\; s_{2,j}\right)=
\sum\limits_{n=1}^{4}\mathrm{e}^{-\beta\varepsilon_n\left(s_{1,j},\;\Xi_{2,j}|\Xi_{1,j},\; s_{2,j}\right)},\\

 {\scriptstyle \overline{\mathbfcal{W}}}_q\left(\Xi_{1,j},\;s_{2,j}|s_{1,j+1},\;\Xi_{2,j+1}\right)=
\sum\limits_{n=1}^{4}\mathrm{e}^{-\beta\overline{\varepsilon}_n\left(\Xi_{1,j},\;s_{2,j}|s_{1,j+1},\;\Xi_{2,j+1}\right)},
\end{array}
\end{equation}
for which $\varepsilon_n$ and $\overline{\varepsilon}_n$ $(n=1,...,4)$  denote  the eigenvalues of the quantum parts of the Hamiltonians  (\ref{HLHamiltonian}).


The eigenvalues $\varepsilon_n$ and $\overline{\varepsilon}_n$ explicitly depend on the values of four classical spin variables of the unit-cells in each block, interacting with the quantum Heisenberg spin dimers $\mathcal{P}_1$-dimer and $\mathcal{P}_2$-dimer. They can be easily found by the straightforward diagonalization of the quantum parts of the Hamiltonians in the standard Ising basis ($|\uparrow \uparrow\rangle,|\uparrow \downarrow\rangle,|\downarrow \uparrow\rangle,|\downarrow \downarrow\rangle$). Thus eigenvalues $\varepsilon_n$ are

\begin{equation}\label{energy1}
\begin{array}{lcl}
 \dfrac{\varepsilon_{1,4}}{J_{\parallel}}= {\dfrac{J}{J_{\parallel}}(1-\Delta)}\mp\Big[\dfrac{2g_1B}{J_{\parallel}}\pm \big(s_{1,j}+\Xi_{2,j}+s_{2,j}+\Xi_{1,j}\big)\Big]\\
 \\
 \dfrac{\varepsilon_{2,3}}{J_{\parallel}}= \dfrac{J}{J_{\parallel}}(\Delta-1)\pm \sqrt{\big(s_{1,j}-\Xi_{2,j}-s_{2,j}+\Xi_{1,j}\big)^2+16\big(\frac{J}{J_{\parallel}}\big)^2},\\
\end{array}
\end{equation}
and for $\overline{\varepsilon}_n$ we have analogously
\begin{equation}\label{energy1}
\begin{array}{lcl}
 \dfrac{\overline{\varepsilon}_{1,4}}{J_{\parallel}}= \dfrac{J}{J_{\parallel}}{(1+\Delta)}\mp\Big[\dfrac{2g_1B}{J_{\parallel}}\pm\big(\Xi_{1,j}+s_{2,j}+\Xi_{1,j+1}+s_{2,j+1}\big)\Big]\\
 \\
 \dfrac{\overline{\varepsilon}_{2,3}}{J_{\parallel}}= \dfrac{J}{J_{\parallel}} {(-1-\Delta)}\pm
 \sqrt{\big(\Xi_{1,j}-s_{2,j}-\Xi_{2,j+1}+s_{1,j+1}\big)^2+16\big(\frac{J}{J_{\parallel}}\big)^2}.
\end{array}
\end{equation}

The transfer matrix of the $\mathcal{P}_1$ has the symmetric form
\begin{equation}\label{TM}
\mathcal{T}= \left(
\begin{array}{cccccc}
 \mathcal{A} & \mathcal{G}  &{\tau} &  {\tau} &\mathcal{K} &  {\mathcal{U}} \\
 \mathcal{G} & \mathcal{B} & {\Omega} & \mathcal{Q} & \mathcal{R} & \mathcal{S} \\
  {\tau} & {\Omega} & \mathcal{C} & \mathcal{V}  & \mathcal{W} & {\gamma} \\
  {\tau} &  \mathcal{Q} & \mathcal{V} & \mathcal{C} & \mathcal{X} & {\gamma} \\
  \mathcal{K}&  \mathcal{R} & \mathcal{W} & \mathcal{X} & \mathcal{D} & \mathcal{J} \\
 \mathcal{U} & \mathcal{S} & {\gamma} & {\gamma} & \mathcal{J} & \mathcal{E}
\end{array} \right),
\end{equation}
where by considering 
\begin{equation}\label{AAA}
\begin{array}{lcl}
z_1=\mathrm{e}^{\beta J_{\perp}},\; \lambda=\mathrm{e}^{\beta K},\; \mu_1=\mathrm{e}^{\beta g_1B},\; \mu_2=\mathrm{e}^{\beta g_2B},\; \mu_3=\mathrm{e}^{\beta g_3B},\; \delta=\mathrm{e}^{\beta{\Delta^{-}}J},\nonumber
\end{array}
\end{equation}
 the following notations are adopted for the all components of the transfer-matrix:

\begin{eqnarray}
&&  \mathcal{A}=4(z_1^{-1}\lambda^{-1}\mu_2^2\mu_3^2)(\varphi_4+\psi_2^0), \;
 \mathcal{G}=4(z_1^{-\frac{1}{2}}\mu_2^{2}\mu_3)(\varphi_3+\psi_2^1), \nonumber \\
&&  \mathcal{K}=4(z_1^{-\frac{1}{2}}\mu_3)(\varphi_1+\psi_2^1), \; 
 \mathcal{U}=4(z_1^{-1}\lambda^{-1})(\varphi_0+\psi_2^0),\nonumber\\
&&  \mathcal{B}=4(\mu_2^{2})(\varphi_2+\psi_2^0), \; 
 \mathcal{Q}=4(z_1^{-\frac{1}{2}}\mu_3)(\varphi_1+\psi_2^3),\nonumber\\
&&  \mathcal{R}=4(\varphi_0+\psi_2^2), \; 
 \mathcal{S}=4(z_1^{-\frac{1}{2}}\mu_3^{-1})(\varphi_1+\psi_2^1),\nonumber\\
&&  \mathcal{C}=4(z_1\lambda^{-1})(\varphi_0+\psi_2^0), \; 
 \mathcal{V}=4(z_1^{1}\lambda^{-1})(\varphi_0+\psi_2^4),\nonumber\\  
&&\mathcal{W}=4(z_1^{\frac{1}{2}}\mu_3^{-1})(\varphi_{-1}+\psi_2^3), \; 
 \mathcal{X}=4(z_1^{\frac{1}{2}}\mu_2^{-2}\mu_3)(\varphi_{-1}+\psi_2^1),\nonumber\\  
 &&  \mathcal{D}=4(\mu_2^{-2})(\varphi_{-2}+\psi_2^0), \; 
 \mathcal{J}=4(z_1^{-\frac{1}{2}}\mu_2^{-2}\mu_3^{-1})(\varphi_{-3}+\psi_2^1),\nonumber\\  
 &&  \mathcal{E}=4(z_1^{-1}\lambda^{-1}\mu_2^{-2}\mu_3^{-2})(\varphi_{-4}+\psi_2^0),\nonumber\\
&& \tau=4(\lambda\mu_2^{2})(\varphi_2+\psi_2^2), \;
\gamma=4(\lambda\mu_3^{-2})(\varphi_{-2}+\psi_2^2),\;\nonumber\\
&& \Omega=4(z_1^{\frac{1}{2}}\mu_2^2\mu_3^{-1})(\varphi_{1}+\psi_2^1),
\end{eqnarray}
for which functions $\varphi$ and  $\psi$ are defined as
\begin{eqnarray}
&& \varphi_n = 2\delta^{-1}\cosh(\beta[2g_1B-nJ]), \nonumber \\
&&\psi_{m}^{m^{\prime}}=2\delta\cosh(\beta\sqrt{(mJ)^2+(m^{\prime}J_{\parallel})^2}).\nonumber
\end{eqnarray}

Analogously, the Boltzmann factors for the $\mathcal{P}_2$-dimer are expressed in a similar way to the $\mathcal{P}_1$-dimer.
The transfer matrix of the $\mathcal{P}_2$ can be easily obtained in an analogous procedure to the $\mathcal{P}_1$, but  by substituting parameter $\Delta^{-}$ with  $\Delta^{+}$. Eventually, the transfer matrix of the unit blocks can be given by $\mathbf{T}=\mathcal{T}\mathcal{T}^{\prime}$, for which $\mathcal{T}^{\prime}$ is the transfer matrix of the Plaquette-2. Due to abbreviate analytical expressions we leave writing this matrix.

\subsection{Gibbs free energy and thermodynamic parameters}
To derive forthright expressions for all thermodynamic parameters of the mixed spin-(1/2,1) Ising-Heisenberg two-leg ladder, the partition function should be calculated by considering unit block Hamiltonians of the model written in Eq. (\ref{HLHamiltonian}). 
 In the thermodynamic limit, the free energy per block can be expressed as
 \begin{equation}\label{FreeE}
 \begin{array}{lcl}
  f=-\frac{1}{\beta}\ln\Lambda_{max},
 \end{array}
 \end{equation}
where, $\Lambda_{max}$ is the largest eigenvalue of the transfer matrix $\mathbf{T}$. Magnetization, entropy and specific heat of the model can be obtained using the Gibbs free energy as follows
\begin{equation}\label{TParameters}
\begin{array}{lcl}
{M}=-\Big(\dfrac{\partial f}{\partial B}\Big)_{T}, \quad {S}=-\Big(\dfrac{\partial f}{\partial T}\Big)_{B}, \quad {C}=-T\Big(\dfrac{\partial^2 f}{\partial T^2}\Big)_{B}.
\end{array}
\end{equation}

 One can exactly obtain $\Lambda_{max}$ and in turn the thermodynamic parameters of the model under consideration using numerical procedure  that we have expressed in our recent publications \cite{Arian42019,Arian52019}.
\section{ Results and discussion}\label{cooling}
\subsection{Magnetization process and discontinuous ground-state phase transition}
\begin{figure*}[t!]
\  \begin{center}
\resizebox{0.3\textwidth}{!}{%
\includegraphics{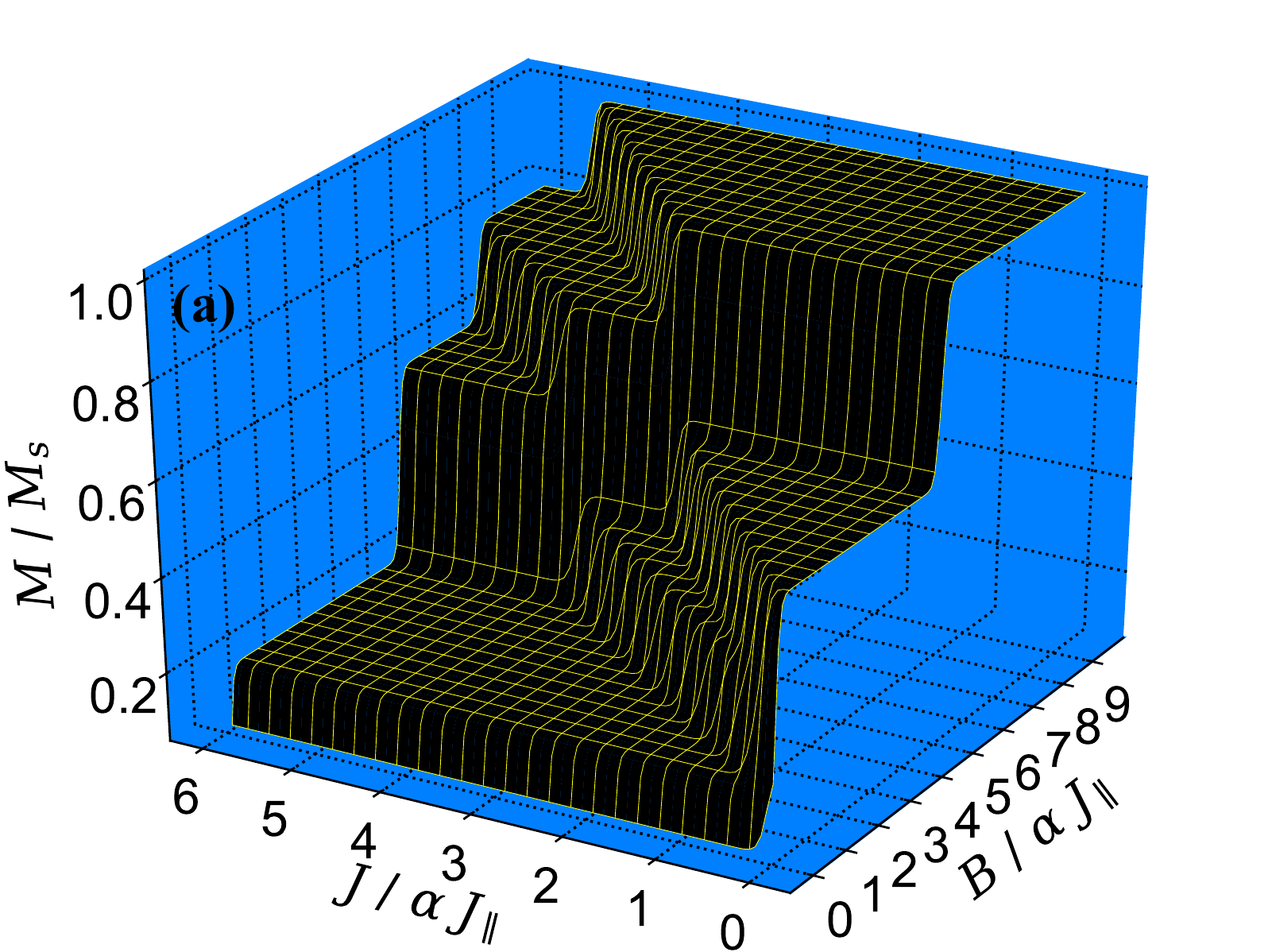}
}
\resizebox{0.3\textwidth}{!}{%
\includegraphics{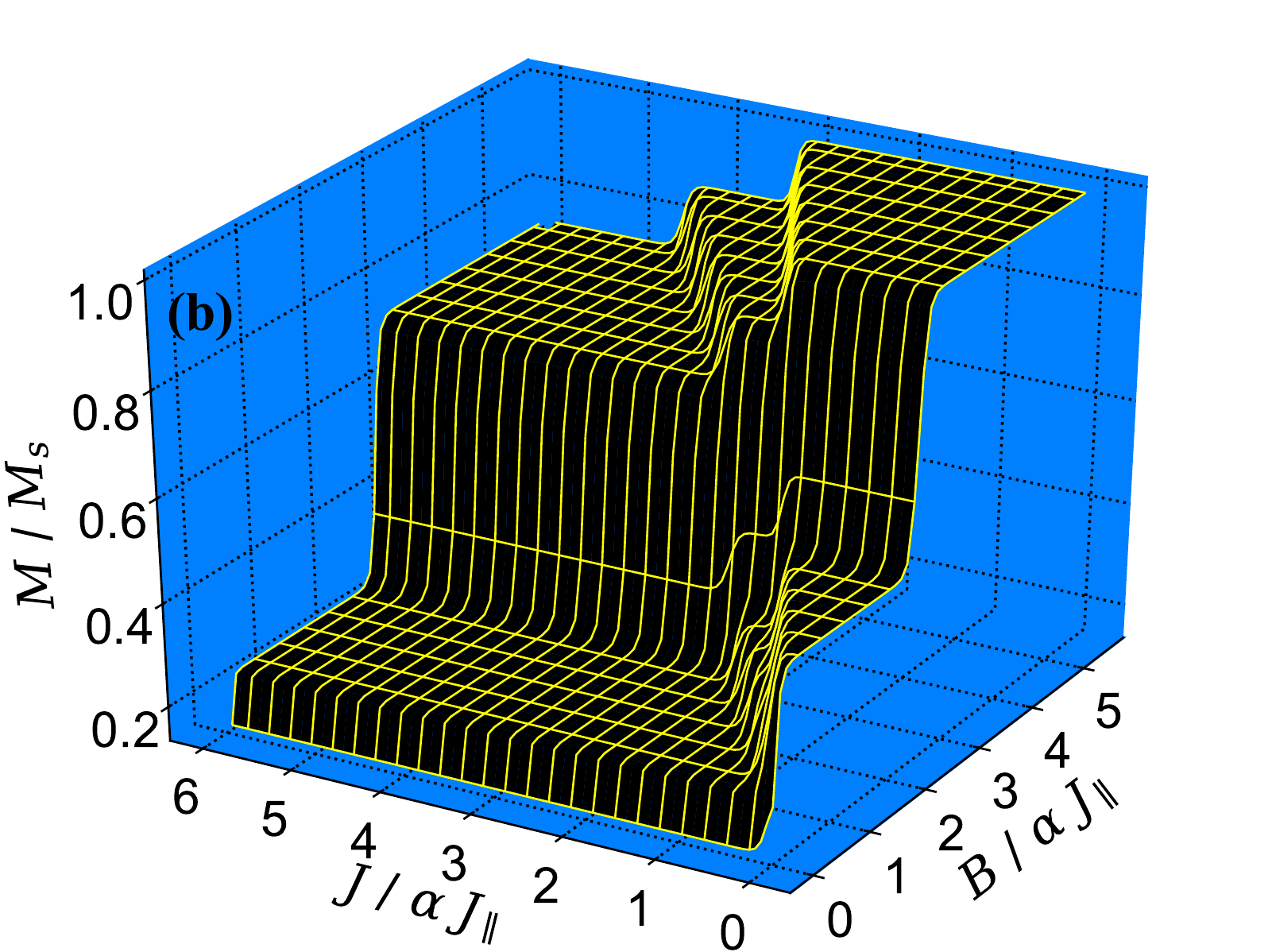}
}
\resizebox{0.3\textwidth}{!}{%
\includegraphics{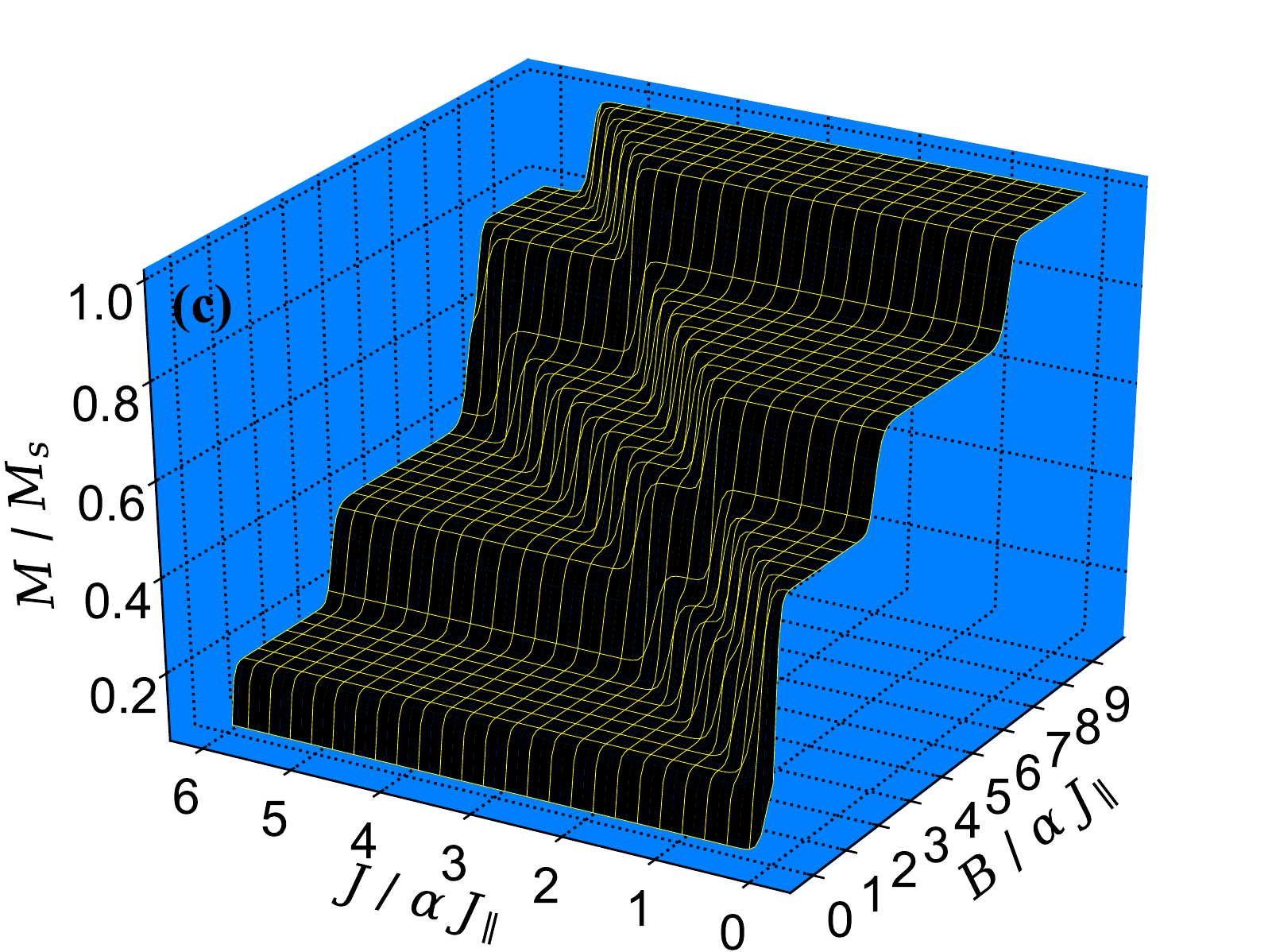}
}
\resizebox{0.3\textwidth}{!}{%
\includegraphics{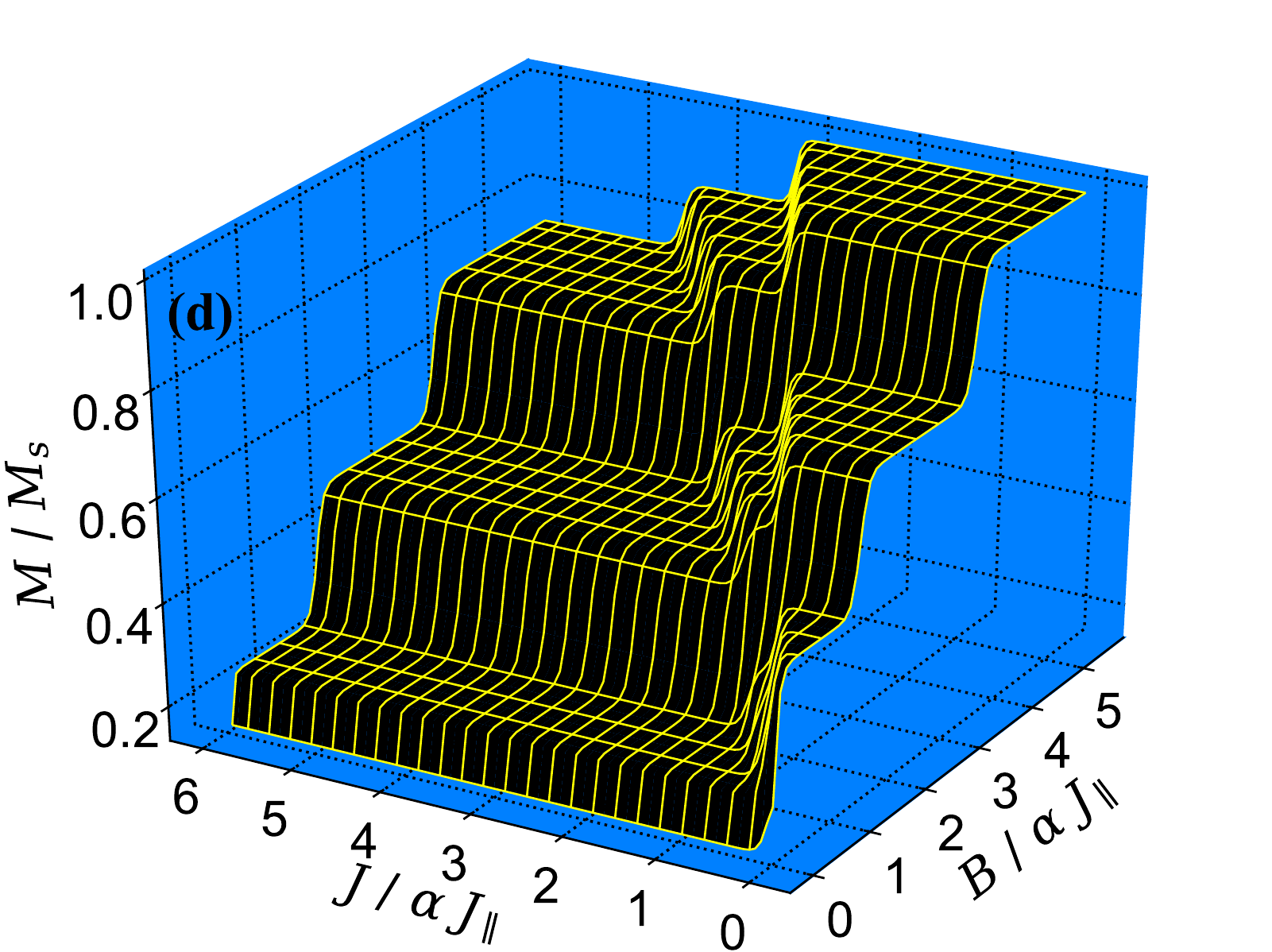}
}
\resizebox{0.3\textwidth}{!}{%
\includegraphics{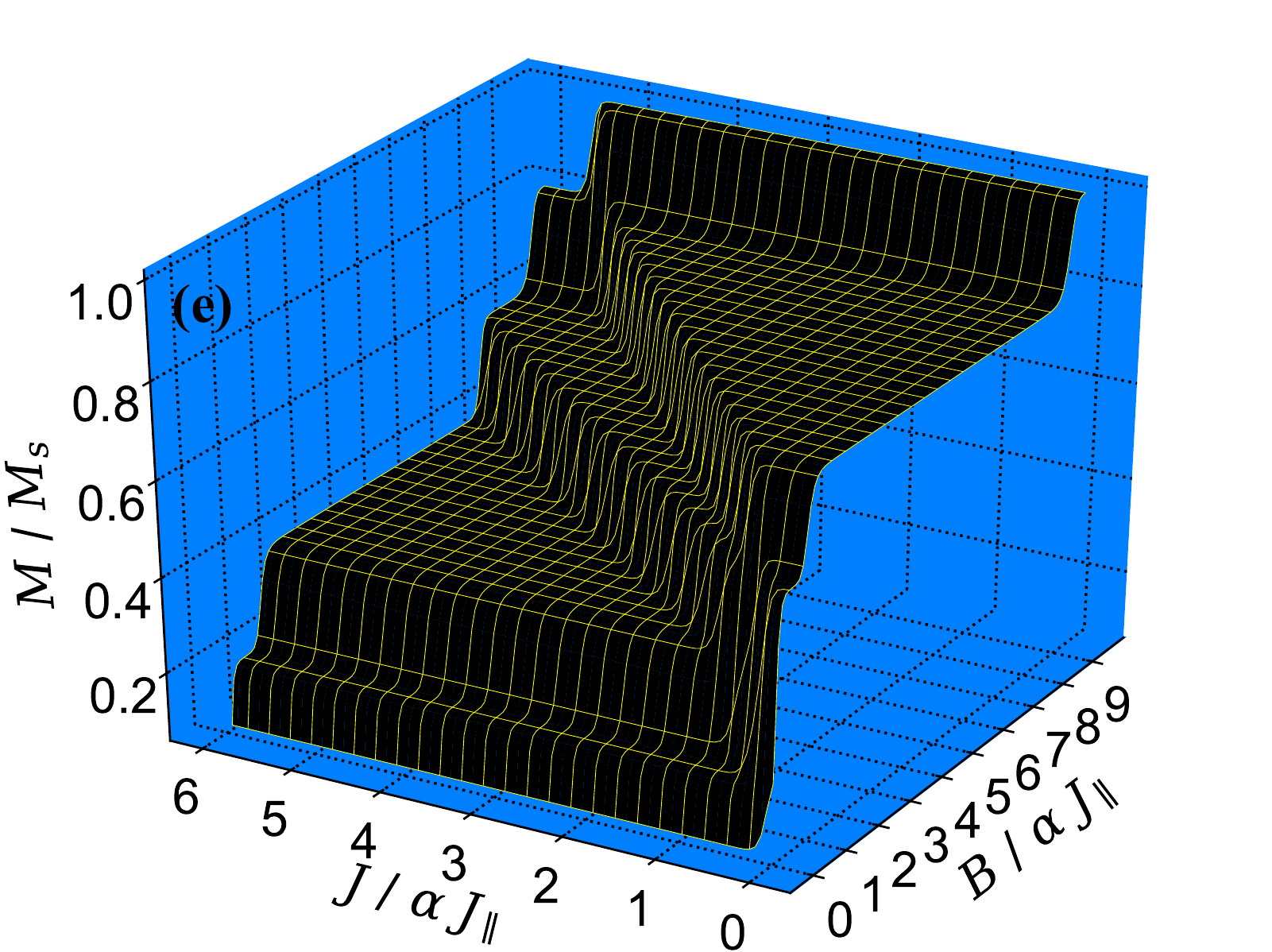}
}
\resizebox{0.3\textwidth}{!}{%
\includegraphics{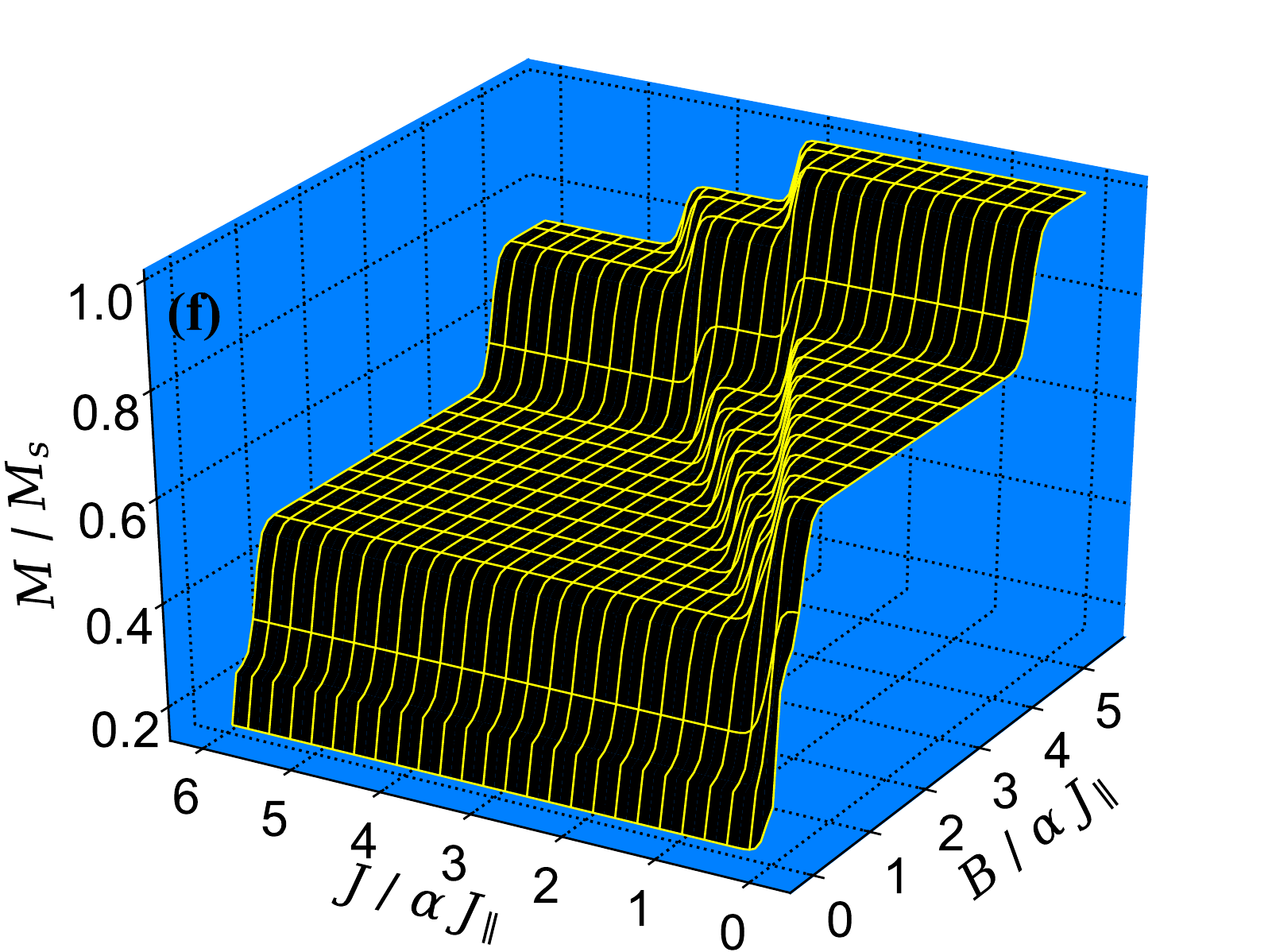}
}
\caption{ 3-D plots of the low-temperature magnetization per its saturation value as a function of the magnetic field $B/\alpha J_{\parallel}$ and the interaction ratio $J/\alpha J_{\parallel}$ where fixed values, $\alpha=0.5$, $\Delta= 0.5$, $\gamma=0.5$, and ${J}_{\perp}/\alpha J_{\parallel}=5$  are assumed. In this plot and next plots we generally consider two different sets for the Land{\'e} g-factors $g_1$, $g_2$ and $g_3$. (a), (c), (e) $\{g_1=1,\; g_2=1,\;g_3=2\}$;  (b), (d), (f) $\{g_1=1,\; g_2=4,\;g_3=2\}$.
Also, three different values of the four-spin Ising term $K/\alpha J_{\parallel}$ have been evaluated. (a) , (b)  $K/\alpha J_{\parallel}=0$; 
 (c) , (d)  $K/\alpha J_{\parallel}=1$;  (e) , (f)  $K/\alpha J_{\parallel}=2$. We here consider $T/\alpha J_{\parallel}=0.12$.}
\label{fig:Mag} 
\end{center}
\end{figure*}

We begin by exploring the low-temperature magnetization process of the mixed spin-(1/2, 1) Ising-XYZ two-leg ladder.
The 3-D magnetization curves in the  ($B/\alpha J-J/\alpha J_{\parallel}$) plane are plotted in Fig. \ref{fig:Mag} for three different fixed values of the special parameter $K/\alpha J_{\parallel}$. In all figures, two distinguished sets of Land{\'e} g-factors $g_1$, $g_2$ and $g_3$ have been optionally assumed. In fact, we formally consider two different versions of spin localization in the body of ladder. First version is assembled, for example, by set $\{g_1=1,\; g_2=1,\;g_3=2\}$, revealing nodal spin-1/2 particles (dark-blue balls in Fig. \ref{figure1}) and Heisenberg dimers  (red balls in Fig. \ref{figure1}) have identical nature. Second version is modulated by set $\{g_1=1,\; g_2=4,\;g_3=2\}$, motivating by consideration all particles in the ladder have different nature. Nevertheless, we enable readers to select a wide range of Land{\'e} g-factors to investigate the model by following our technical procedure.

Panels \ref{fig:Mag}(a), \ref{fig:Mag}(c), \ref{fig:Mag}(e) display the normalized magnetization $M$ with respect to its saturation value $M_s$ for the set $\{g_1=1,\; g_2=1,\;g_3=2\}$, denoting spin-1/2 particles localized on the legs and on the rungs have identical g-factors. Panels \ref{fig:Mag}(b), \ref{fig:Mag}(d), \ref{fig:Mag}(f) illustrate the magnetization for different set of g-factors, i.e.,
$\{g_1=1,\; g_2=4,\;g_3=2\}$ revealing all particles interacted together have different  g-factors. Meanwhile, three different values of the cyclic four-spin Ising interaction $K/\alpha J_{\parallel}$ have been considered. Panels \ref{fig:Mag}(a) , \ref{fig:Mag}(b)  represent 3-D magnetization curve of the model for two different sets of Land{\'e} g-factors such that $K/\alpha J_{\parallel}=0$. In panels \ref{fig:Mag}(c) , \ref{fig:Mag}(d) we consider fixed value $K/\alpha J_{\parallel}=1$, and in panels  \ref{fig:Mag}(e) , \ref{fig:Mag}(f) we have assumed $K/\alpha J_{\parallel}=2$.

\begin{figure*}[t!]
\  \begin{center}
\resizebox{0.45\textwidth}{!}{%
\includegraphics{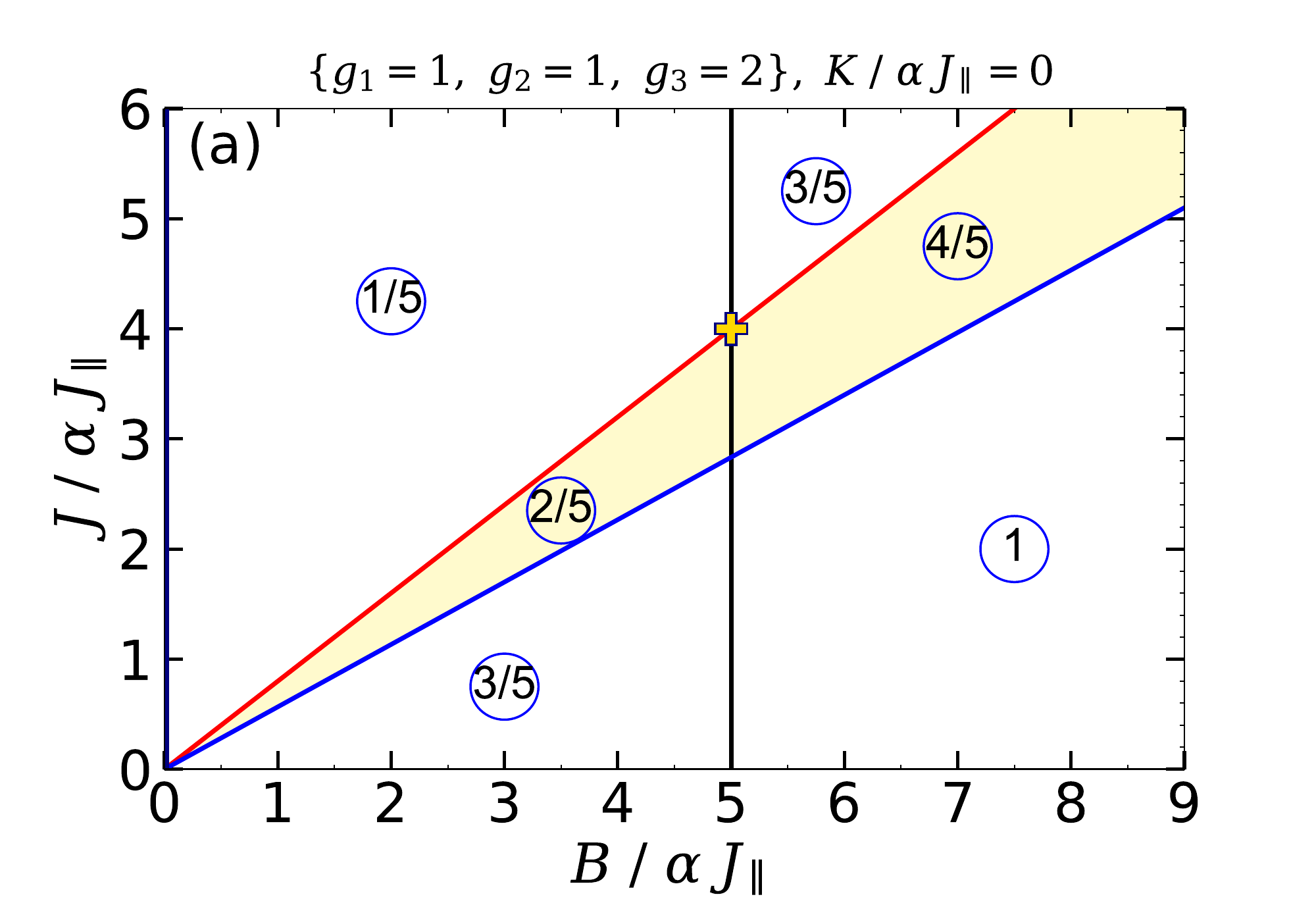}
}
\resizebox{0.45\textwidth}{!}{%
\includegraphics{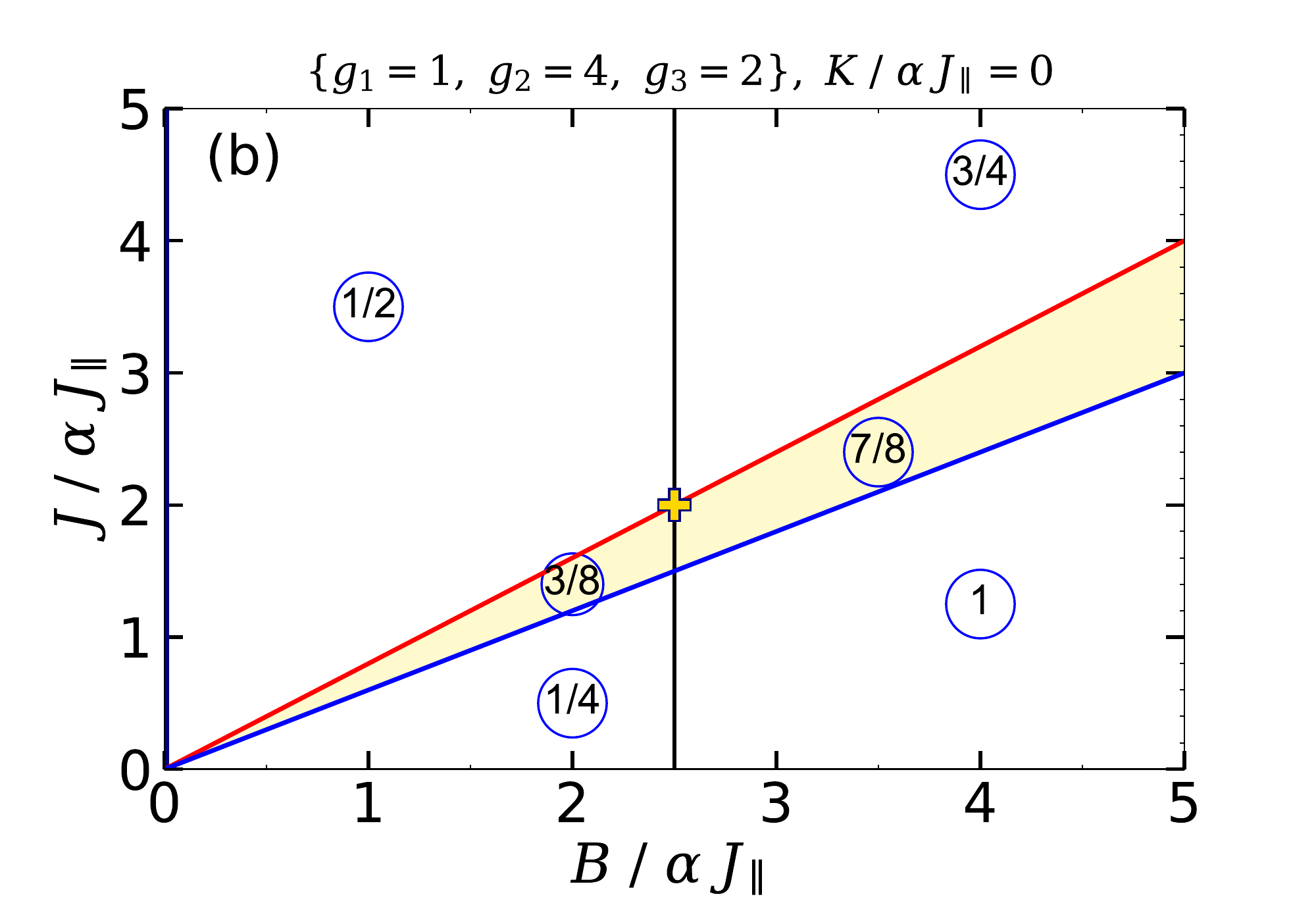}
}
\resizebox{0.45\textwidth}{!}{%
\includegraphics{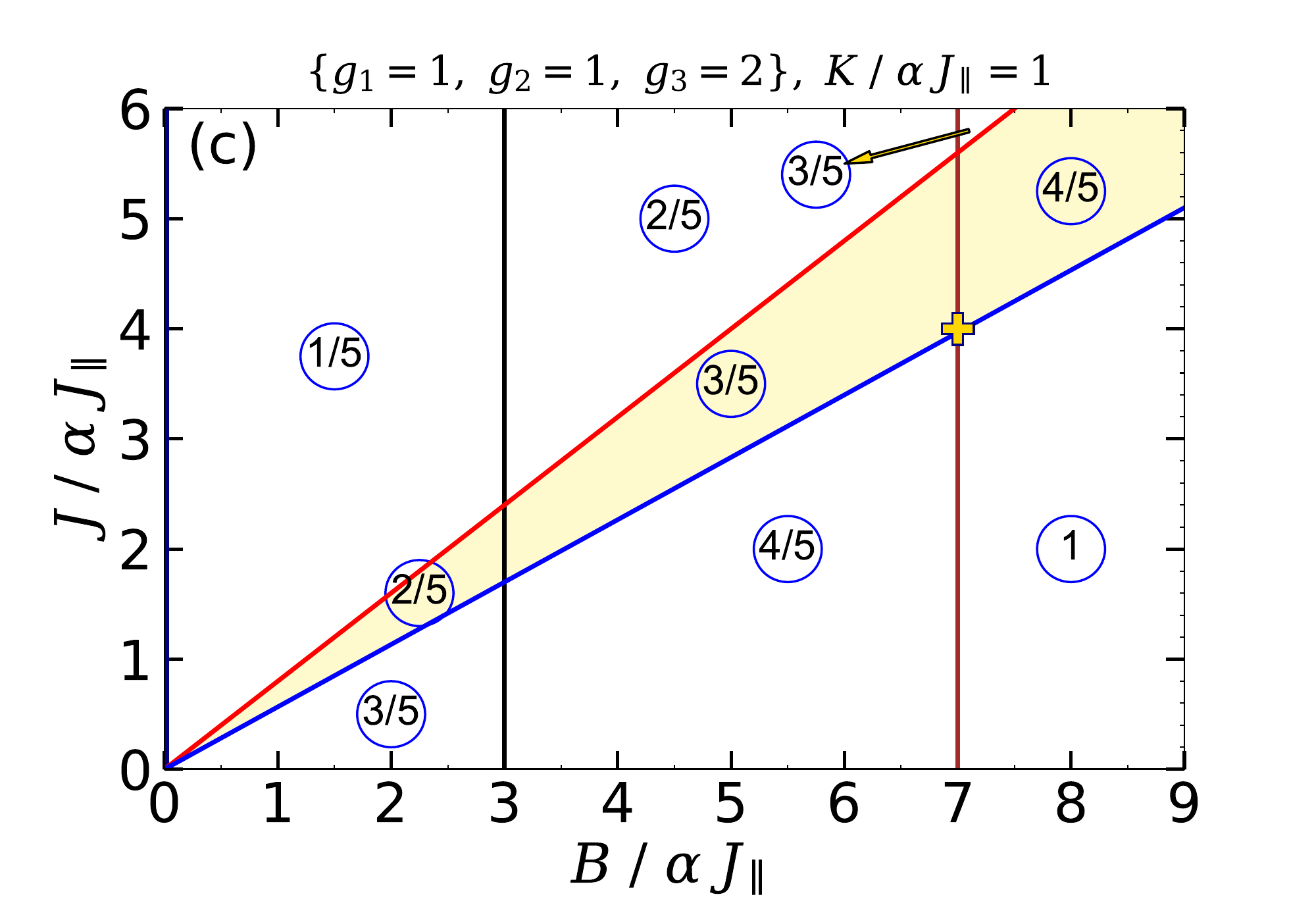}
}
\resizebox{0.45\textwidth}{!}{%
\includegraphics{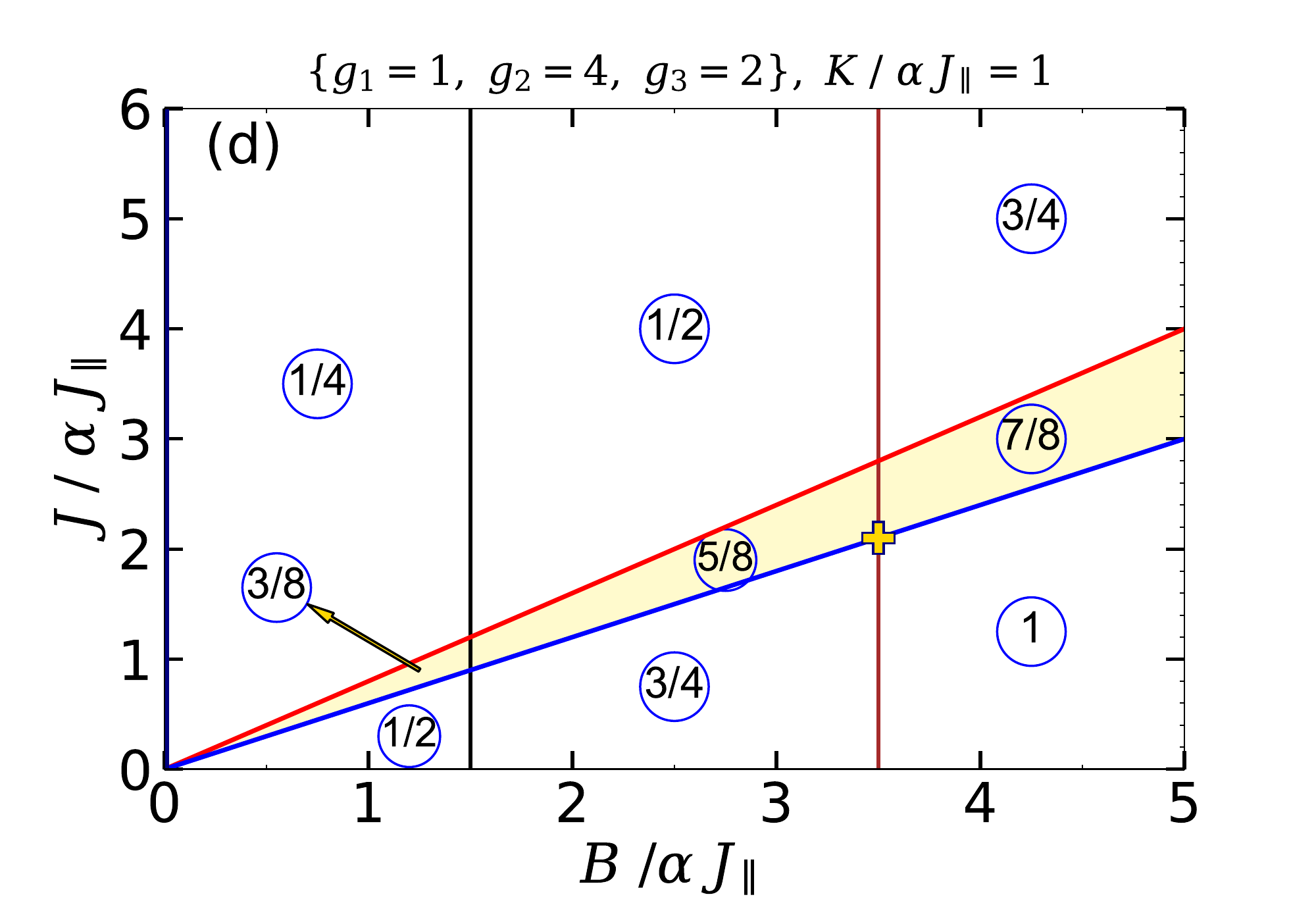}
}
\resizebox{0.45\textwidth}{!}{%
\includegraphics{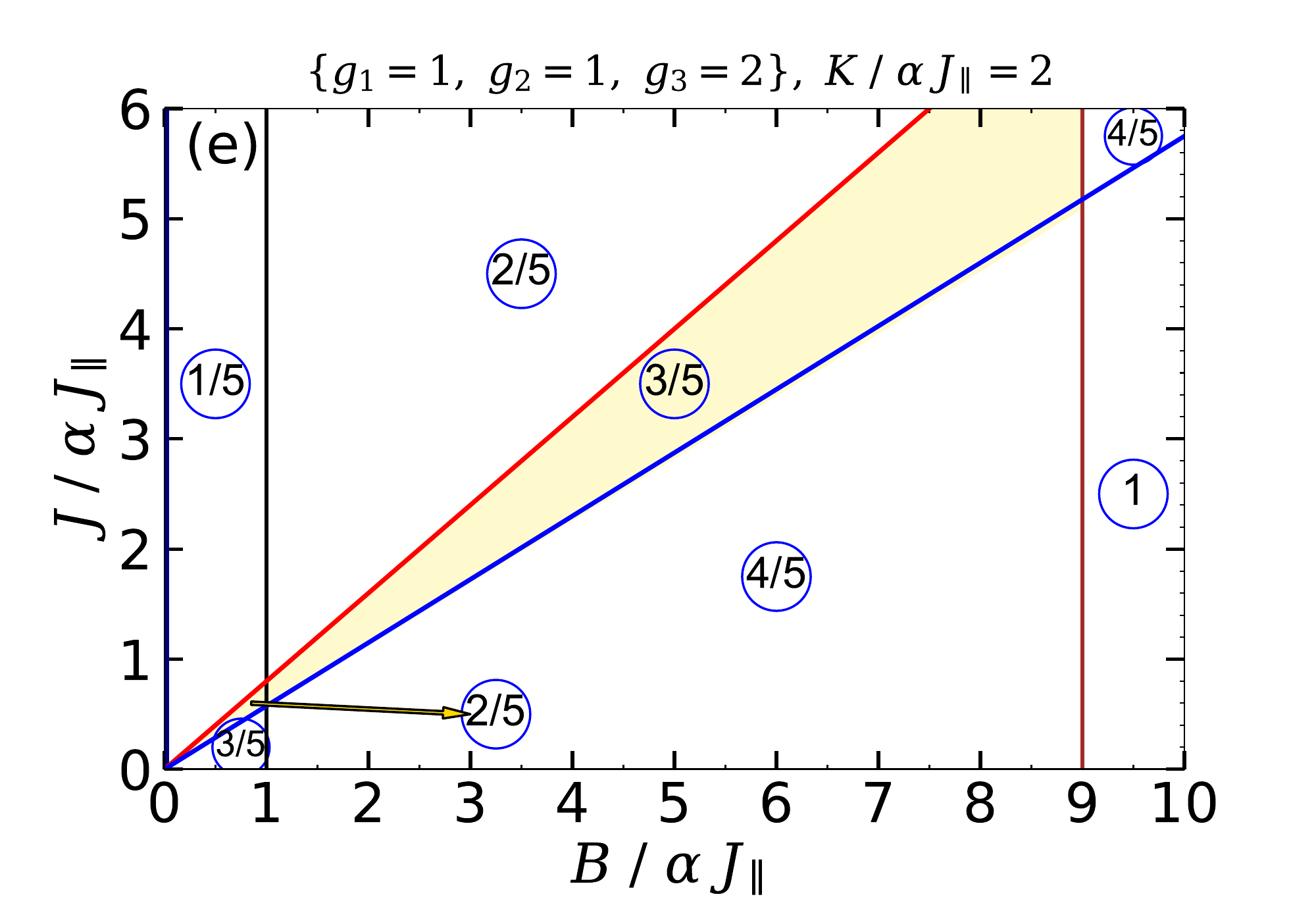} 
}
\resizebox{0.45\textwidth}{!}{%
\includegraphics{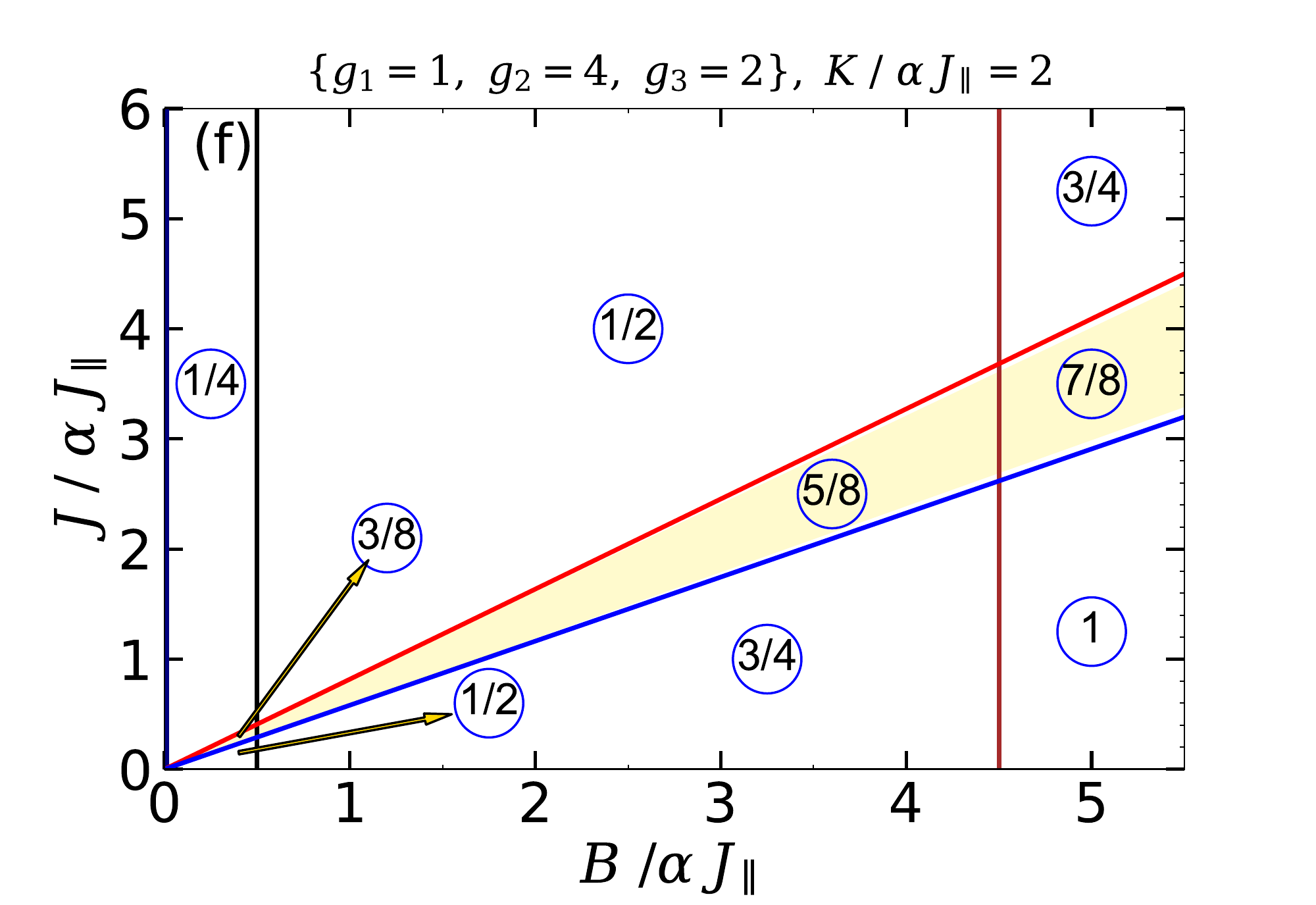}
}
\caption{The ground-state phase diagram of the mixed-spin Ising-XYZ two-leg ladder within the ($B/\alpha J_{\parallel}-J/\alpha J_{\parallel}$) plane for three different values of the ratio $K/\alpha J_{\parallel}$. Solid lines with different colors label discontinuous phase transitions. Shaded area in all figures are the same. Fractional numbers rounded by circles indicate the corresponding magnetization plateau to a given ground-state in units of saturated magnetization $M_s$. For example, 1/5 denotes  intermediate one-fifth plateau, whereas 1/2 represents magnetization one-half plateau, and so on.
 (a)  $\{g_1=1$, $g_2=1$, $g_3=2\}$ and $K/\alpha J_{\parallel}=0$,  (b)  $\{g_1=1$, $g_2=4$, $g_3=2\}$ and $K/\alpha J_{\parallel}=0$.
 (c)  $\{g_1=1$, $g_2=1$, $g_3=2\}$ and $K/\alpha J_{\parallel}=1$,  (d)  $\{g_1=1$, $g_2=4$, $g_3=2\}$ and $K/\alpha J_{\parallel}=1$. 
 (e)  $\{g_1=1$, $g_2=1$, $g_3=2\}$ and $K/\alpha J_{\parallel}=2$,  (f)  $\{g_1=1$, $g_2=4$, $g_3=2\}$ and $K/\alpha J_{\parallel}=2$.
    The same set of other parameters to Fig. \ref{fig:Mag} have been conceived.}
\label{fig:QPT_BJ}
\end{center}
\end{figure*}

\begin{figure*}[t!]
\begin{center}
\resizebox{0.45\textwidth}{!}{%
\includegraphics{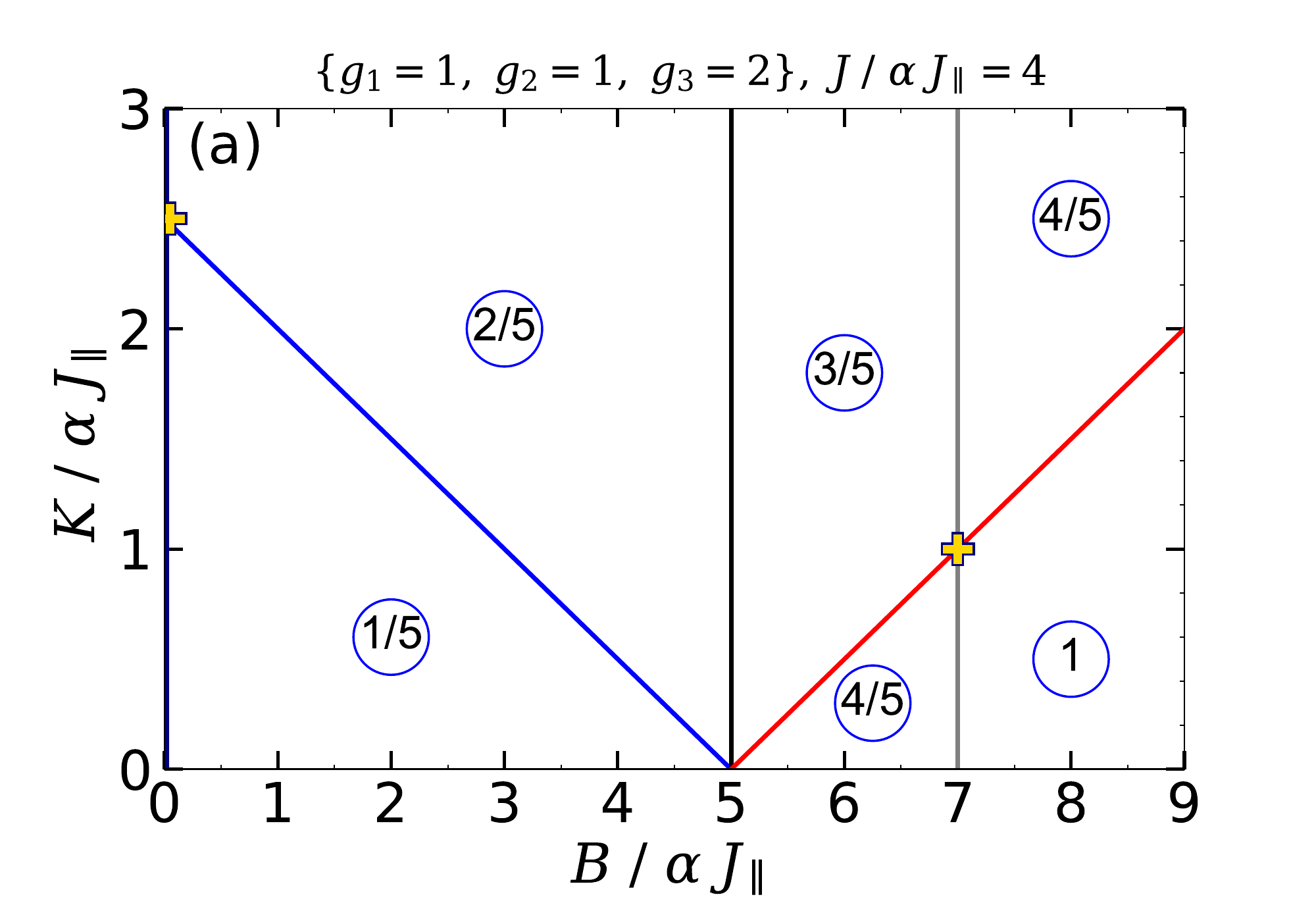}
}
\resizebox{0.45\textwidth}{!}{%
\includegraphics{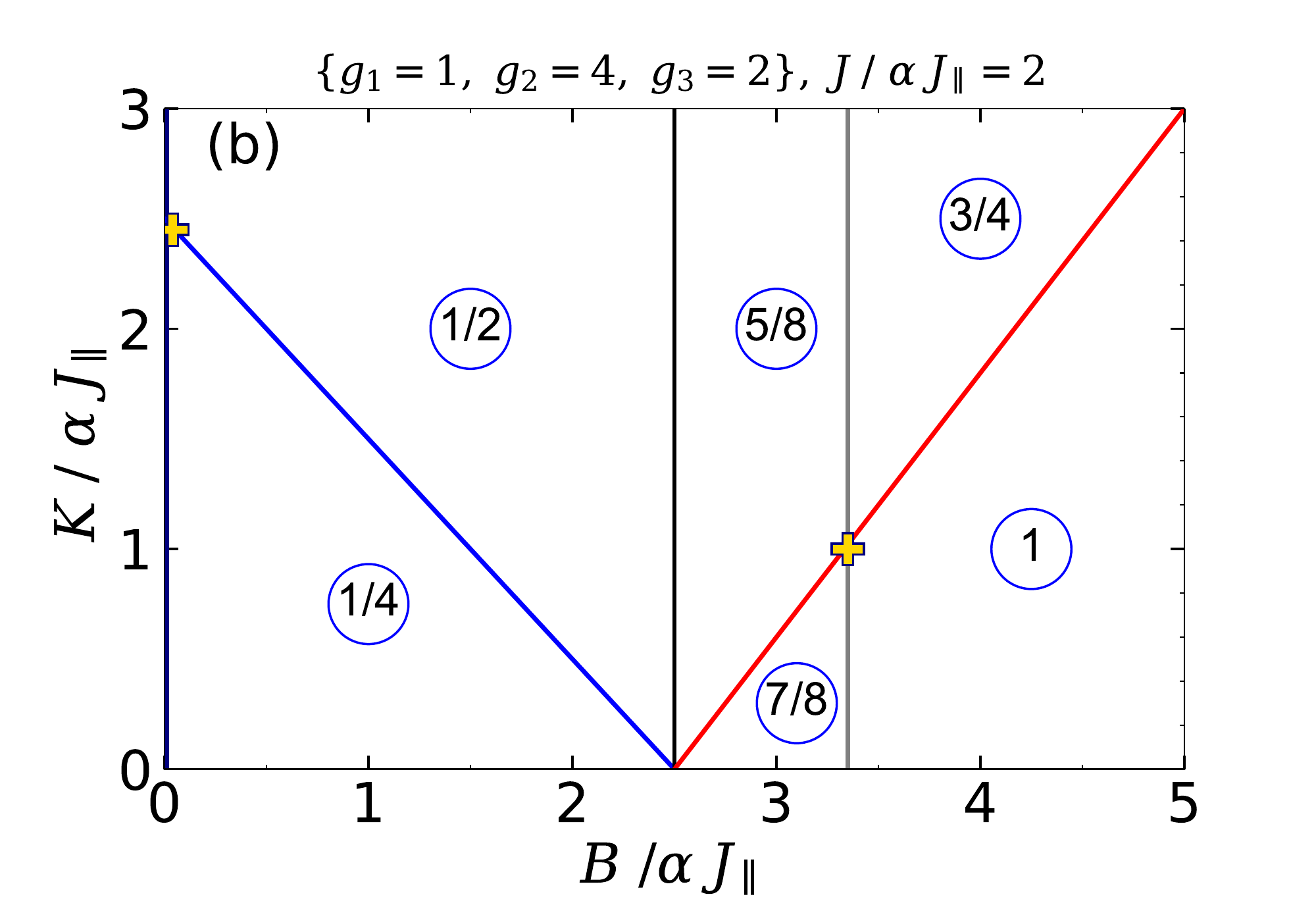}
}
\caption{First-order zero-temperature phase transition of the model within the ($B/\alpha J_{\parallel}-K/\alpha J_{\parallel}$) plane. (a)  $\{g_1=1$, $g_2=1$, $g_3=2\}$ and $J/\alpha J_{\parallel}=4$,  (b)  $\{g_1=1$, $g_2=4$, $g_3=2\}$ and $J/\alpha J_{\parallel}=2$. Other parameters of the Hamiltonian have been taken as for Fig. \ref{fig:Mag}. Filled pluses display the intersection of multiple ground states, where the degeneracy between them occurs.}
\label{fig:QPT_BK}
\end{center}
\end{figure*}

Generally speaking, as it is illustrated in Fig. \ref{fig:Mag}, the alterations of cyclic four-spin Ising interaction has substantial influences on the magnetization behavior in  ($B/\alpha J_{\parallel}-J/\alpha J_{\parallel}$) plane. 
We discuss this stimulating medium in Fig. \ref{fig:QPT_BJ}.
Possible magnetic ground states of the mixed spin-(1/2,1) Ising-Heisenberg two-leg ladder can be found by changing in magnetization behavior, depending on the mutual interplay between the model parameters $\Delta^{-}$, $\Delta^{+}$, $J_{\perp}/\alpha J_{\parallel}$ and $K/\alpha J_{\parallel}$. It is argued in Figs. \ref{fig:QPT_BJ} and \ref{fig:QPT_BK}, the ground-state phase diagram in the ($B/\alpha J_{\parallel}-J/\alpha J_{\parallel}$)  plane and  ($B/\alpha J_{\parallel}-K/\alpha J_{\parallel}$) plane, respectively. In this study, we focus on the interplay between the ratio $K/\alpha J_{\parallel}$  and zero-temperature phase spectra of the model. 
 
Figure \ref{fig:QPT_BJ} (a) illustrates the zero-temperature phase diagram of the  ladder  in   ($B/\alpha J_{\parallel}-J/\alpha J_{\parallel}$) plane for $K/\alpha J_{\parallel}=0$ and g-factors set $\{g_1=1$, $g_2=1$, $g_3=2\}$. Other parameters have been assumed to be $\alpha=0.5$, $\Delta= 0.5$, $\gamma=0.5$, and ${J}_{\perp}/\alpha J_{\parallel}=5$. The later assumption denotes the interaction between spin-1/2 and spin-1 particles localized on the Ising rungs of the ladder to be strong-rung  ferromagnetic interaction. A notable remark from Fig.  \ref{fig:QPT_BJ} is that  level-crossing magnetic field shown by red and blue lines in all panels has identical gradients with a linear dependence on the exchange interaction $J/\alpha J_{\parallel}$. 
The model, independent of the quantity $K/\alpha J_{\parallel}$, presents in the low-temperature magnetization curve an instant magnetization jump from zero to an intermediate plateau normalized with respect to the saturation magnetization.  

It can be seen from Fig. \ref{fig:QPT_BJ} (a) that for the case $K/\alpha J_{\parallel}=0$ when the set of g-factors $\{g_1=1$, $g_2=1$, $g_3=2\}$ is considered, the model may also exhibits several intermediate plateaux such as $(1/5)-$plateau, $(2/5)-$plateau, $(3/5)-$plateau and $(4/5)-$plateau of the saturation value $M_s$. On the other hand, as illustrated in Fig. \ref{fig:QPT_BJ} (b), when the set $\{g_1=1$, $g_2=4$, $g_3=2\}$ is assumed, the model reproduces magnetic ground-state phase spectra corresponding to the magnetization $(1/4)-$plateau, $(3/8)-$plateau, $(1/2)-$plateau, $(3/4)-$plateau and $(7/8)-$plateau of the saturation value (the same fixed values of all parameters to Fig. \ref{fig:Mag}(b) are supposed). Filled-plus marks in both panels \ref{fig:QPT_BJ} (a) and \ref{fig:QPT_BJ} (b) demonstrate the co-ordinates of critical exchange interactions $J/\alpha J_{\parallel}=4$ and $J/\alpha J_{\parallel}=2$ at which four ground-states become
degenerate. Although, each point of the both blue and red lines are fascinating to count in our investigations,  on an optional basis we choose two introduced quadruple points such that they play the most important role to continue our studies on the quantum correlation and the thermodynamics of the model. 

A deep insight into the nature of different phase boundaries can be obtained by considering the typical cyclic four-spin Ising interaction in the magnetization process. With this in mind, we have plotted in Figs. \ref{fig:QPT_BJ} (c)-\ref{fig:QPT_BJ} (f) the ground-state phase diagram when the system involves an additional Ising term $K/\alpha J_{\parallel}\neq 0$. Surprisingly, the phase boundaries undergo substantial changes. For instance, when $K/\alpha J_{\parallel}> 0$ for the set $\{g_1=1$, $g_2=4$, $g_3=2\}$ an intermediate $(5/8)-$plateau is visible in the magnetization curve (Fig. \ref{fig:QPT_BJ} (d)).

To bring an insight into how the cyclic four-spin Ising term play its  rule to confine  magnetic ground-state phase boundaries, we plot in Fig. \ref{fig:QPT_BK} the possible ground-state phase diagram in the  ($B/\alpha J_{\parallel}-K/\alpha J_{\parallel}$) plane, where the co-ordinates of two aforementioned quadruple points have been optionally taken as fixed exchange interactions. To verify this point,  in Fig.  \ref{fig:QPT_BK}(a), is plotted the ground-state phase diagram when the set of  Land{\'e} g-factors $\{g_1=1$, $g_2=1$, $g_3=2\}$ and fixed
 $J/\alpha J_{\parallel}= 4$ are assumed. Figure \ref{fig:QPT_BK}(b) demonstrates the same theme but for the set of  $\{g_1=1$, $g_2=4$, $g_3=2\}$ and fixed $J/\alpha J_{\parallel}= 2$.

  As a result, one perceives from both panels of Fig. \ref{fig:QPT_BK} that there are three critical points at which the boundaries of two or more ground states cut each other off, revealing that they become degenerate. The co-ordinates of these fancy points are as $\{B/\alpha J_{\parallel}=0,\;K/\alpha J_{\parallel}=2.5\}$, $\{B/\alpha J_{\parallel}=5,\;K/\alpha J_{\parallel}=0\}$ and $\{B/\alpha J_{\parallel}=7,\;K/\alpha J_{\parallel}=1\}$ (see panel \ref{fig:QPT_BK}(a)). In a different fashion, as shown in panel \ref{fig:QPT_BK}(b), we find the same critical points for the cyclic four-spin Ising term at the special co-ordinates $\{B/\alpha J_{\parallel}=0,\;K/\alpha J_{\parallel}=2.5\}$, $\{B/\alpha J_{\parallel}\approx 3.4,\;K/\alpha J_{\parallel}=1\}$. In what follows, we will also focus on the special critical point $K/\alpha J_{\parallel}=1$ as a quadruple point to investigate the thermodynamics of the model in different situations.
\subsection{Specific heat}
\begin{figure*}[t!]
\begin{center}
\resizebox{0.45\textwidth}{!}{%
 \includegraphics[trim=30 1 100 1, clip]{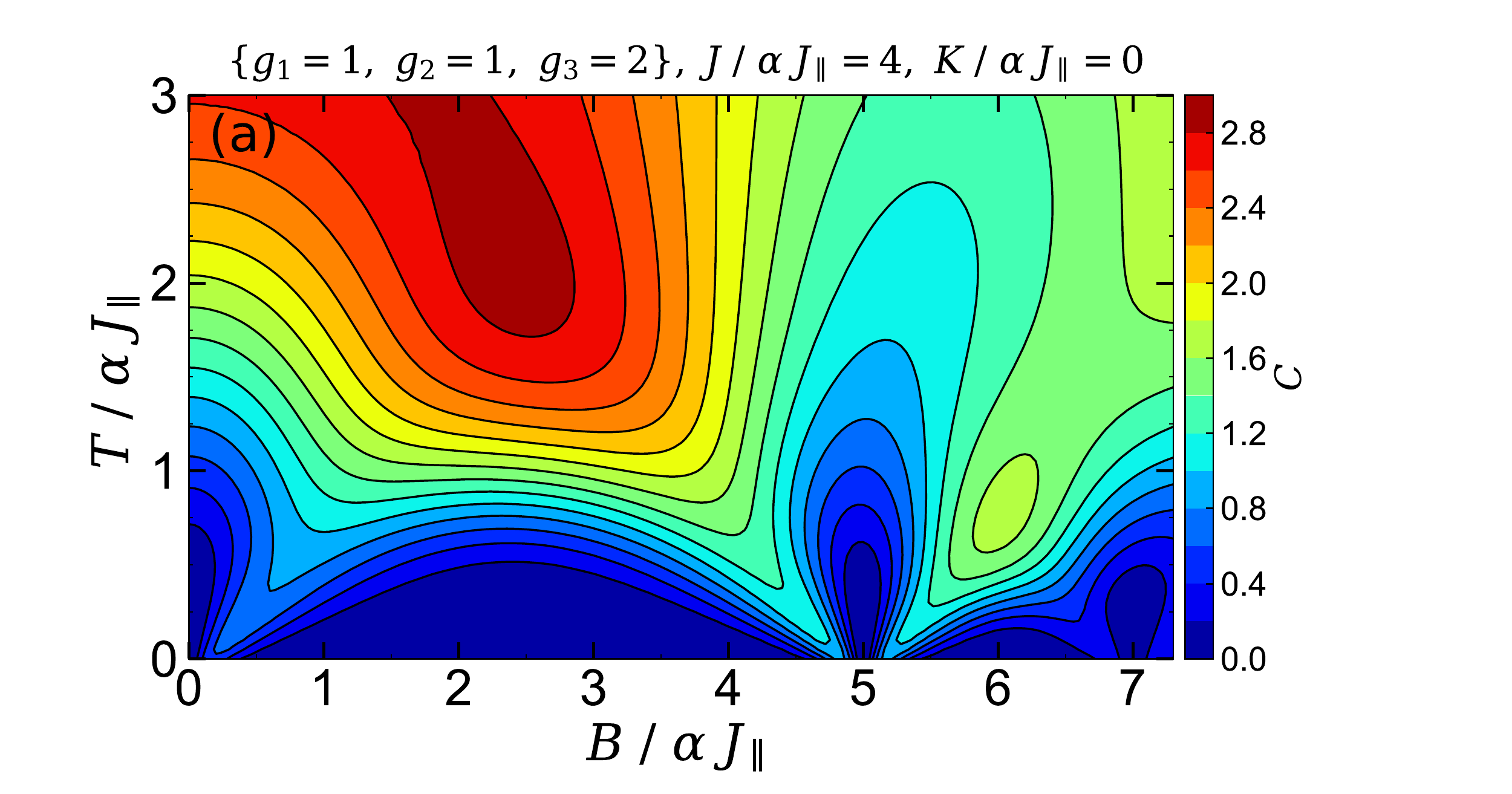} 
}
\resizebox{0.45\textwidth}{!}{%
\includegraphics[trim=30 1 100 1, clip]{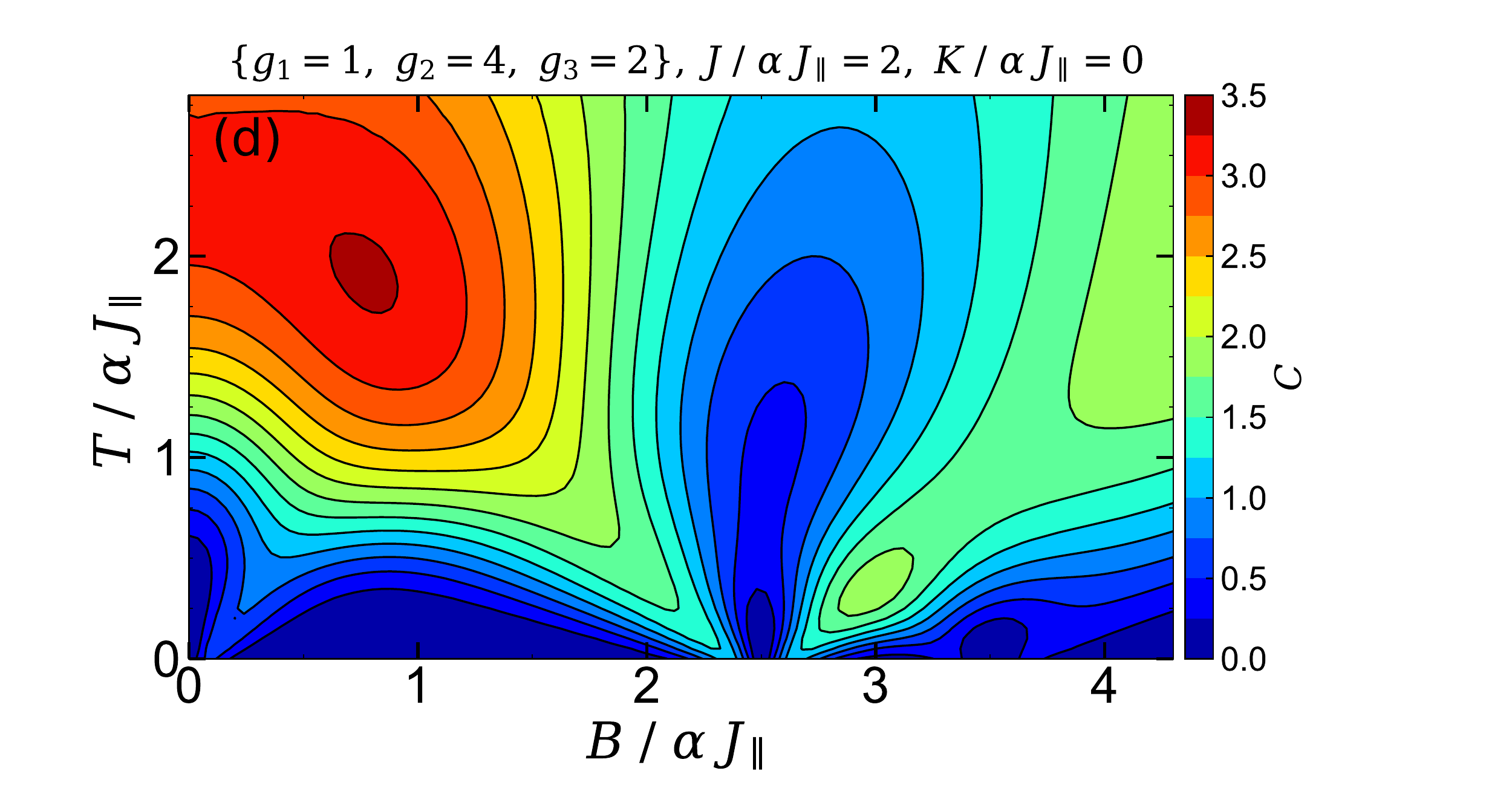}
}
\resizebox{0.45\textwidth}{!}{%
\includegraphics[trim=30 1 50 1, clip]{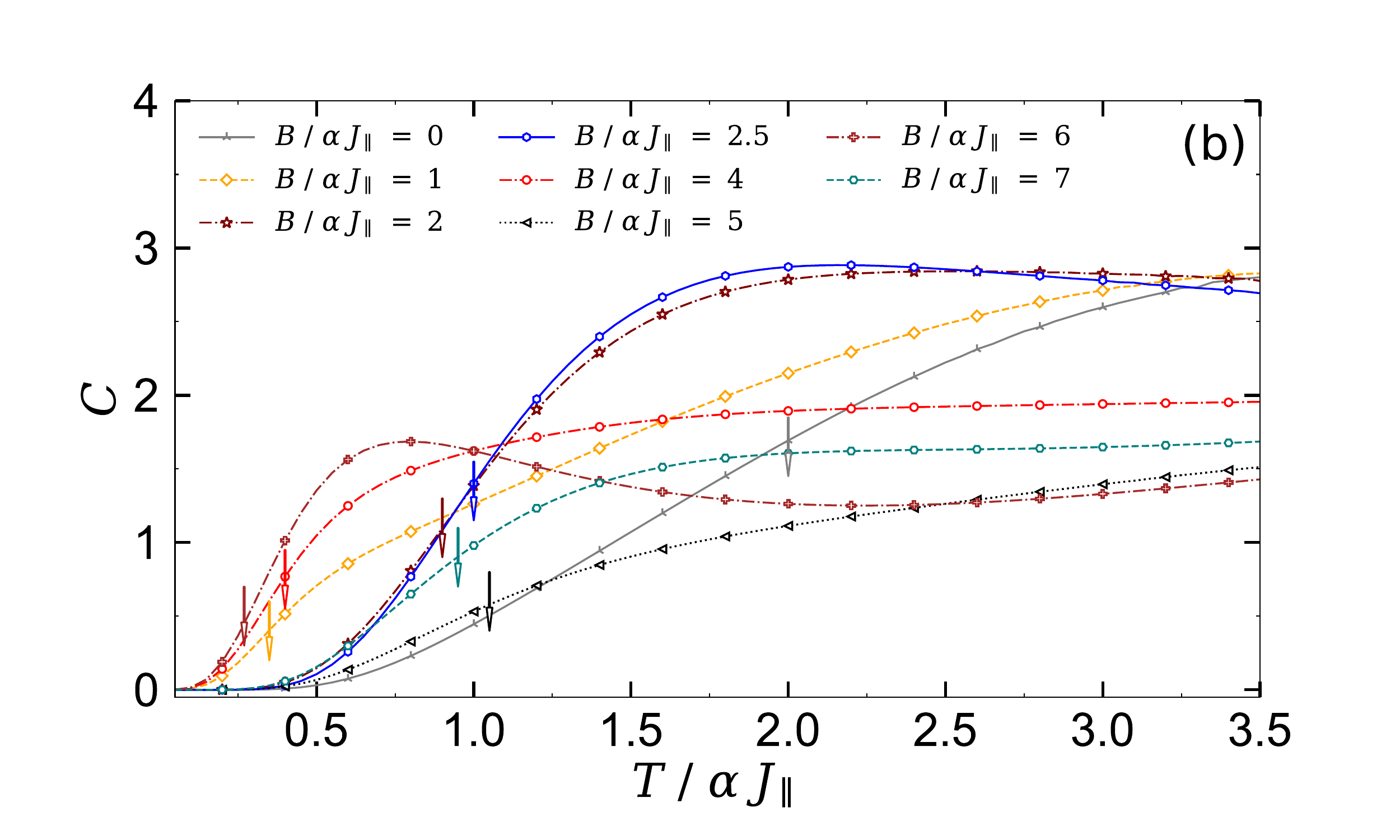}
}
\resizebox{0.45\textwidth}{!}{%
\includegraphics[trim=30 1 50 1, clip]{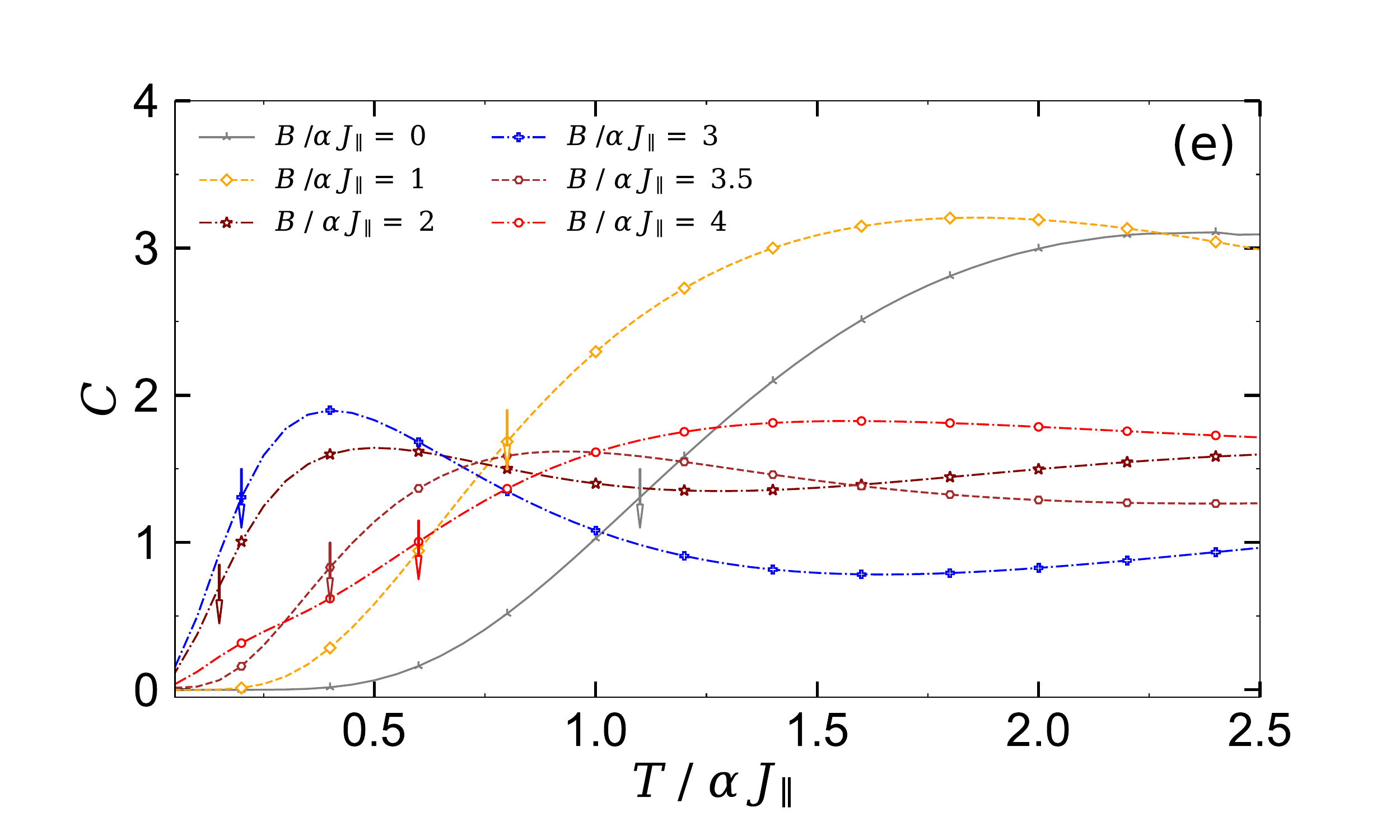}
}
\resizebox{0.45\textwidth}{!}{%
\includegraphics[trim=20 1 50 1, clip]{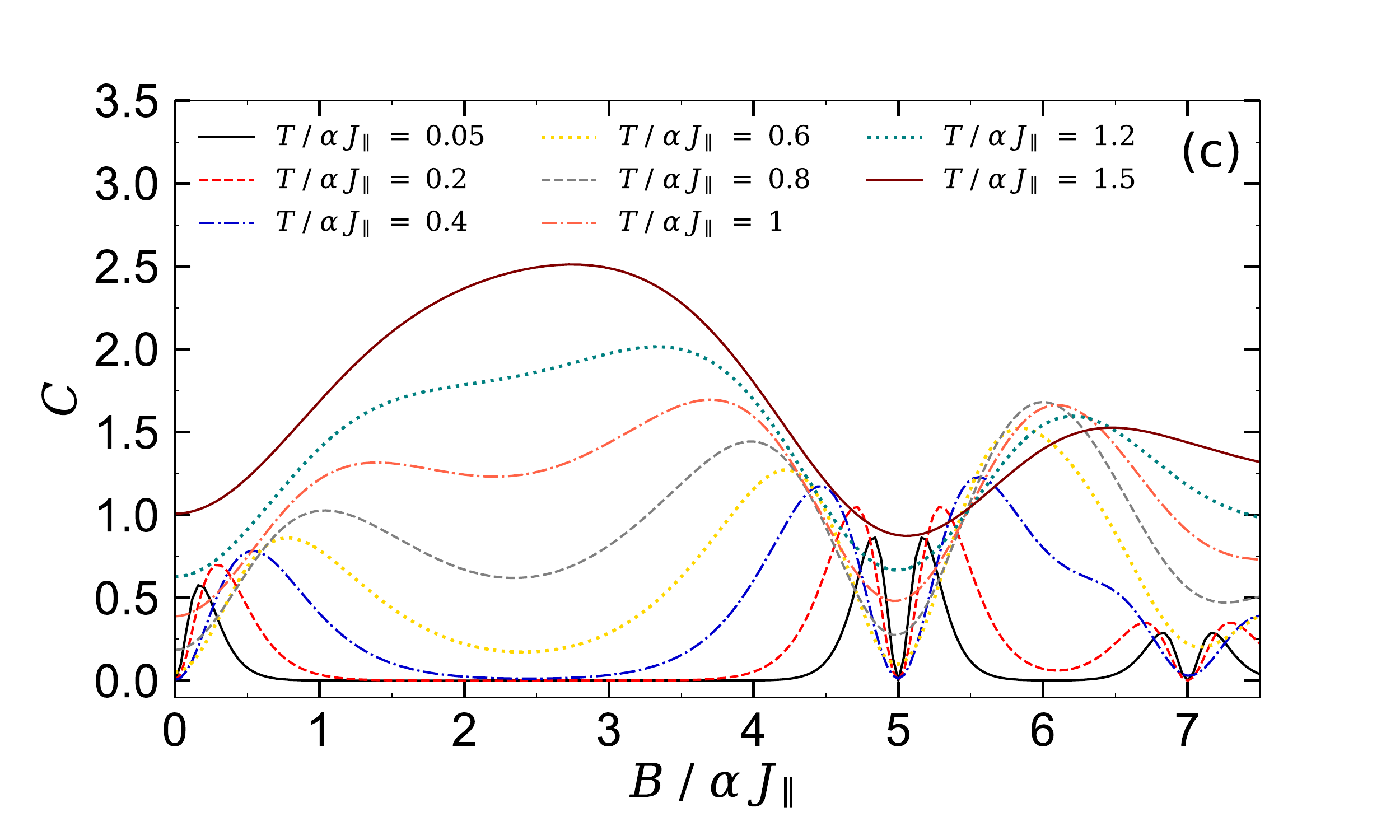}
}
\resizebox{0.45\textwidth}{!}{%
\includegraphics[trim=20 1 50 1, clip]{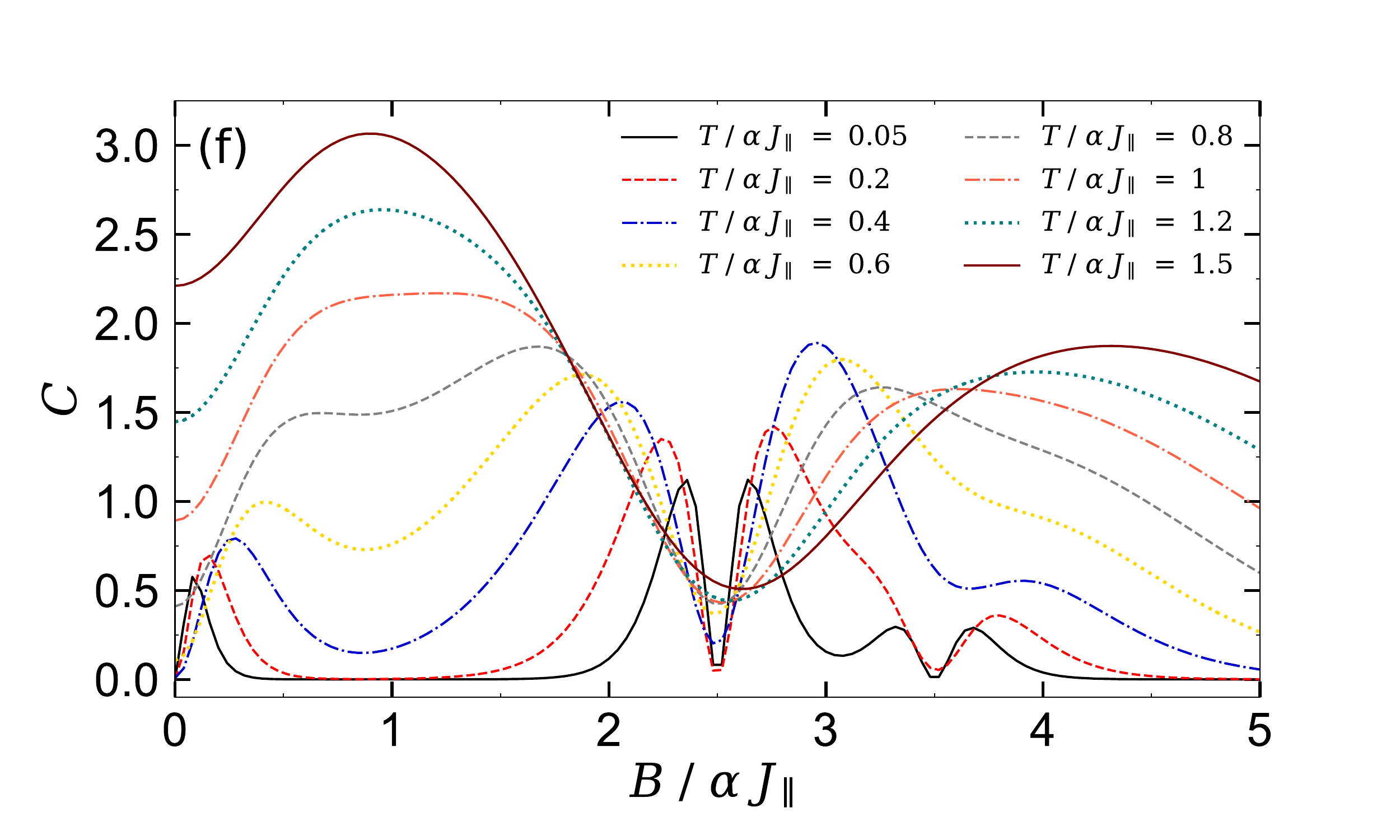} 
} 
\caption{ (a) Contour plot of the specific heat in the  ($B/\alpha J_{\parallel}-T/\alpha J_{\parallel}$) plane for the set  $\{g_1=1$, $g_2=1$, $g_3=2\}$ and fixed value $J/\alpha J_{\parallel}=4$. (b) The temperature dependence of the specific heat for several fixed magnetic fields when the condition $\{g_1=1$, $g_2=1$, $g_3=2\}$ and  $J/\alpha J_{\parallel}=4$ is assumed. Arrows illustrate the transition temperature at which the specific heat has a steep increasing to make a maximum.
(c) The specific heat as a function of the magnetic field for a number of selected temperatures under the condition $\{g_1=1$, $g_2=1$, $g_3=2\}$ and  $J/\alpha J_{\parallel}=4$. (d) Contour plot of the specific heat for the arbitrary set $\{g_1=1$, $g_2=4$, $g_3=2\}$ and fixed value $J/\alpha J_{\parallel}=2$.  (e) The specific heat curve versus temperature for various fixed values of the magnetic field when the condition $\{g_1=1$, $g_2=4$, $g_3=2\}$ and fixed $J/\alpha J_{\parallel}=2$ are hypothesized. (f) The specific heat as a function of the magnetic field for a number of selected temperatures, visualizing the set $\{g_1=1$, $g_2=4$, $g_3=2\}$ and fixed value $J/\alpha J_{\parallel}=2$. 
In all panels is assumed the zero value for the cyclic four-spin Ising term, i.e., $K/\alpha J_{\parallel}=0$, and other parameters have been taken as Fig. \ref{fig:Mag}. }
\label{fig:SHeat1}
\end{center}
\end{figure*}

Let us continue our discussion with  the all dependencies of the specific heat of the mixed-spin  Ising-XYZ  ladder in the
 ($B/\alpha J_{\parallel}-T/\alpha J_{\parallel}$) plane by considering the co-ordinates of the aforedescribed critical points as fixed values for interaction parameters  $J/\alpha J_{\parallel}$ and $K/\alpha J_{\parallel}$. To this end, we display in Figs. \ref{fig:SHeat1} (a) and  \ref{fig:SHeat1} (d) contour plot of the specific heat as a function of the temperature and the magnetic field with $K/\alpha J_{\parallel}=0$ by supposing two conditions  $\{g_1=1$, $g_2=1$, $g_3=2\}$, $J/\alpha J_{\parallel}=4$ , and $\{g_1=1$, $g_2=4$, $g_3=2\}$, 
$J/\alpha J_{\parallel}=2$, respectively. By inspecting Fig. \ref{fig:SHeat1} (b)  one can realize that there is an anomalous Schottky maximum arisen at high temperatures ($T/\alpha J_{\parallel}> 1$) and moderate magnetic fields  ($1<B/\alpha J_{\parallel}< 3$). 
With increase in the magnetic field is demonstrated a smaller peak at, respectively, low temperatures. The transition temperature could be fixed in the temperature intervals in which the specific heat curve has a steep increase and changes much, as we illustrated by upside down arrows.

Under the situation $\{g_1=1$, $g_2=4$, $g_3=2\}$, $J/\alpha J_{\parallel}=2$ (Fig. \ref{fig:SHeat1} (e)), the typical  Schottky maximum appears at lower magnetic fields.

Another interesting phenomenon observed at sufficiently low temperatures is that the contour lines of the specific heat are remarkably accumulated nearby the critical magnetic fields at which magnetization jump occurs. It is quite noteworthy that we witness a huge accumulation of contour lines close to the co-ordinates of quadruple point (filled-plus marks in Fig. \ref{fig:QPT_BJ}) rather than other critical points. For more clarity, we plot in Figs. \ref{fig:SHeat1} (c) and \ref{fig:SHeat1} (f) the specific heat as a function of the magnetic field for several fixed values of the temperature where other parameters have been set as panels \ref{fig:SHeat1} (a) and \ref{fig:SHeat1} (d), respectively. One can find that the specific heat becomes minimum for a wide range of the temperature at the critical magnetic fields, for example $B_c/\alpha J_{\parallel}=\{0,\; 5,\; 7\}$ when the situation $\{g_1=1$, $g_2=1$, $g_3=2\}$, $J/\alpha J_{\parallel}=4$, $K/\alpha J_{\parallel}=0$ is considered. Under different condition $\{g_1=1$, $g_2=4$, $g_3=2\}$, $J/\alpha J_{\parallel}=2$, $K/\alpha J_{\parallel}=0$, the specific heat minima occur at the critical points  $B_c/\alpha J_{\parallel}=\{0,\; 2.5,\; 3.5\}$.

\begin{figure*}[t!]
\begin{center}
\resizebox{0.45\textwidth}{!}{%
 \includegraphics[trim=30 1 100 1, clip]{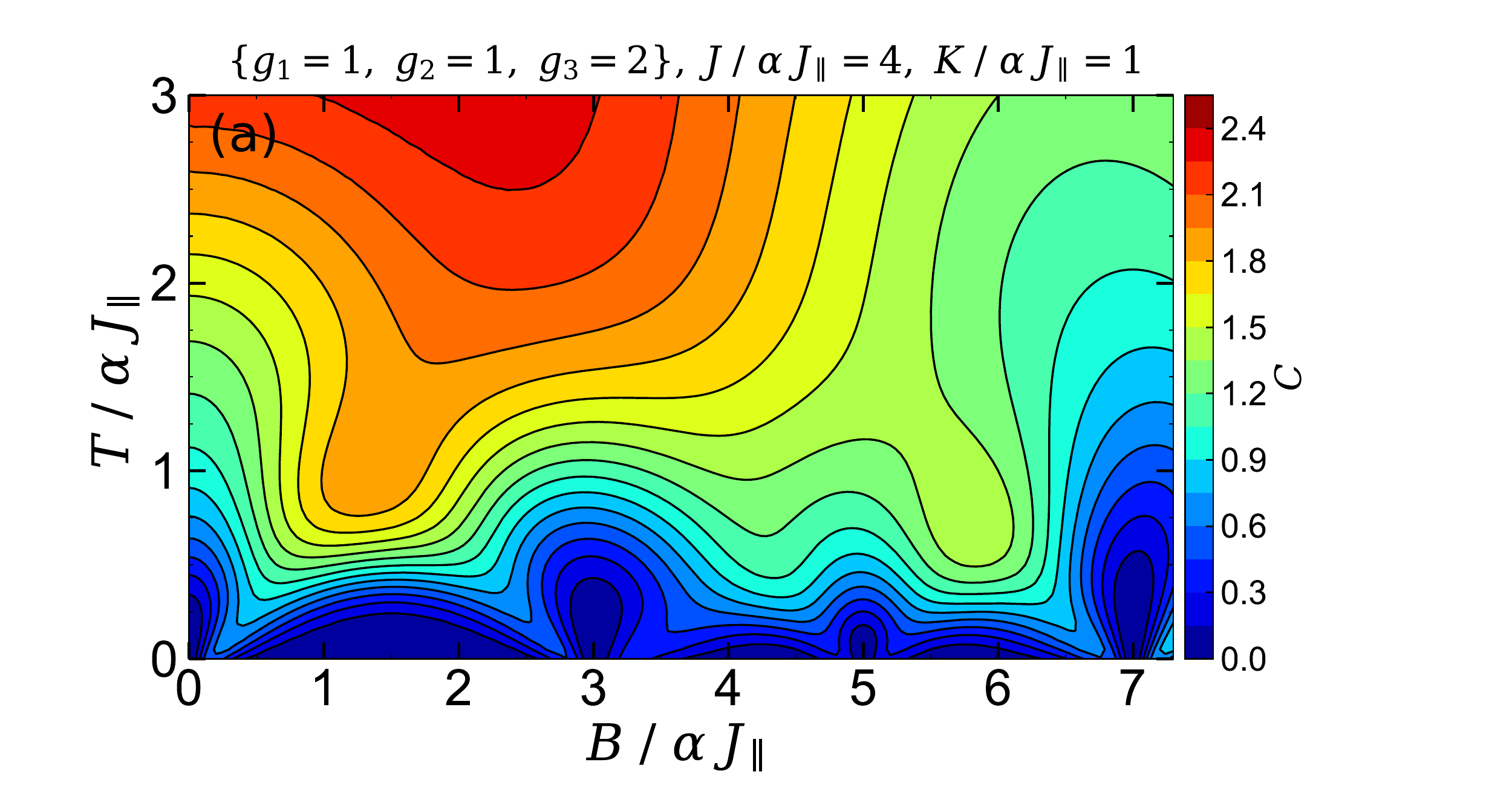} 
}
\resizebox{0.45\textwidth}{!}{%
\includegraphics[trim=30 1 100 1, clip]{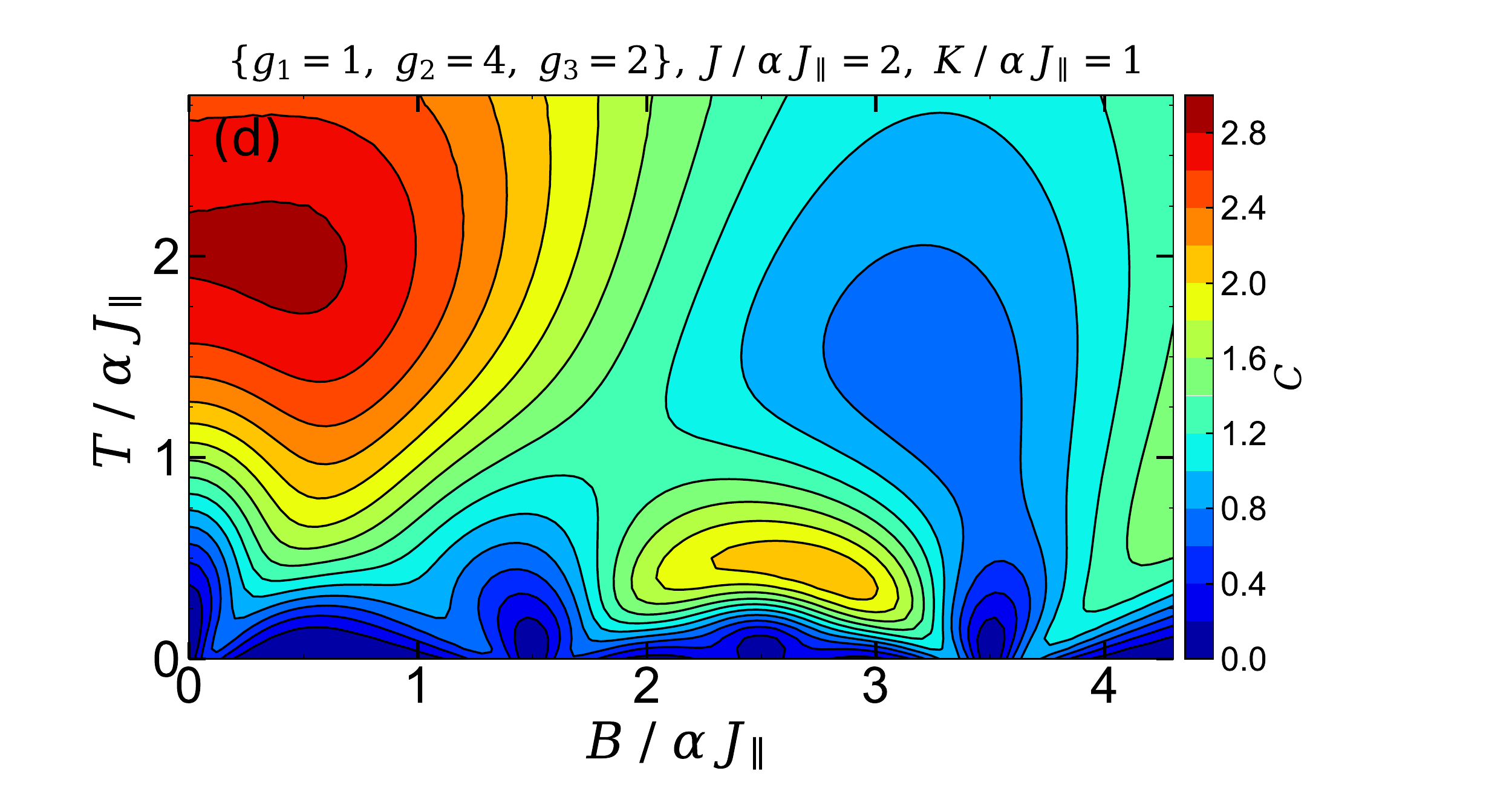}
}
\resizebox{0.45\textwidth}{!}{%
\includegraphics[trim=30 1 50 1, clip]{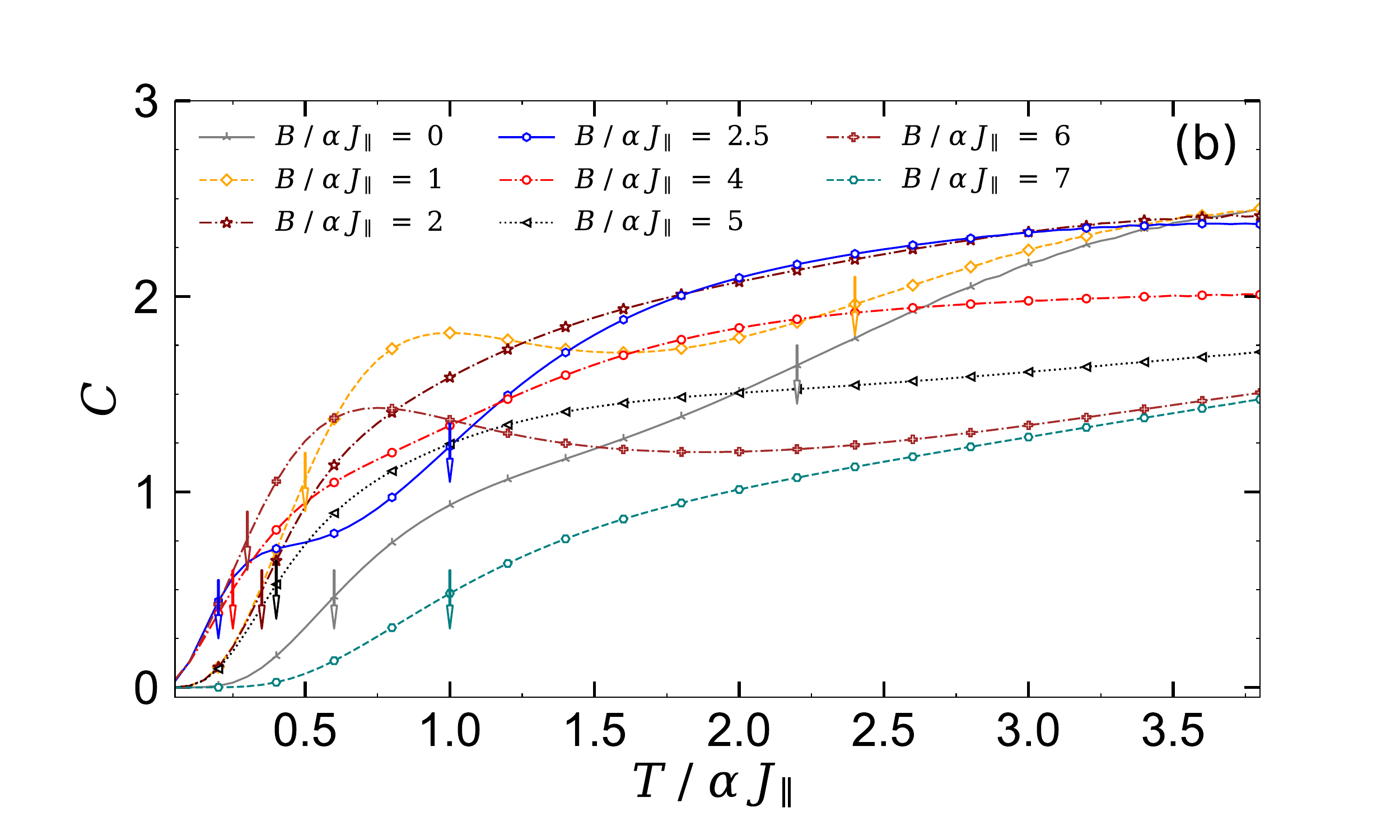}
}
\resizebox{0.45\textwidth}{!}{%
\includegraphics[trim=30 1 50 1, clip]{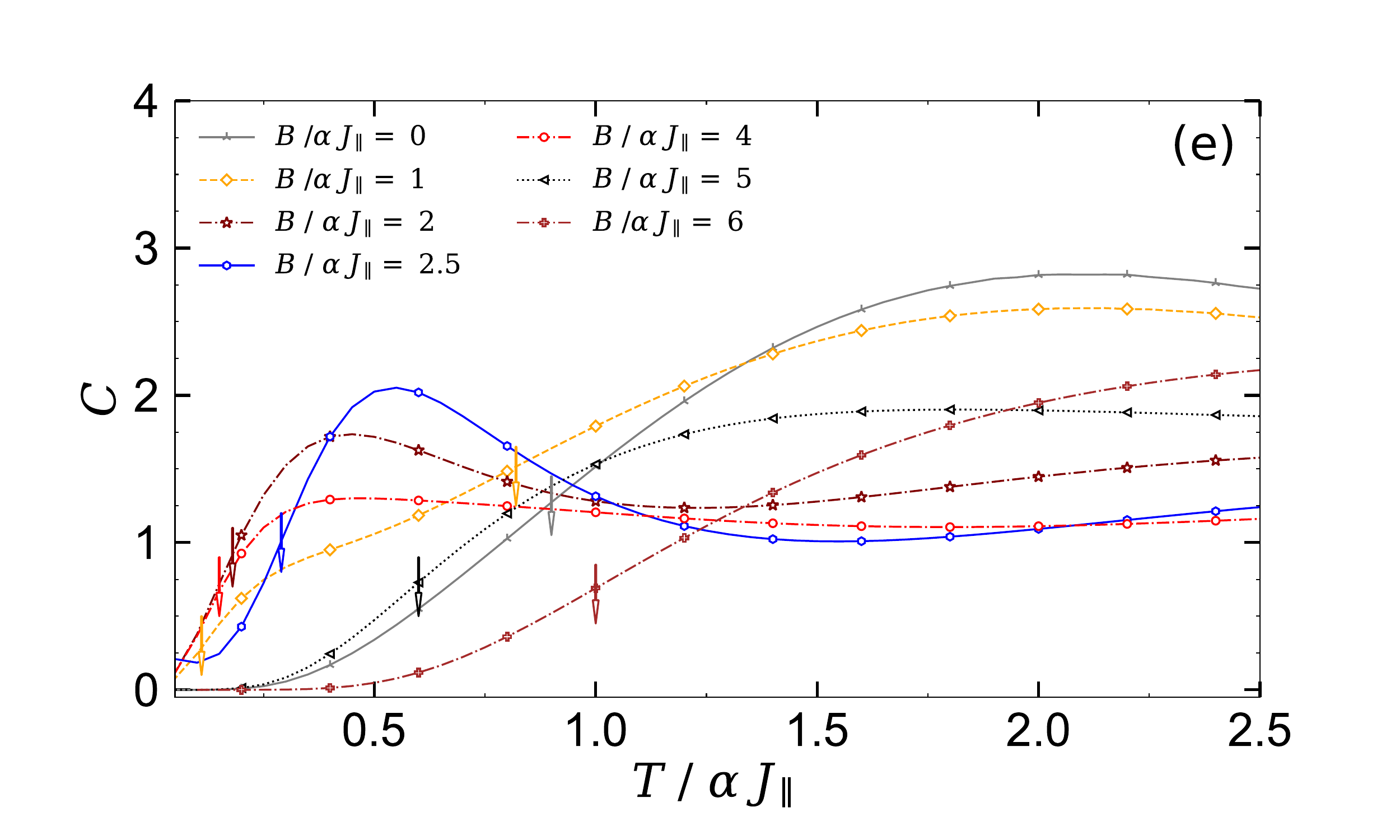}
}
\resizebox{0.45\textwidth}{!}{%
\includegraphics[trim=20 1 50 1, clip]{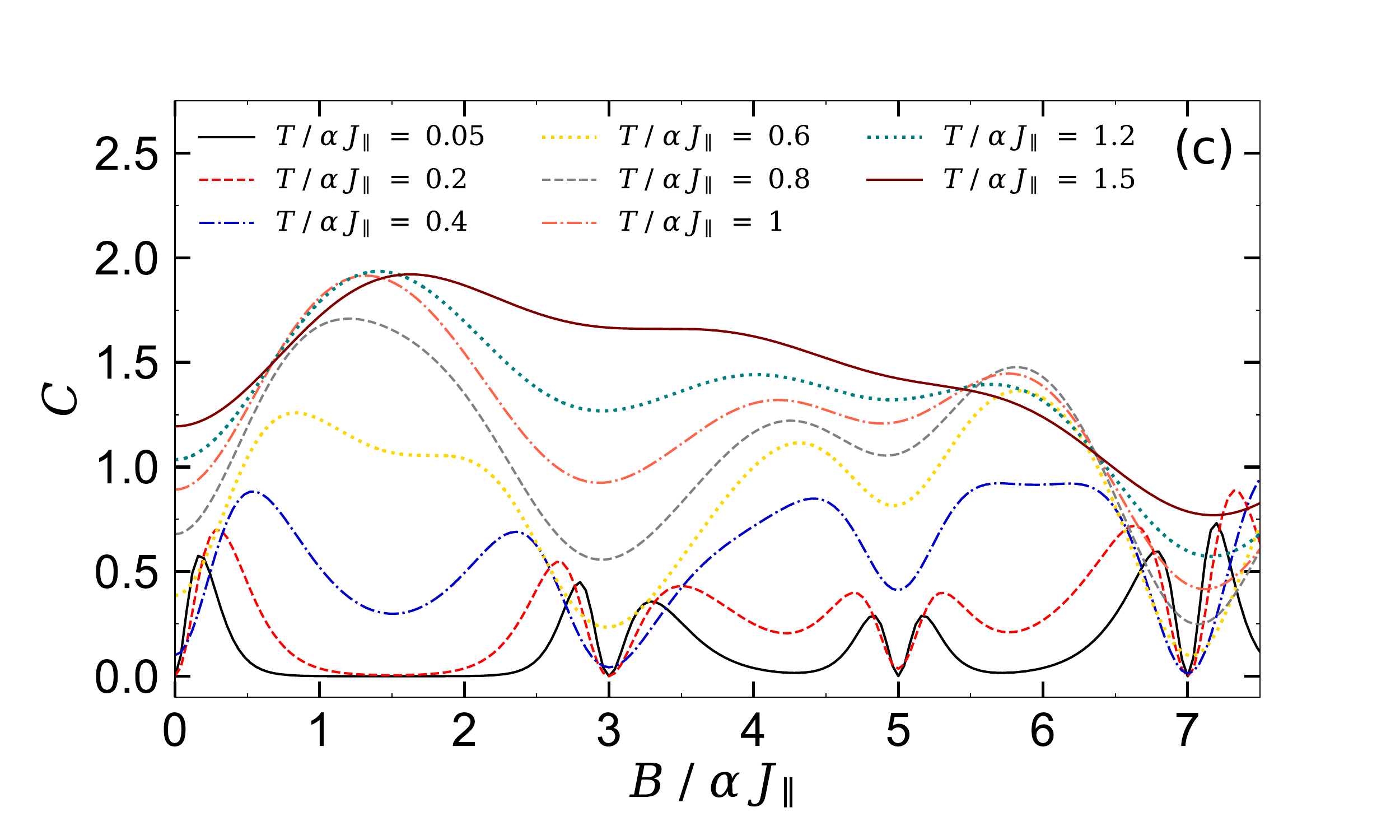}
}
\resizebox{0.45\textwidth}{!}{%
\includegraphics[trim=20 1 50 1, clip]{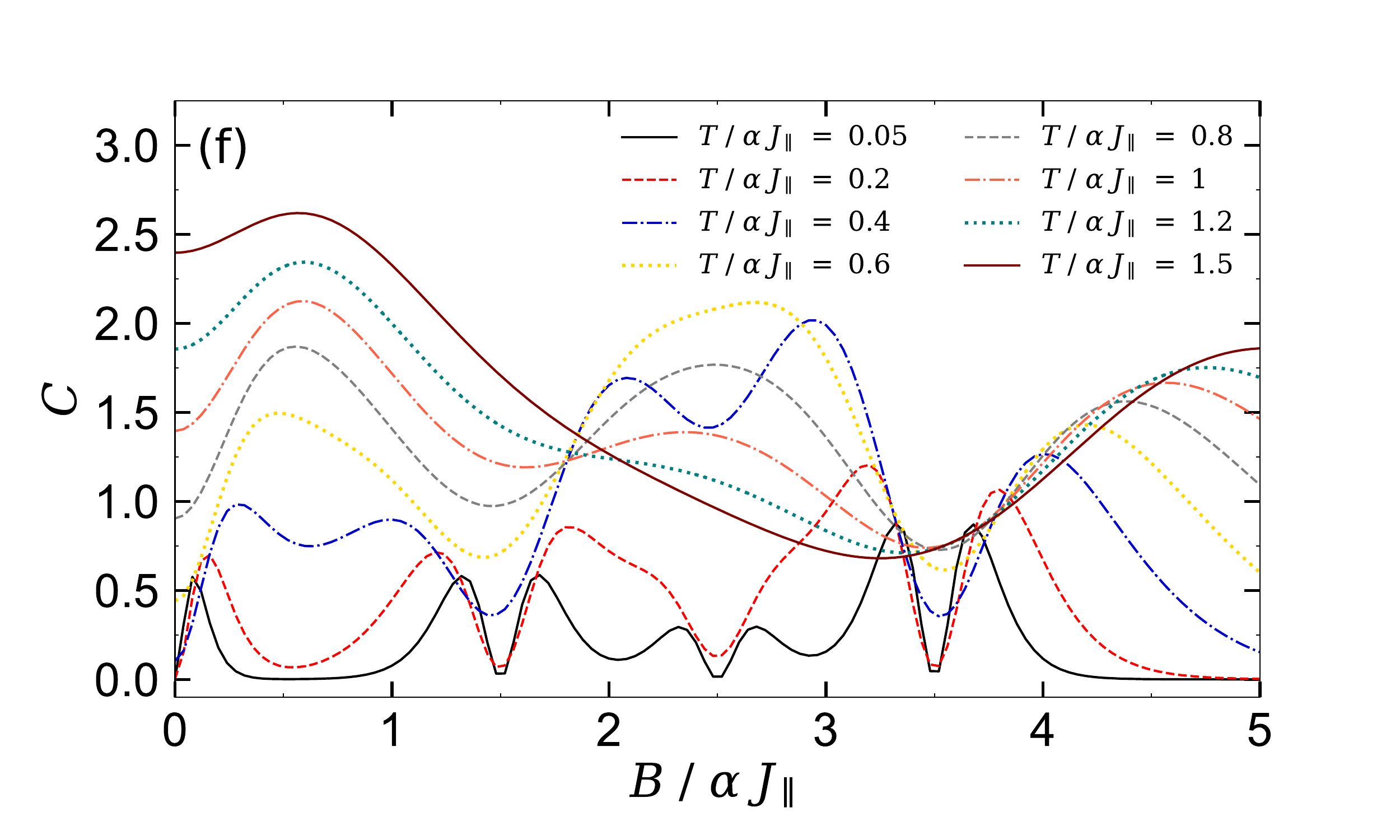}
}   
\caption{ (a) The temperature and magnetic field dependencies of the specific heat of the model under the same situation to Fig. \ref{fig:SHeat1} but for nonzero  value of the cyclic four-Ising interaction  $K/\alpha J_{\parallel}=1$. (a), (b), (c) display  the specific heat for the set $\{g_1=1$, $g_2=1$, $g_3=2\}$,  $J/\alpha J_{\parallel}=4$, and (d), (e), (f) correspond to the set $\{g_1=1$, $g_2=4$, $g_3=2\}$ and  fixed value $J/\alpha J_{\parallel}=2$, fitting with co-ordinates of the quadruple point marked in Fig. \ref{fig:QPT_BK} (b).  Arrows imply the same definition to Figs. \ref{fig:SHeat1} (b) and \ref{fig:SHeat1} (e).
}
\label{fig:SHeat2}
\end{center}
\end{figure*}
To uncover the effect of  cyclic four-spin Ising term $K/\alpha J_{\parallel}$ on the specific heat behavior, in Fig. \ref{fig:SHeat2}, is shown the 
specific heat in the  ($B/\alpha J_{\parallel}-T/\alpha J_{\parallel}$) plane by assuming fixed value $K/\alpha J_{\parallel}=1$ where other parameters have been taken as Figs. \ref{fig:SHeat1}(a)$-$ \ref{fig:SHeat1}(f). In panels \ref{fig:SHeat2}(a) and \ref{fig:SHeat2}(d),
we observe significant evolution in the specific heat curve as we consider non-zero cyclic four-spin Ising term $K/\alpha J_{\parallel}\neq 0$.
More importantly, the term  $K/\alpha J_{\parallel}$ results in changing the field-temperature position of the Schottky peak (compare Figs. \ref{fig:SHeat1}(b) and \ref{fig:SHeat1}(e) with Figs. \ref{fig:SHeat2}(b) and \ref{fig:SHeat2}(e)). Analogously, the density of contour lines of the specific heat remarkably increases close to the critical magnetic fields $B_c/\alpha J_{\parallel}=\{0,\; 3,\;  5,\; 7\}$ (Fig. \ref{fig:SHeat2}(a)) and  $B_c/\alpha J_{\parallel}=\{0,\; 1.5,\; 2. 5,\; 3.5\}$ (Fig. \ref{fig:SHeat2}(d)), at which a magnetization jump occurs between two ground states. The specific heat becomes minimum at these special critical magnetic fields (see panels \ref{fig:SHeat2}(c) and \ref{fig:SHeat2}(f)).

\subsection{Magnetocaloric effect}
In this part, let us examine MCE properties of the mixed-spin (1/2, 1) Ising-XYZ two-leg  ladder and present the most interesting results obtained for the isentropy lines together with magnetic Gr{\" u}neisen parameter multiplied by the magnetic field $B\Gamma_B$ for two arbitrary sets of Land{\'e} g-factors in the  limit of quadruple points co-ordinates depicted in Figs. \ref{fig:QPT_BJ} and \ref{fig:QPT_BK}.
We would note that one of the most important results obtained from our numerical calculations and simulations is uncovering the fact that the adiabatic demagnetization process strongly depends on the cyclic four-spin Ising interaction parameter $K/\alpha J_{\parallel}$.
In what follows, we try to pave a way to understand this fact through plotting intelligible figures in different perspectives. 


\begin{figure*}[]
\begin{center}
\resizebox{0.45\textwidth}{!}{%
 \includegraphics[trim=20 1 80 1, clip]{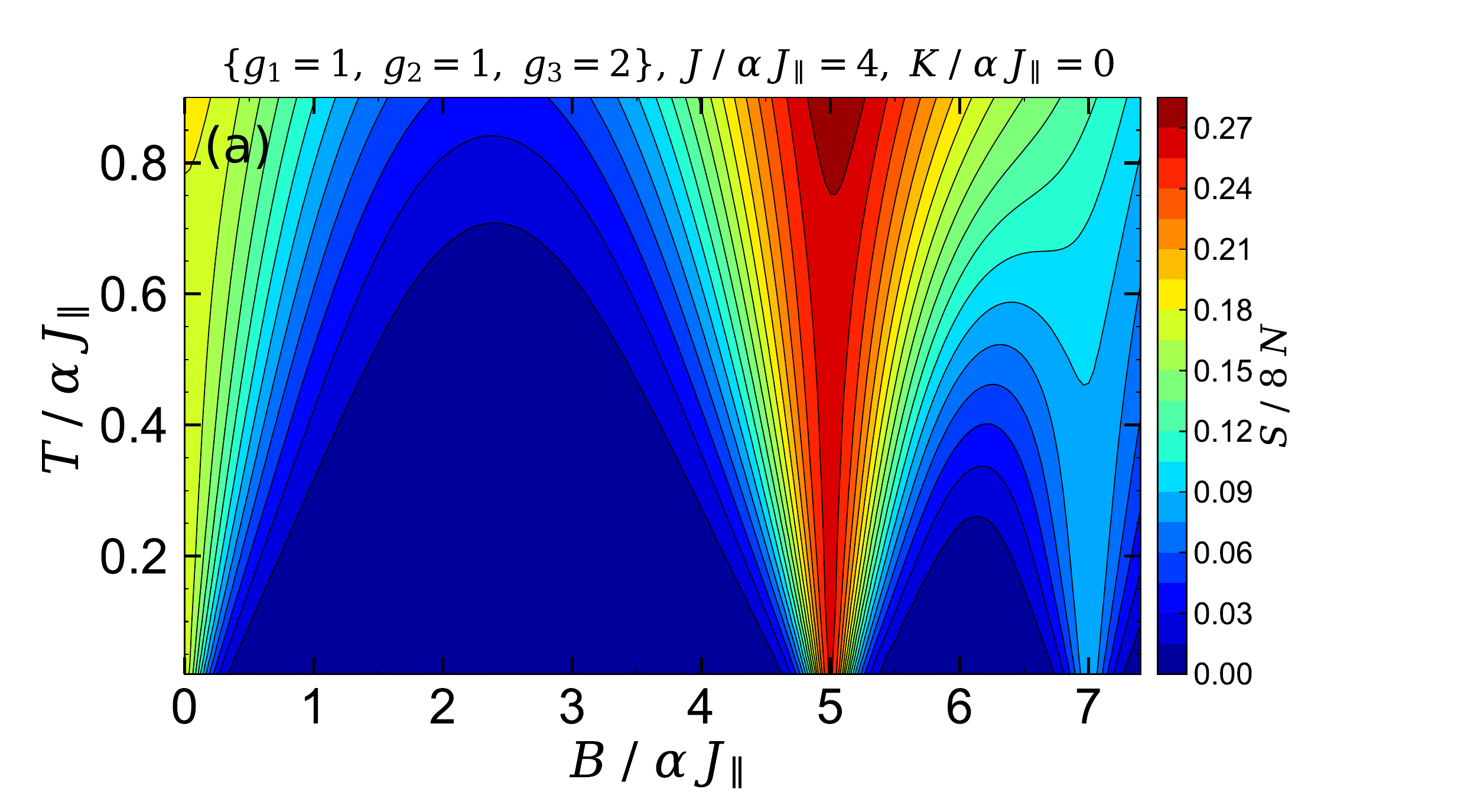} 
}
\resizebox{0.45\textwidth}{!}{%
\includegraphics[trim=20 1 80 1, clip]{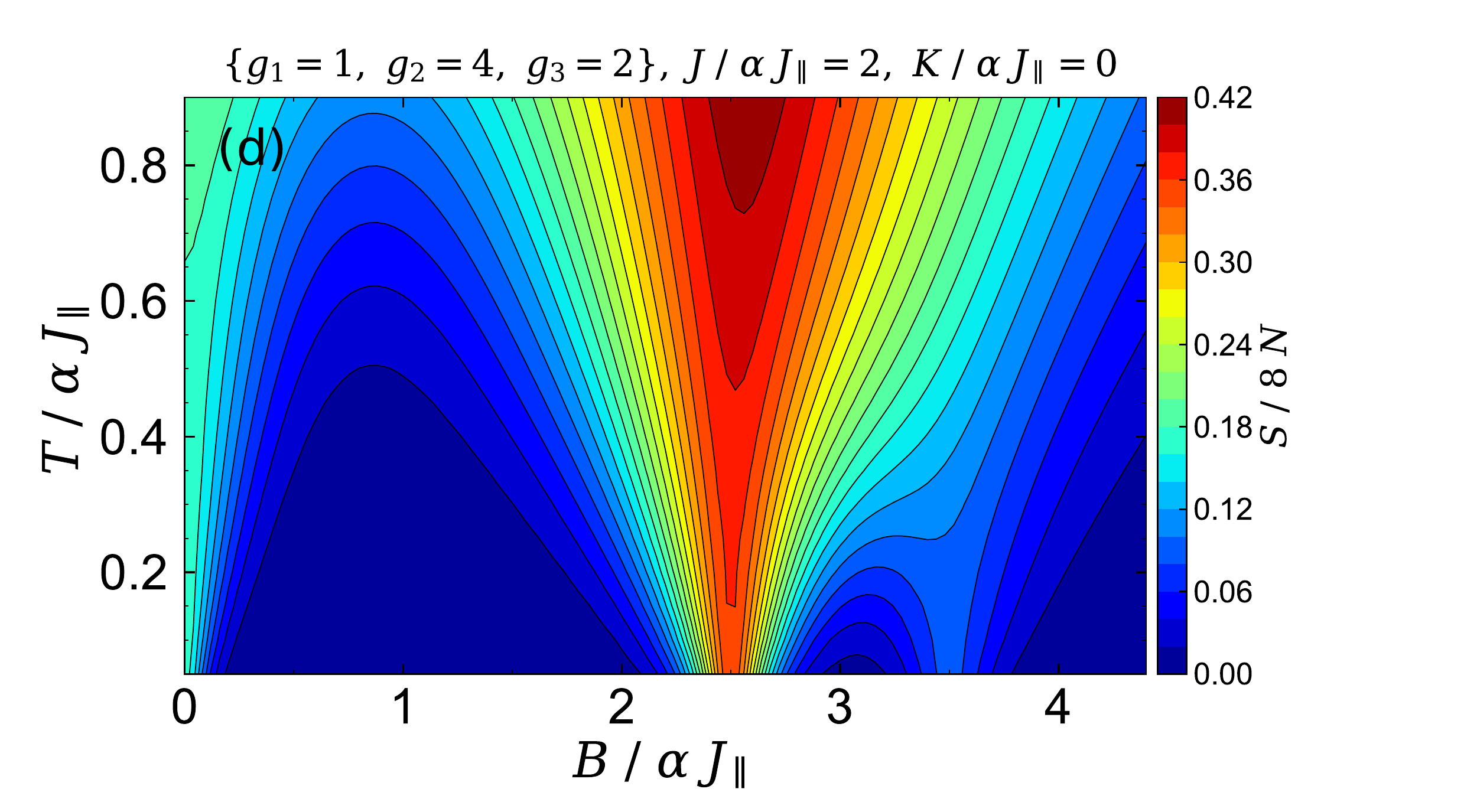}
}
\resizebox{0.45\textwidth}{!}{%
\includegraphics[trim=30 1 50 1, clip]{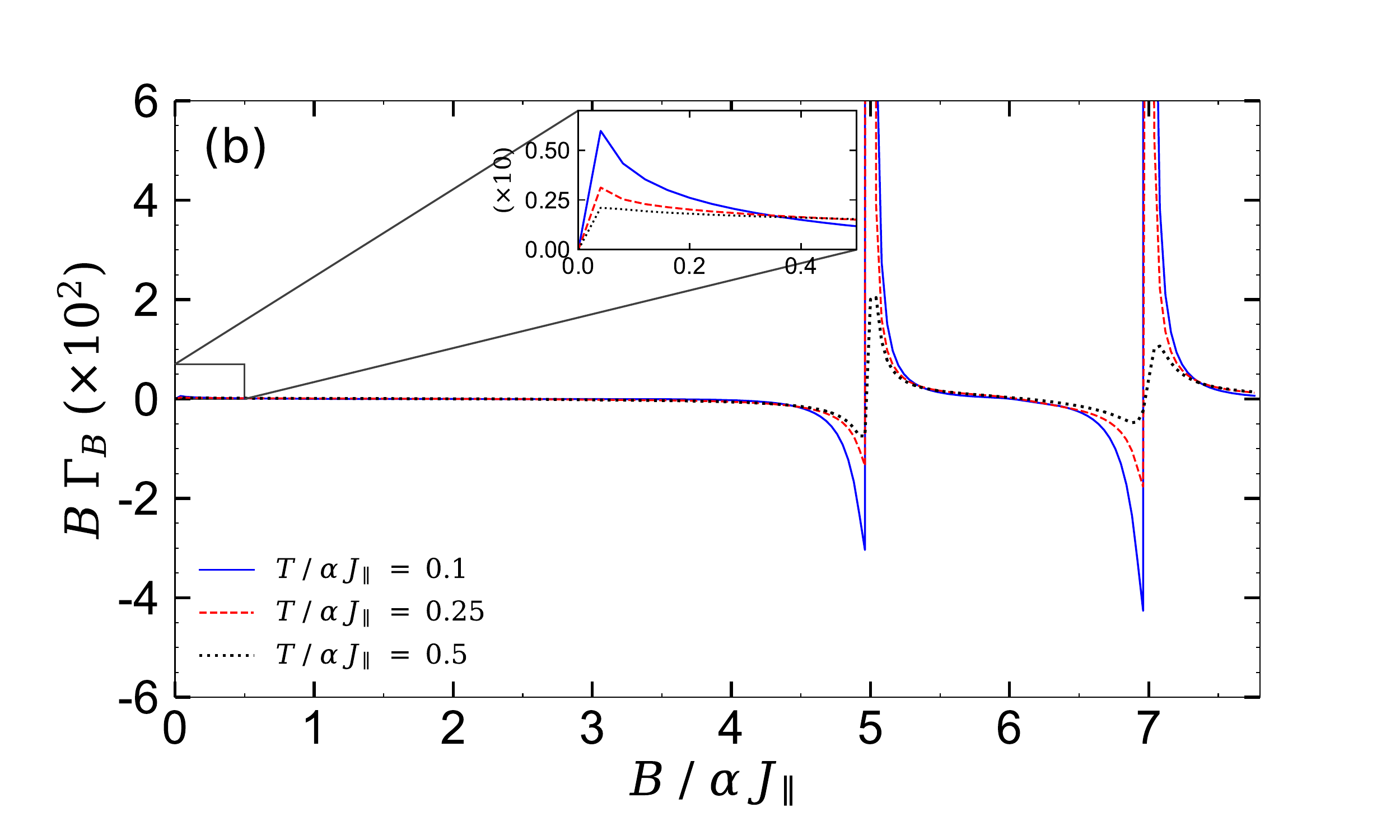}
}
\resizebox{0.45\textwidth}{!}{%
\includegraphics[trim=30 1 50 1, clip]{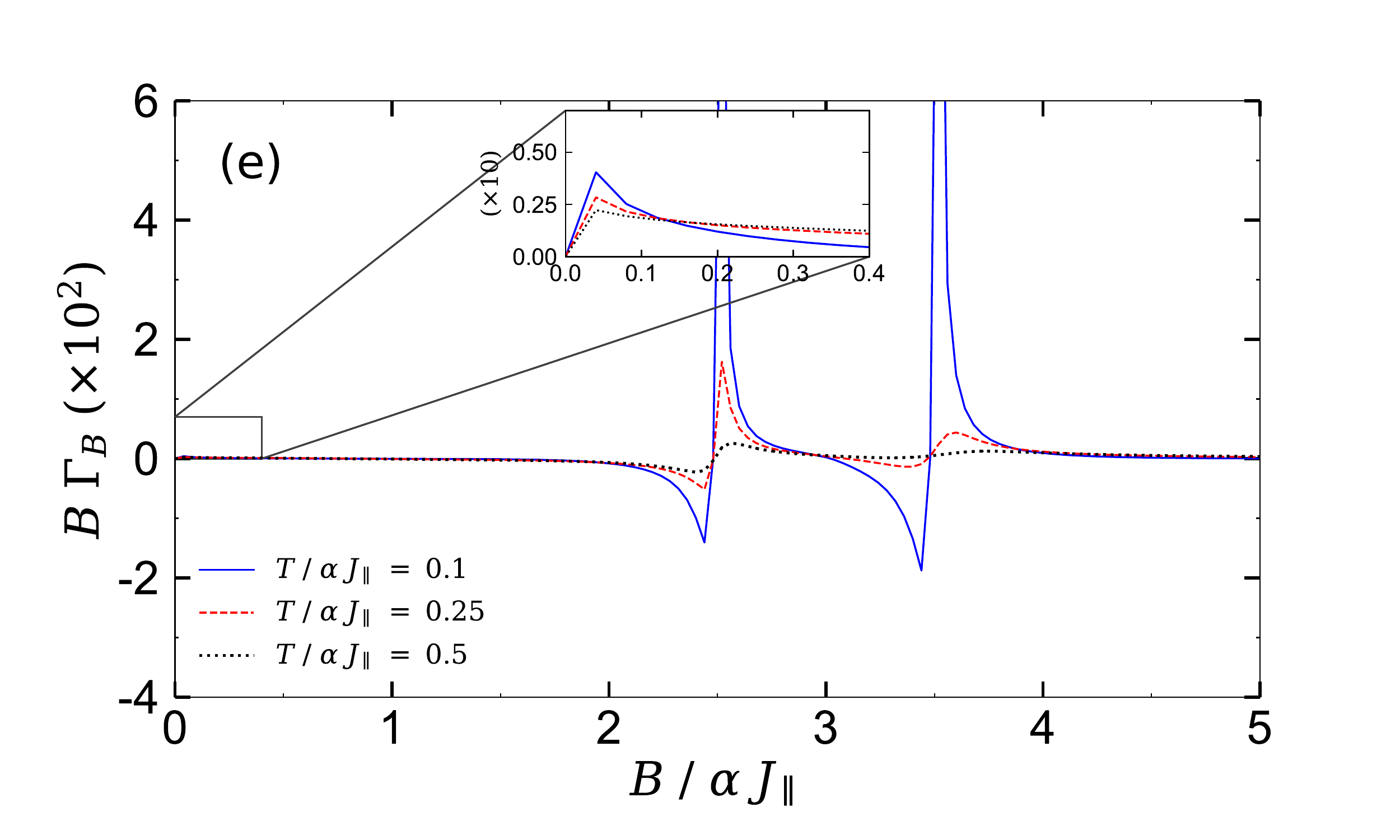}
}
\resizebox{0.45\textwidth}{!}{%
\includegraphics[trim=20 1 50 1, clip]{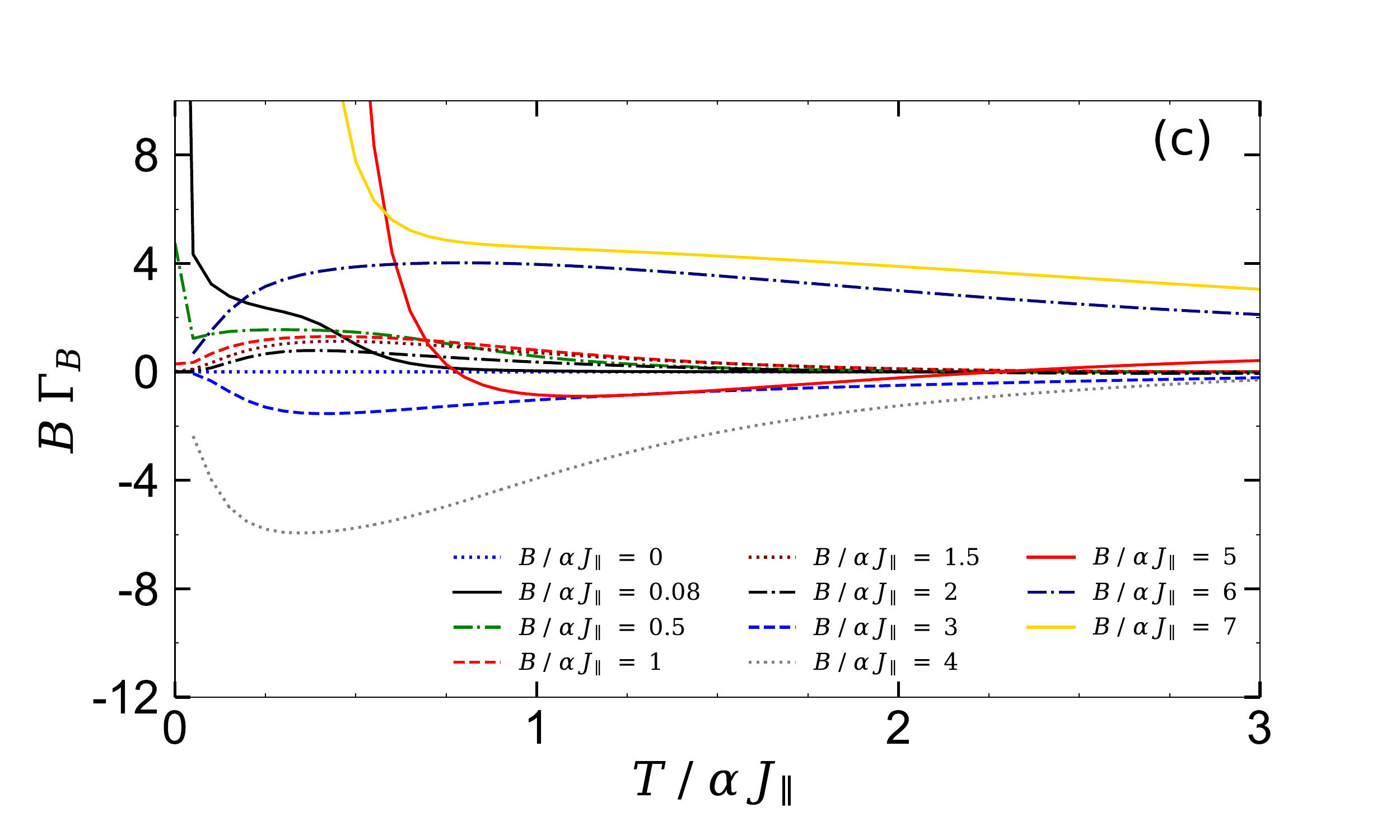}
}
\resizebox{0.45\textwidth}{!}{%
\includegraphics[trim=20 1 50 1, clip]{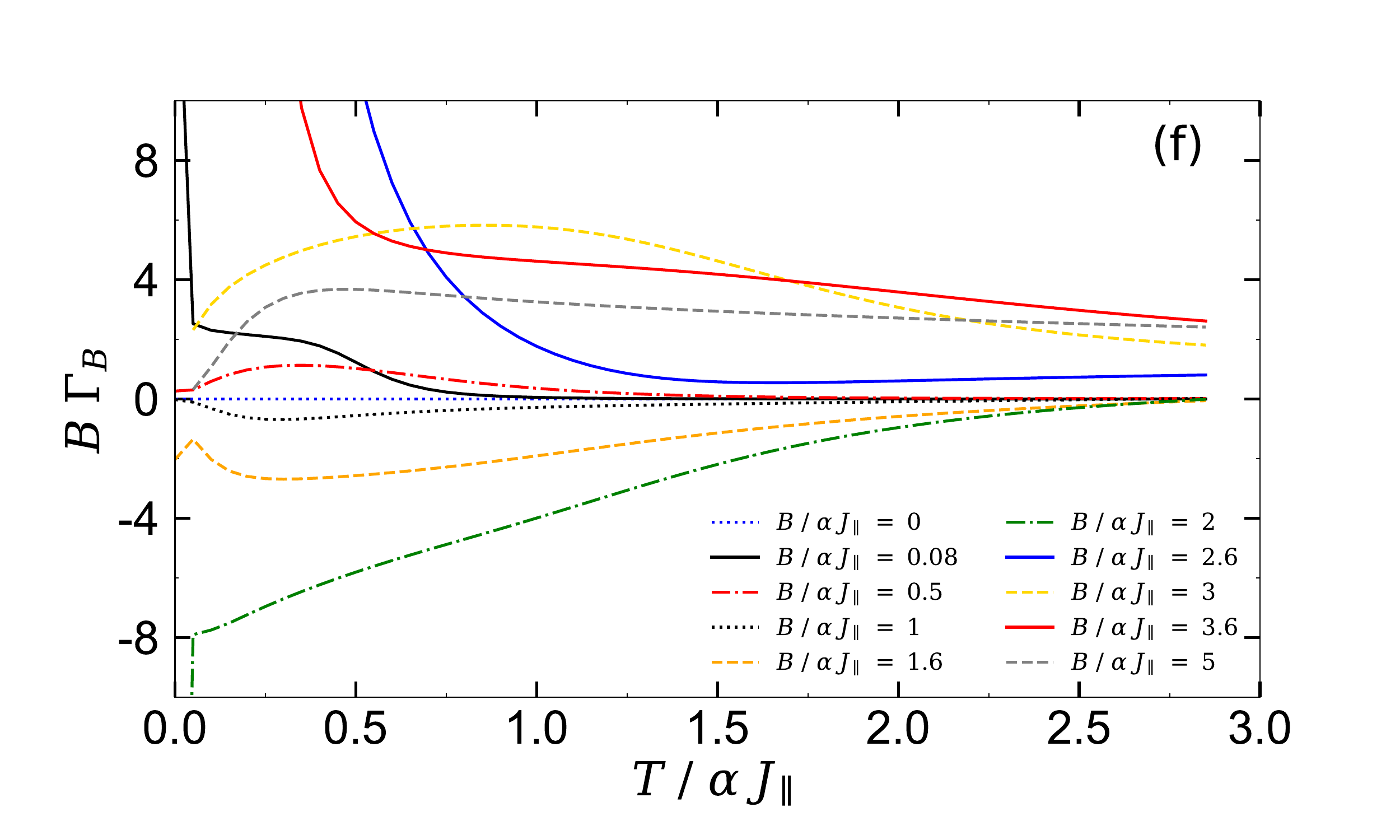}
}
\caption{Entropy of the mixed-spin (1,1/2) Ising-XYZ two-leg  ladder and the corresponding magnetic Gr{\" u}neisen parameter times the magnetic field $B\Gamma_B$ as a functions of the temperature and the magnetic field for the zero cyclic four-Ising interaction 
$K/\alpha J_{\parallel}=0$. 
(a) The entropy  in the field-temperature plane for the case when the set $\{g_1=1$, $g_2=1$, $g_3=2\}$ and fixed $J/\alpha J_{\parallel}=4$ are assumed. 
(b) The magnetic field dependencies of the dimensionless parameter $B\Gamma_B$  for three selected  temperatures $T/\alpha J_{\parallel}=0.1$, $T/\alpha J_{\parallel}=0.25$ and $T/\alpha J_{\parallel}=0.5$. 
(c) The temperature dependencies of the parameter $B\Gamma_B$  for several selected values of ratio $B/\alpha J_{\parallel}$ and the same set of other parameters to panel (a).
(d) Isentropic curve in the field-temperature plane such that the set $\{g_1=1$, $g_2=4$, $g_3=2\}$ and fixed $J/\alpha J_{\parallel}=2$ are considered.
(e) The corresponding Gr{\" u}neisen parameter as function of the magnetic field under the circumstance $\{g_1=1$, $g_2=4$, $g_3=2\}$, where fixed $J/\alpha J_{\parallel}=2$ is supposed.
(f) Gr{\" u}neisen parameter as function of the temperature under the same circumstances to panel (d) such that various fixed values of the magnetic field are supposed.
 Remaining parameters have been taken as Fig. \ref{fig:Mag}. 
 }
\label{fig:EntropyK0}
\end{center}
\end{figure*}

 
Figure \ref{fig:EntropyK0}(a) shows the isothermal dependence of entropy $S/8N$ in the field-temperature plane including adiabatic demagnetization curves (solid contour lines) for the case when we consider the situation $\{g_1=1$, $g_2=1$, $g_3=2\}$ and fixed value $J/\alpha J_{\parallel}=4$.
It is quite evident that,  the isentropy lines are suddenly accumulated  nearby the critical magnetic fields $B_c/\alpha J_{\parallel}=\{0,\; 5,\; 7\}$. 
The density of the isentropy lines close to the field-position of quadruple point with the corresponding co-ordinates 
$(B/\alpha J_{\parallel},\; J/\alpha J_{\parallel},\; K/\alpha J_{\parallel})\equiv (5,\;4,\;0)$ is much more than other places.

 In Fig.  \ref{fig:EntropyK0}(d), we display the isentropic changes of temperature $T/\alpha J_{\parallel}$ as a function of the external magnetic field $B/\alpha J_{\parallel}$ for the set $\{g_1=1$, $g_2=4$, $g_3=2\}$, assuming fixed values of  $J/\alpha J_{\parallel}=2$ and $K/\alpha J_{\parallel}=0$. It can be seen enhanced regions of MCE at critical points $B_c/\alpha J_{\parallel}=\{0,\; 2. 5,\; 3.5\}$ due to the ground-state phase transition and/or magnetization jump from one plateau to another one. In this case, the maximum value of  entropy is $S_{\max}/8N \approx 2.07$, where $T>0$.
 
 To identify the cooling rate of the model in the vicinity of particular field-induced  phase transitions,
in Fig. \ref{fig:EntropyK0}(b), is depicted the quantitative fingerprint of magnetic Gr{\" u}neisen parameter multiplied by the applied field with respect to the original magnetic field  when the set $\{g_1=1$, $g_2=1$, $g_3=2\}$ and fixed interaction ratios $J/\alpha J_{\parallel}=4$  and $K/\alpha J_{\parallel}=0$ are used.
In accordance with general expectations, we witness very sharp peaks containing positive and negative values close to the critical magnetic fields at which a ground-state phase transition occurs. When the temperature increases, the magnitude of these peaks rapidly changes. This  behavior is directly connected to the anomalous zero-temperature entropy of the model under consideration at critical points.

Accordingly, a similar scenario is presented in Fig. \ref{fig:EntropyK0}(e) but for the case $\{g_1=1$, $g_2=4$, $g_3=2\}$ and fixed interaction ratio $J/\alpha J_{\parallel}=2$, assuming the same set of other parameters as used in Fig. \ref{fig:EntropyK0}(d). This imagination reveals an identical enhanced MCE at respective ground-state phase transitions, namely, at critical points $B_c/\alpha J_{\parallel}=\{0,\; 2. 5,\; 3.5\}$. 
In both plots Fig. \ref{fig:EntropyK0}(b) and Fig. \ref{fig:EntropyK0}(e), is seen a steep decreasing of the cooling rate in the vicinity of zero magnetic field point $B_c/\alpha J_{\parallel}\approx 0$ (zoomed insets), reminiscing a quick magnetization jump from $M/M_s=0$ to the magnetization intermediate $(1/5)-$plateau of saturation value.

The temperature dependence of the parameter $B\Gamma_B$ for various fixed values of the magnetic field is presented in Figs. \ref{fig:EntropyK0}(c) and \ref{fig:EntropyK0}(f) under the same circumstances to, respectively, Fig. \ref{fig:EntropyK0}(a) and Fig. \ref{fig:EntropyK0}(d). One sees that by cooling the system, close to the critical magnetic field at which ground-state phase transition occurs, parameter $B\Gamma_B$ goes to infinity (solid lines), while for other values of the magnetic field, by ultra-cooling the system 
($T/\alpha J_{\parallel}\ll 1$), $B\Gamma_B$ tends to zero. 

\begin{figure*}[t!]
\begin{center}
\resizebox{0.45\textwidth}{!}{%
 \includegraphics[trim=20 1 80 1, clip]{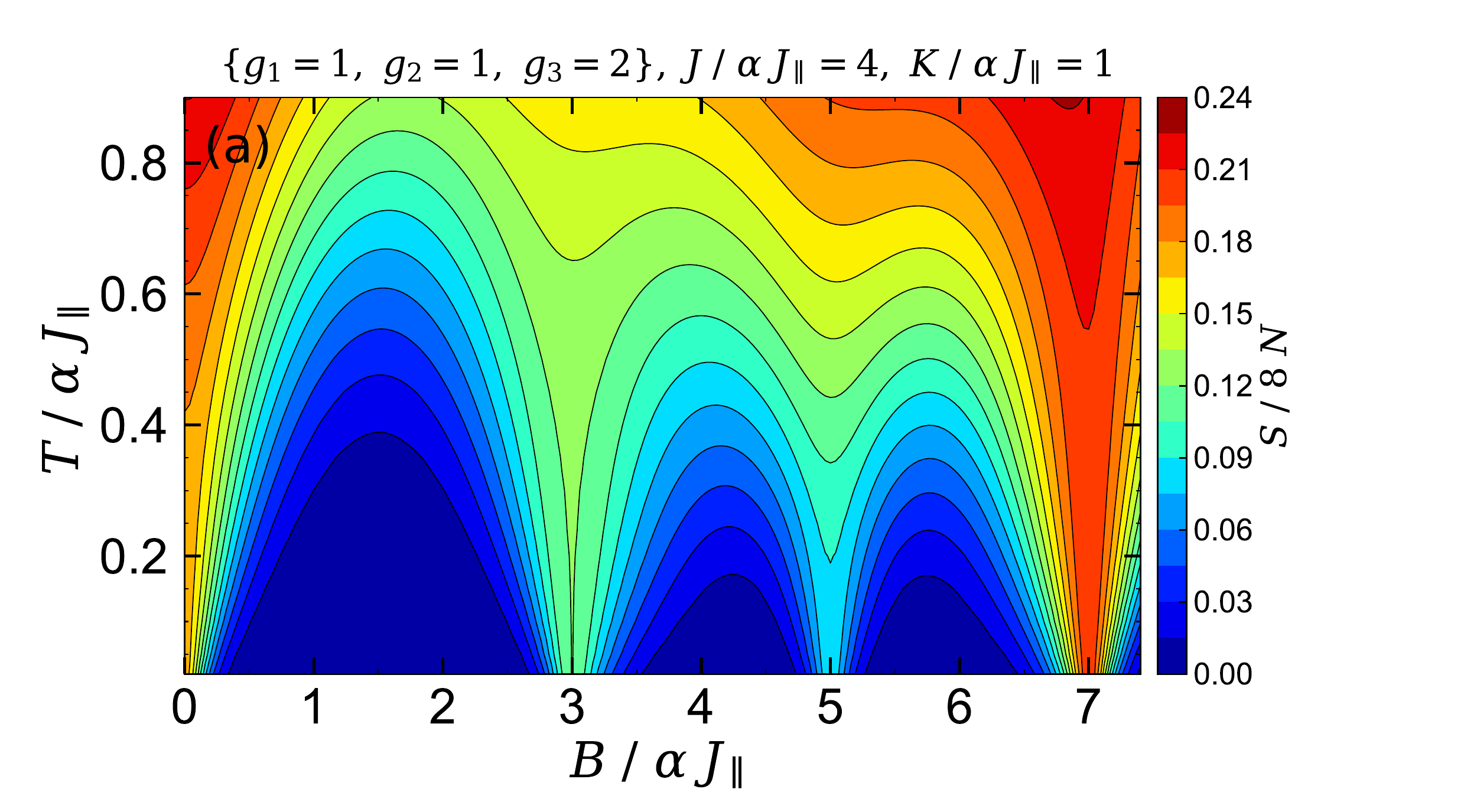} 
}
\resizebox{0.45\textwidth}{!}{%
\includegraphics[trim=20 1 80 1, clip]{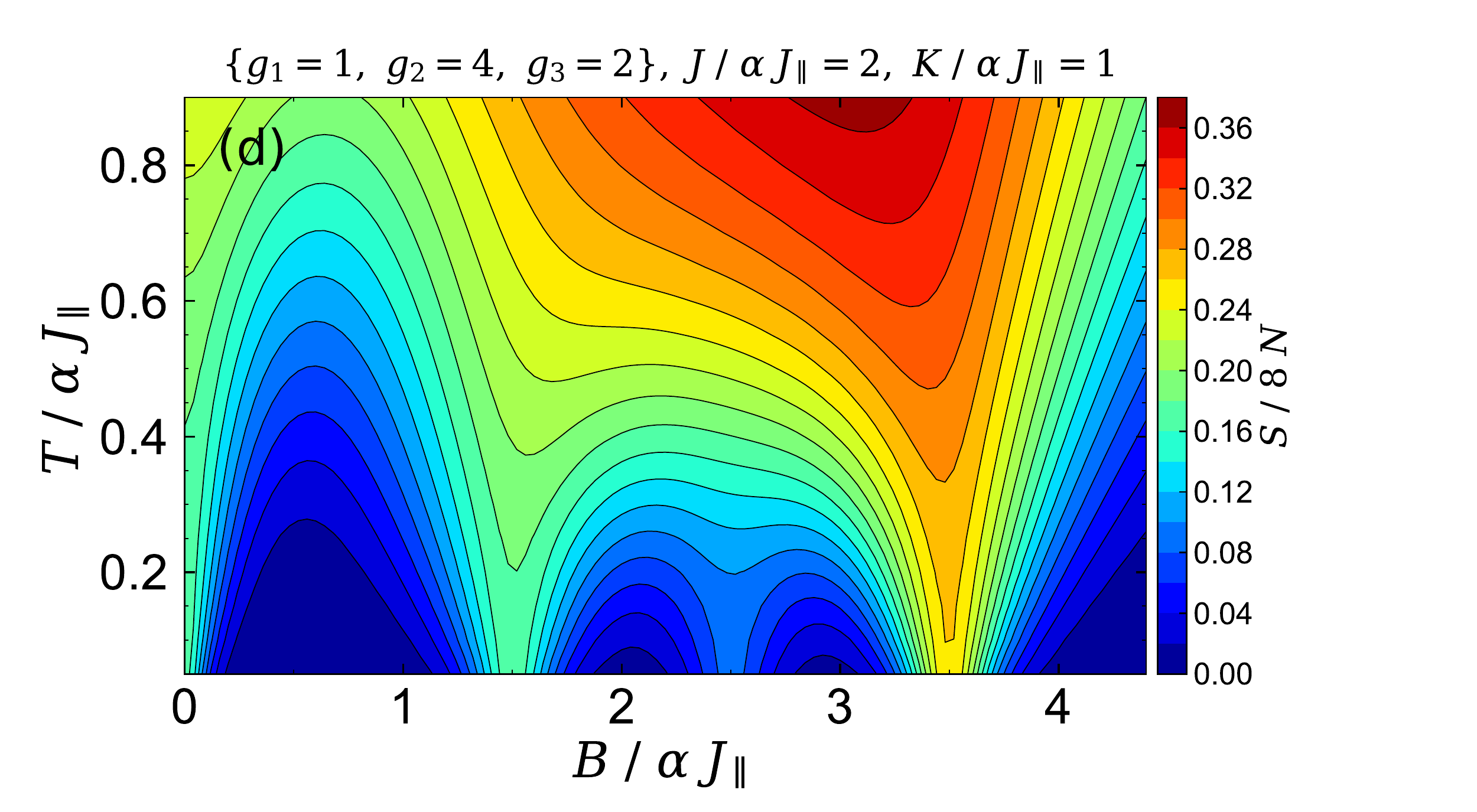}
}
\resizebox{0.45\textwidth}{!}{%
\includegraphics[trim=30 1 50 1, clip]{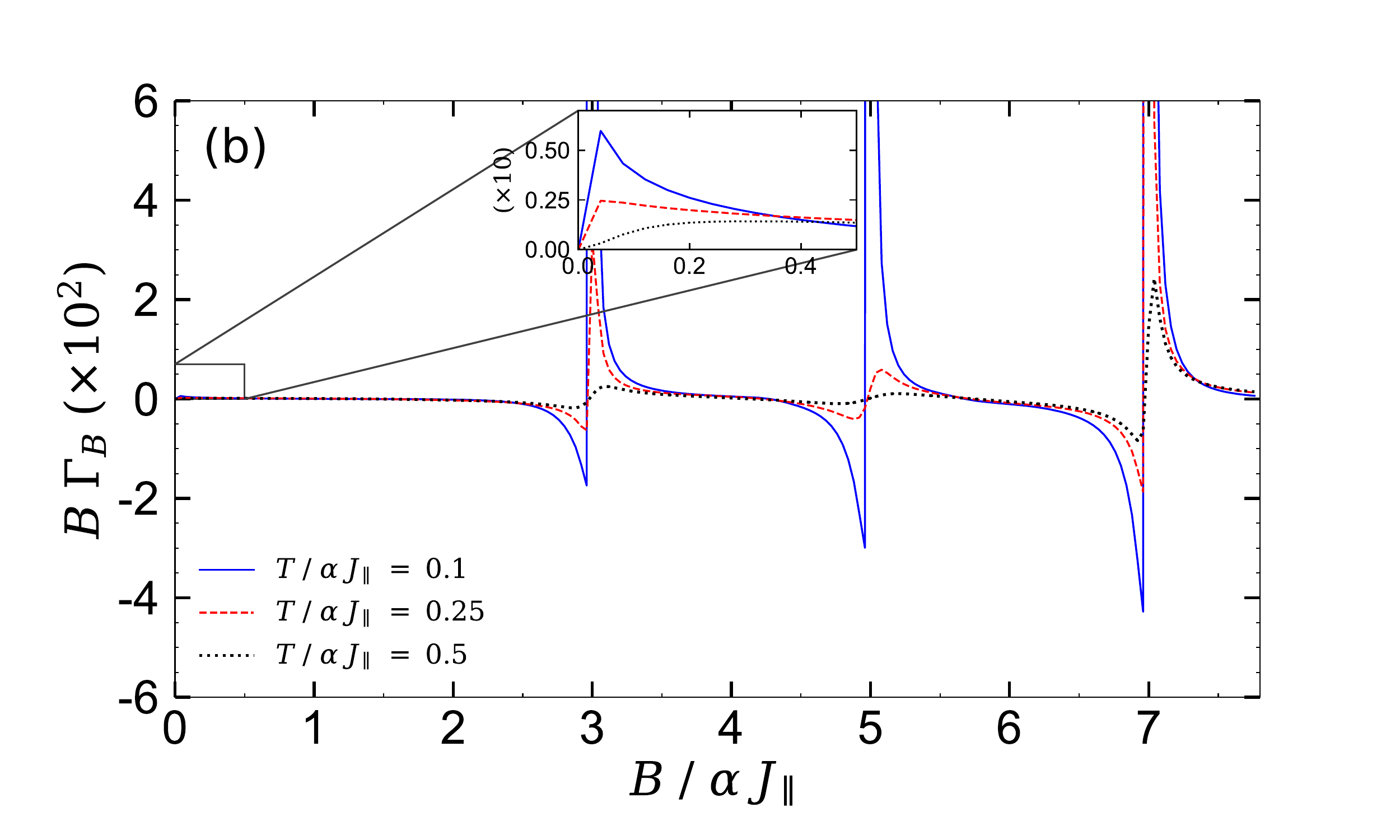}
}
\resizebox{0.45\textwidth}{!}{%
\includegraphics[trim=30 1 50 1, clip]{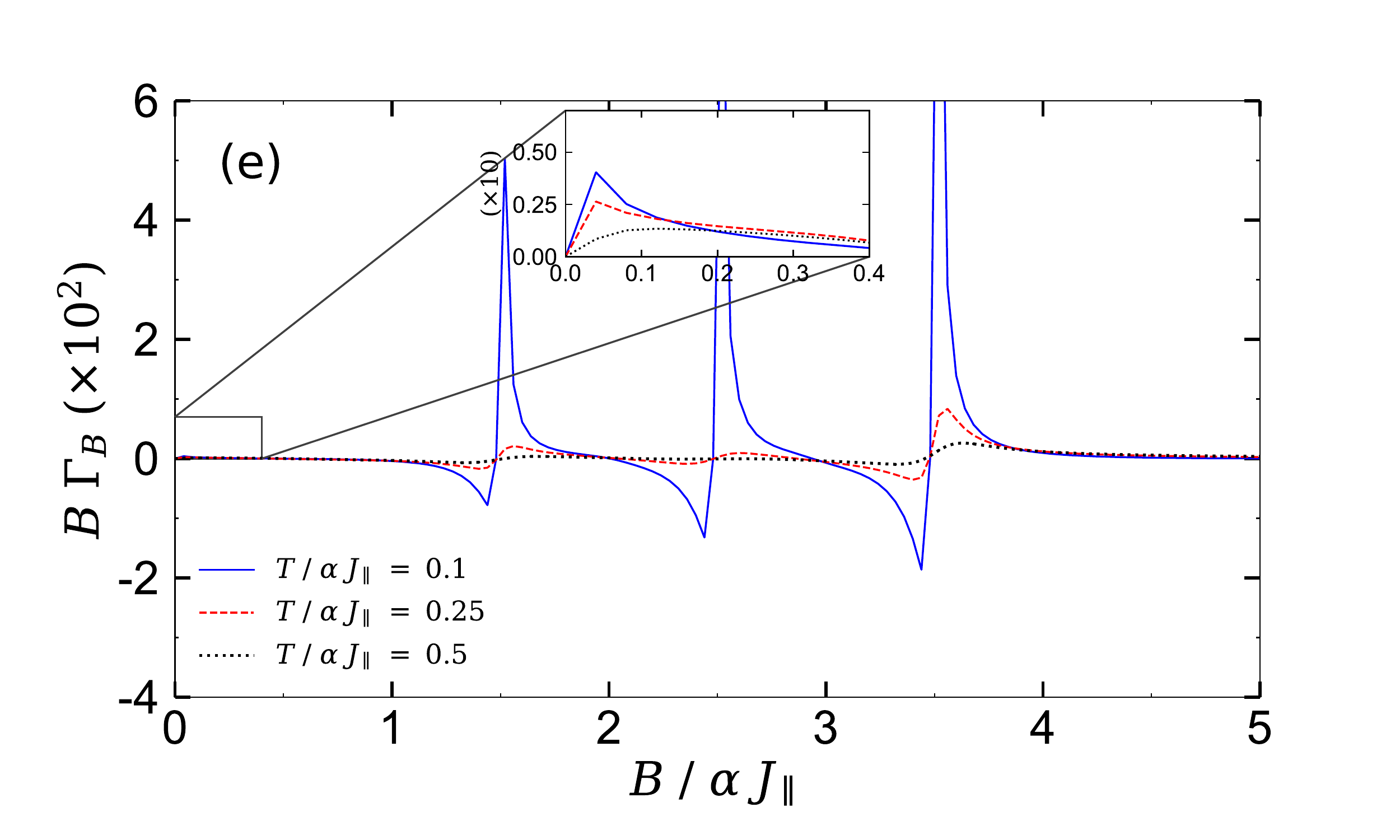} 
}
\resizebox{0.45\textwidth}{!}{%
\includegraphics[trim=20 1 50 1, clip]{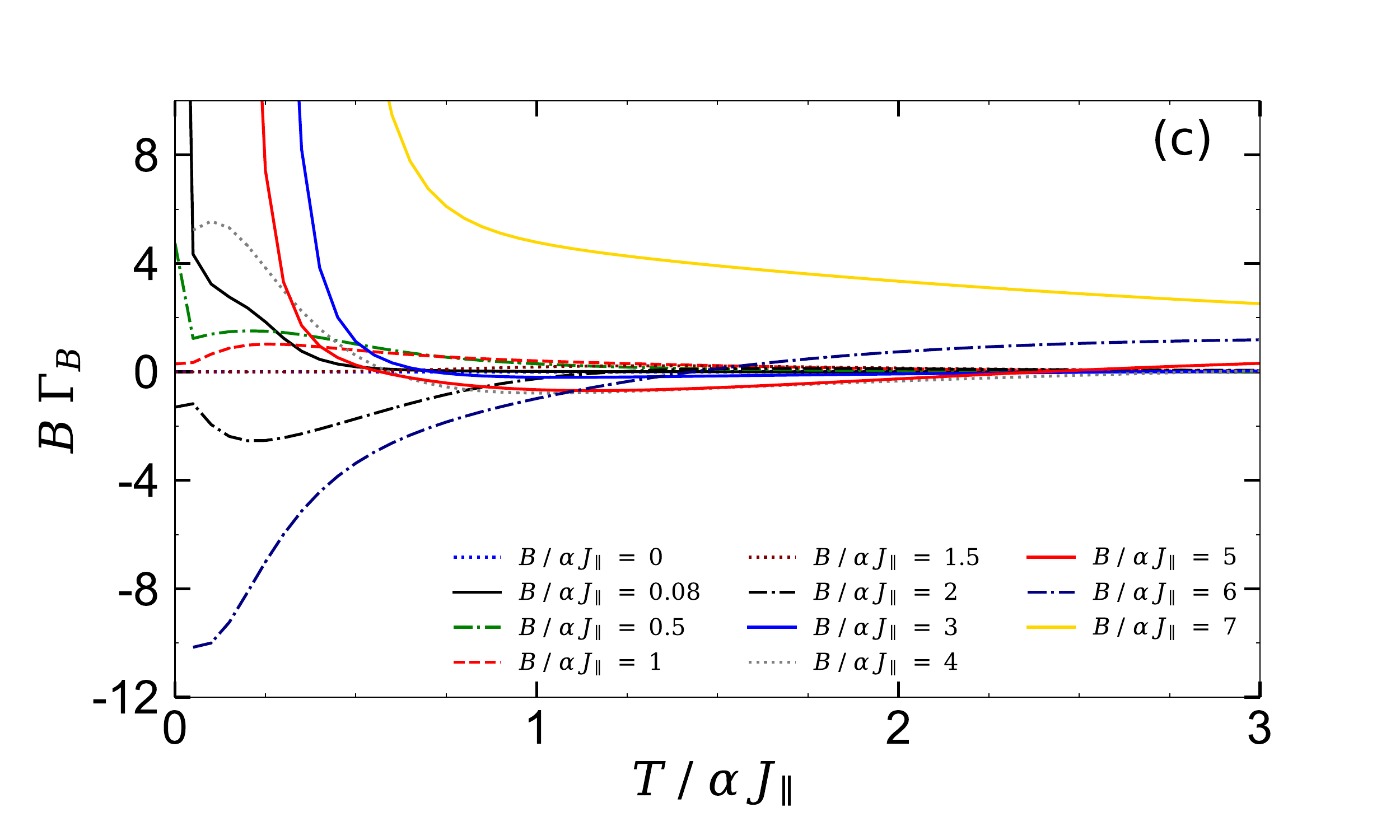} 
}
\resizebox{0.45\textwidth}{!}{%
\includegraphics[trim=20 1 50 1, clip]{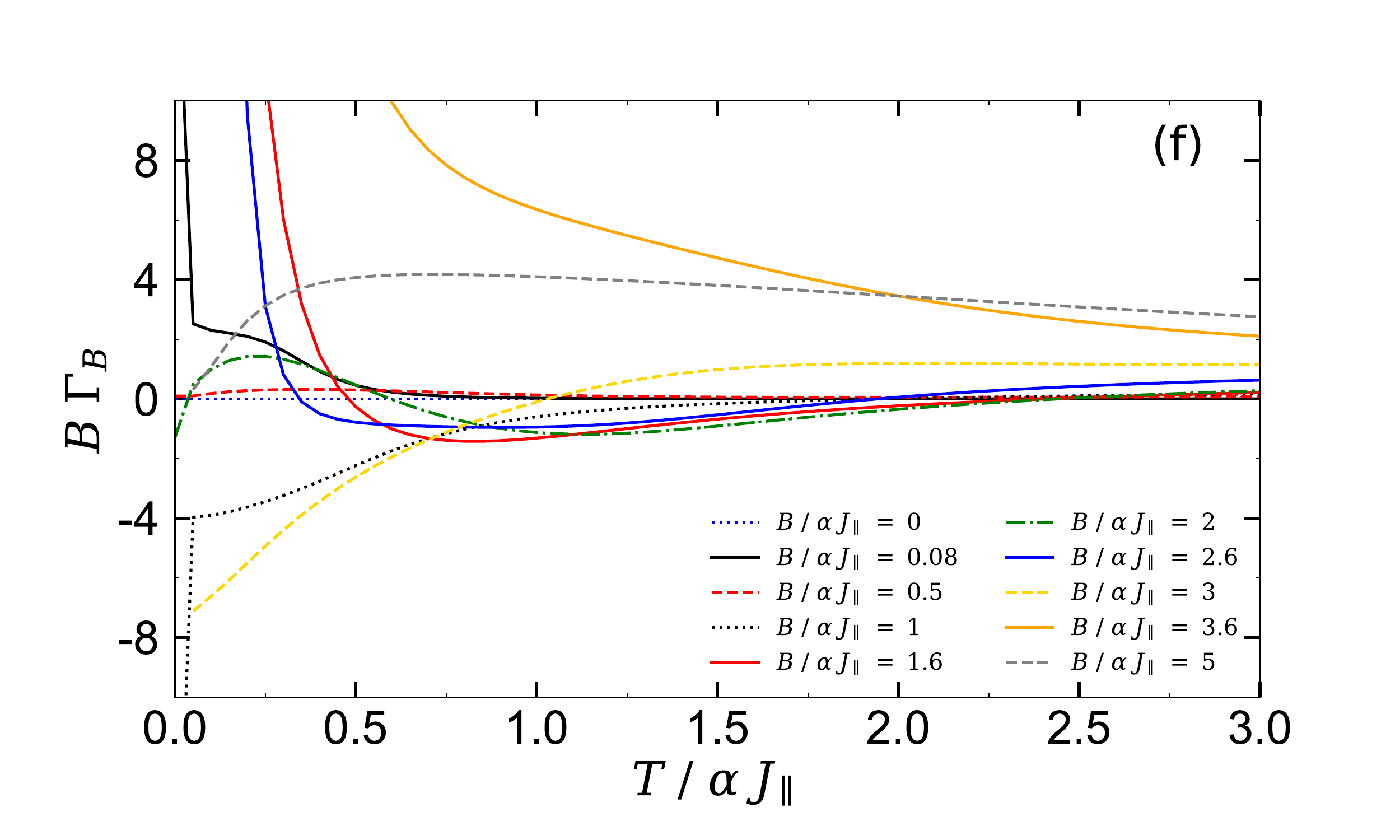} 
}
\caption{The entropy and Gr{\" u}neisen parameter as functions of the temperature and the magnetic field for the nonzero cyclic four-spin Ising interaction $K/\alpha J_{\parallel}=1$.
(a) The contour plot of the entropy together with a number of isentropy lines for the case $\{g_1=1$, $g_2=1$, $g_3=2\}$ and fixed $J/\alpha J_{\parallel}=4$.
(b)  The corresponding magnetic Gr{\" u}neisen parameter $B\Gamma_B$ versus ratio $B/\alpha J_{\parallel}$ for three selected  temperatures $T/\alpha J_{\parallel}=0.1$, $T/\alpha J_{\parallel}=0.25$ and $T/\alpha J_{\parallel}=0.5$. 
(c)  Dimensionless parameter $B\Gamma_B$ as a function of the ratio $T/\alpha J_{\parallel}$ for the same parameters set to panel (a), while several fixed magnetic fields are assumed.
(d) Isentropy lines for the set  $\{g_1=1$, $g_2=4$, $g_3=2\}$ and fixed $J/\alpha J_{\parallel}=2$. 
(e)  The corresponding magnetic Gr{\" u}neisen parameter as a function of magnetic field at three different  temperatures $T/\alpha J_{\parallel}=0.1$, $T/\alpha J_{\parallel}=0.25$ and $T/\alpha J_{\parallel}=0.5$ under the same condition to panel (d), i.e.,  $\{g_1=1$, $g_2=4$, $g_3=2\}$ and fixed $J/\alpha J_{\parallel}=2$.
(f) The Gr{\" u}neisen parameter multiplied by the magnetic field as a function of the temperature for the same parameter set to panel (d), where different fixed values of the magnetic field are selected. 
 Remaining parameters such as $\alpha$, $\gamma$, $\Delta$ and $J_{\perp}/\alpha J_{\parallel}$ have been assumed as Fig. \ref{fig:Mag}.
 }
\label{fig:EntropyK1}
\end{center}
\end{figure*}

Figure \ref{fig:EntropyK1}(a) shows the isentropy lines in the field-temperature plane under the same circumstances to Fig. \ref{fig:EntropyK0}(a) but for nonzero value $K/\alpha J_{\parallel}=1$. By inspecting this figure, one can see that  by imposing a nonzero value of cyclic four-spin Ising interaction, an enhanced MCE will appear nearby one another critical magnetic field $B/\alpha J_{\parallel}=3$, denoting phase transition from the ground-state with magnetization $M/M_s=1/5$ to that of with magnetization $M/M_s=2/5$ (see Fig. \ref{fig:QPT_BJ}(c)). In Fig.  \ref{fig:EntropyK1}(d), we plot contour plot of the entropy and a number of isentropy lines for the case $\{g_1=1$, $g_2=4$, $g_3=2\}$ and fixed $J/\alpha J_{\parallel}=2$ and $K/\alpha J_{\parallel}=1$. Again, we observe an extensive variation in the entropy of the model specifically close to the point $B/\alpha J_{\parallel}=1.5$, where the magnetization jump happens from $(1/4)-$plateau to $(1/2)-$plateau normalized with the saturation magnetization (see Fig. \ref{fig:QPT_BJ}(d)).
In result, the four-spin Ising interaction $K/\alpha J_{\parallel}$ has a great influence on the entropy behavior regardless of what values we choose for Land{\'e} g-factors.

Last but not least, let us discuss the effects of parameter  $K/\alpha J_{\parallel}$ on the cooling rate. For this purpose, we plot in Figs. \ref{fig:EntropyK1}(b) and \ref{fig:EntropyK1}(e), the magnetic Gr{\" u}neisen parameter times the field against ratio $B/\alpha J_{\parallel}$ at three different temperatures  $T/\alpha J_{\parallel}=\{0.1, 0.25, 0.5\}$, by keeping other parameters of the Hamiltonian as in panels \ref{fig:EntropyK1}(a) and \ref{fig:EntropyK1}(d), respectively. 
The blue solid curve in Fig. \ref{fig:EntropyK1}(b) starts at almost zero magnetic field with a steep slope (see zoomed inset) and crosses the first transition at  $B/\alpha J_{\parallel}=3$, and cuts two ground states with the corresponding magnetization values $M/M_s=1/5$ and $M/M_s=2/5$ numbered in  Fig. \ref{fig:QPT_BJ}(c). Third peak arise at $B/\alpha J_{\parallel}=5$, where the boundary between ground states with magnetization $M/M_s=2/5$ and $M/M_s=3/5$ exists. Final peak arises at $B/\alpha J_{\parallel}=7$ which is the magnetic field position of the quadruple point marked in Fig. \ref{fig:QPT_BJ}(c). 
By assuming the set $\{g_1=1$, $g_2=4$, $g_3=2\}$ and utilizing fixed value of $J/\alpha J_{\parallel}=2$ (Fig. \ref{fig:EntropyK1}(e)), the magnetic-position of the Gr{\" u}neisen peaks will change. This phenomenon is in accordance with the change in magnetization steps and jumps as shown in Figs. \ref{fig:Mag}(d) and  \ref{fig:QPT_BJ}(d).

The temperature dependence of the parameter $B\Gamma_B$ for several fixed values of the magnetic field is depicted in Figs. \ref{fig:EntropyK1}(c) and \ref{fig:EntropyK1}(f), where the other parameters have been taken as Fig. \ref{fig:EntropyK1}(a) and Fig. \ref{fig:EntropyK1}(d), respectively. Nearby the critical magnetic fields, we observe that the behavior of $B\Gamma_B$ against the temperature  is similar to the case when $K/\alpha J_{\parallel}=0$ is assumed. Nonetheless, it is understandable that by considering  
$K/\alpha J_{\parallel}>0$, during ultra-cooling process ( $T/\alpha J_{\parallel}\ll 1$), parameter $B\Gamma_B$ changes remarkably unlike the case when $K/\alpha J_{\parallel}=0$. One can optionally allocate different values to the g-factors and creates another g-factor sets and repeat the same procedure. Hence, different outcomes may be achieved.

It could be expected from ground-state phase diagram plotted in Fig. \ref{fig:QPT_BJ}(e) that by considering higher values of the interaction parameter $K/\alpha J_{\parallel}$ (for instance,  $K/\alpha J_{\parallel}=2$ and fixed  $J/\alpha J_{\parallel}=4$) the magnetic Gr{\" u}neisen curve would has  an extra peak at critical magnetic field $B/\alpha J_{\parallel}=1$ when the g-factors are set as  $\{g_1=1$, $g_2=1$, $g_3=2\}$. By inspecting Fig. \ref{fig:QPT_BJ}(f), one immediately finds that  an enhanced MCE will occur at the critical field $B/\alpha J_{\parallel}=0.5$, as long as, the g-factors are set as  $\{g_1=1$, $g_2=4$, $g_3=2\}$ and fixed value $J/\alpha J_{\parallel}=2$ is assumed.

\subsection{Correlation function}
General expression of the nearest neighbor correlation function between Heisenberg dimers of each plaquette relies on the exchange interaction derivative of the Gibbs free energy (Eq. (\ref{FreeE})). Since we have assumed all interaction parameters between Heisenberg dimers are identical, namely, $J_{j-1}=J_{j}=J_{j+1}=\cdots =J_{2N}=J$, to get an overall introduction of  the first derivative of free energy $f$, and evoke the pair correlation function $\mathcal{G}_j^{xx}$ of the $j$-th Heisenberg dimer-rung, we need to consider 
\begin{equation}\label{CorrFun}
\begin{array}{lcl}
\mathcal{G}_j^{xx}=\langle \sigma_{1,j}^x\sigma_{2,j}^x\rangle=\langle \sigma_{1,j}^{\prime x}\sigma_{2,j}^{\prime x}\rangle=
-\dfrac{1}{2\gamma}\dfrac{\partial f}{\partial J}.
 \end{array}
 \end{equation}
 
\begin{figure*}[t!]
\begin{center}
\resizebox{0.47\textwidth}{!}{%
\includegraphics{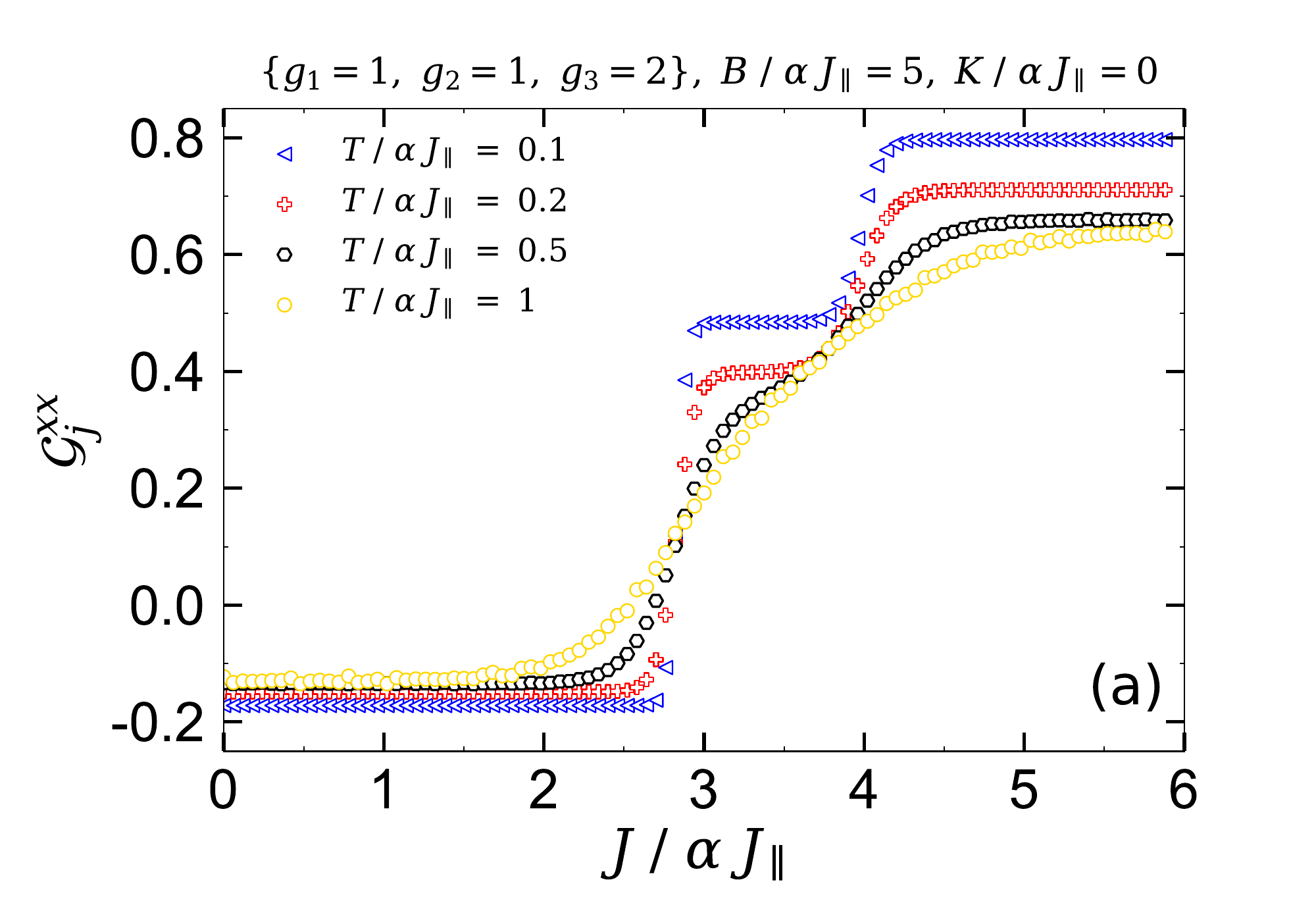}
}
\resizebox{0.47\textwidth}{!}{%
\includegraphics{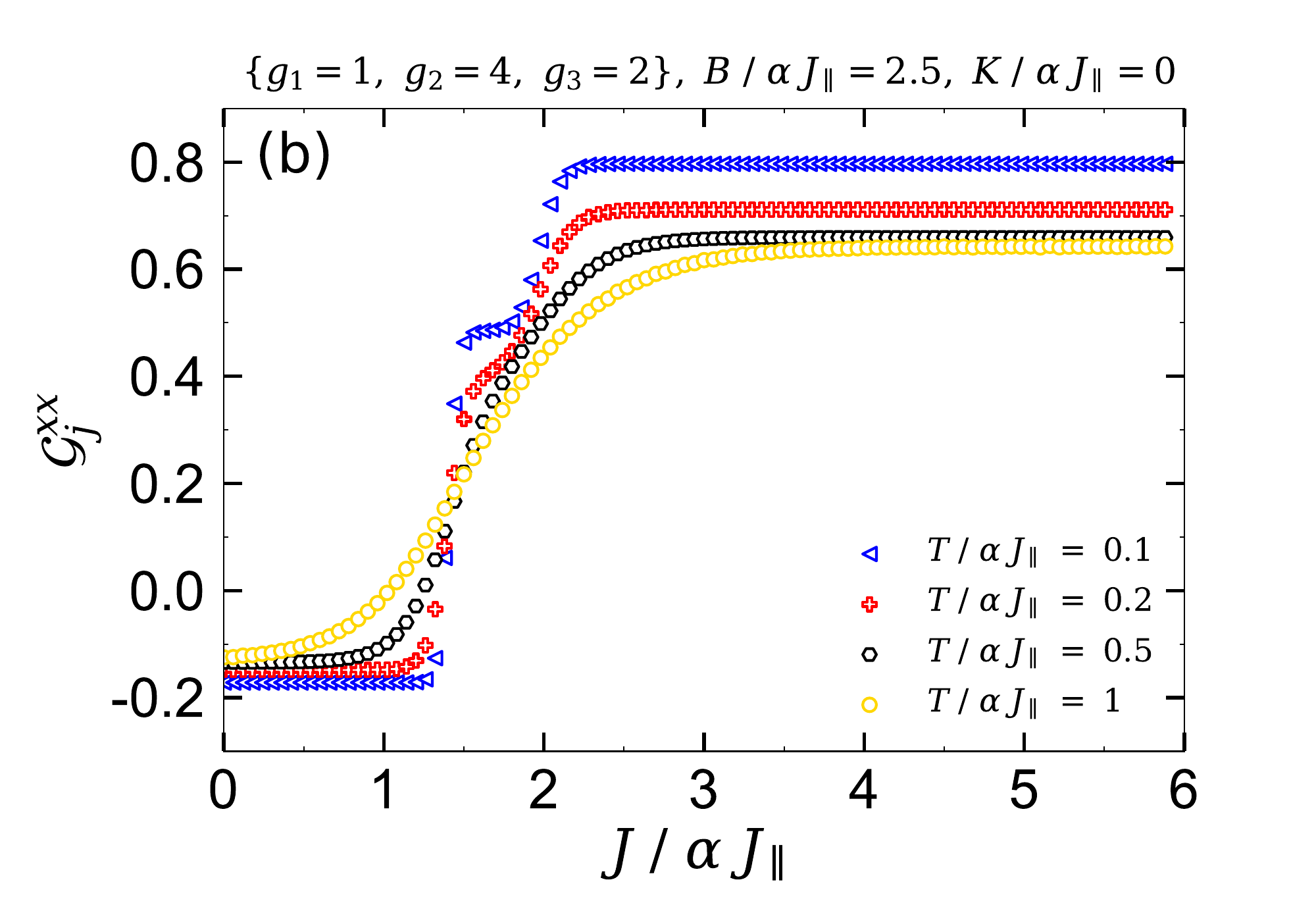}
}
\resizebox{0.47\textwidth}{!}{%
\includegraphics{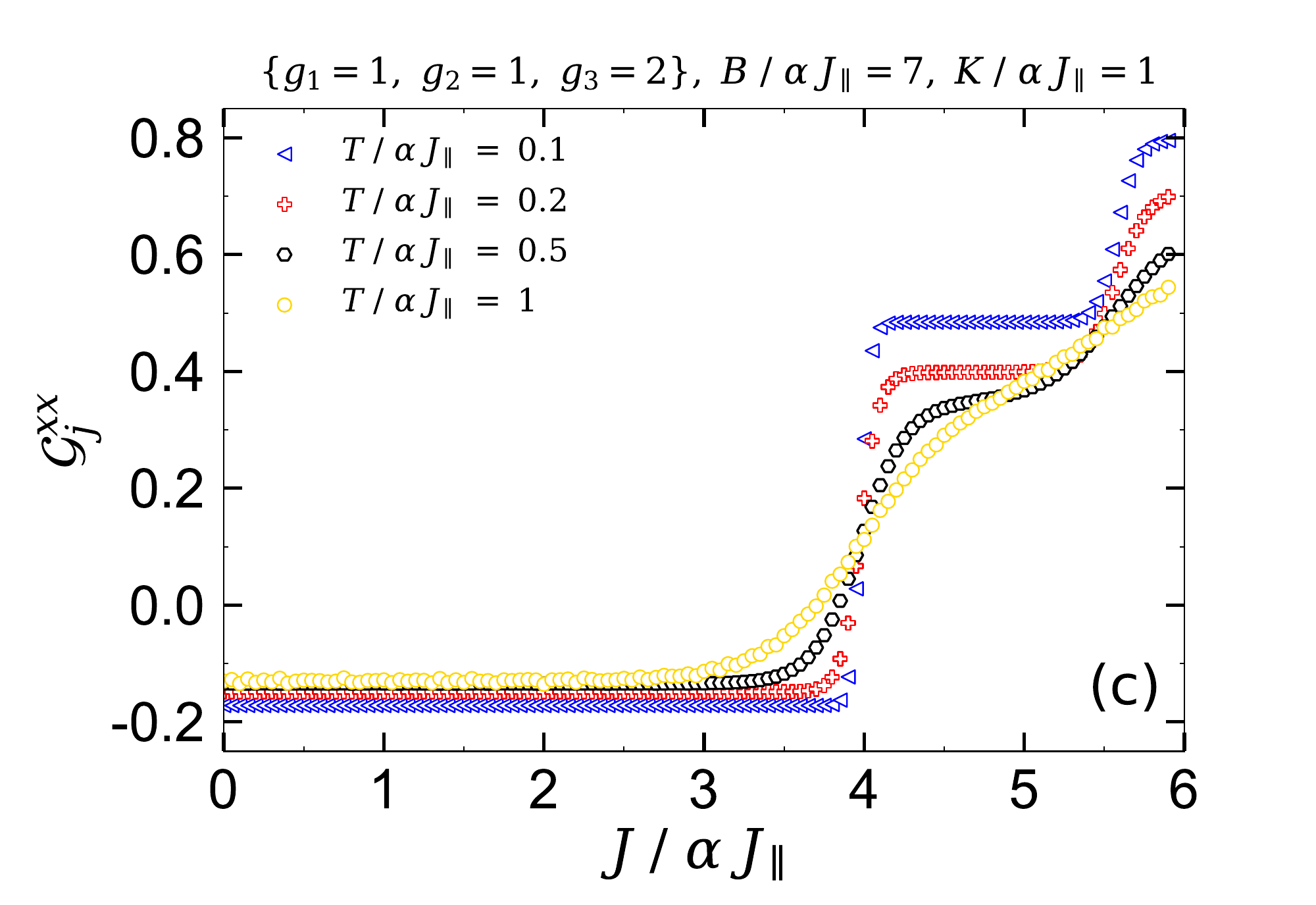}
}
\resizebox{0.47\textwidth}{!}{%
\includegraphics{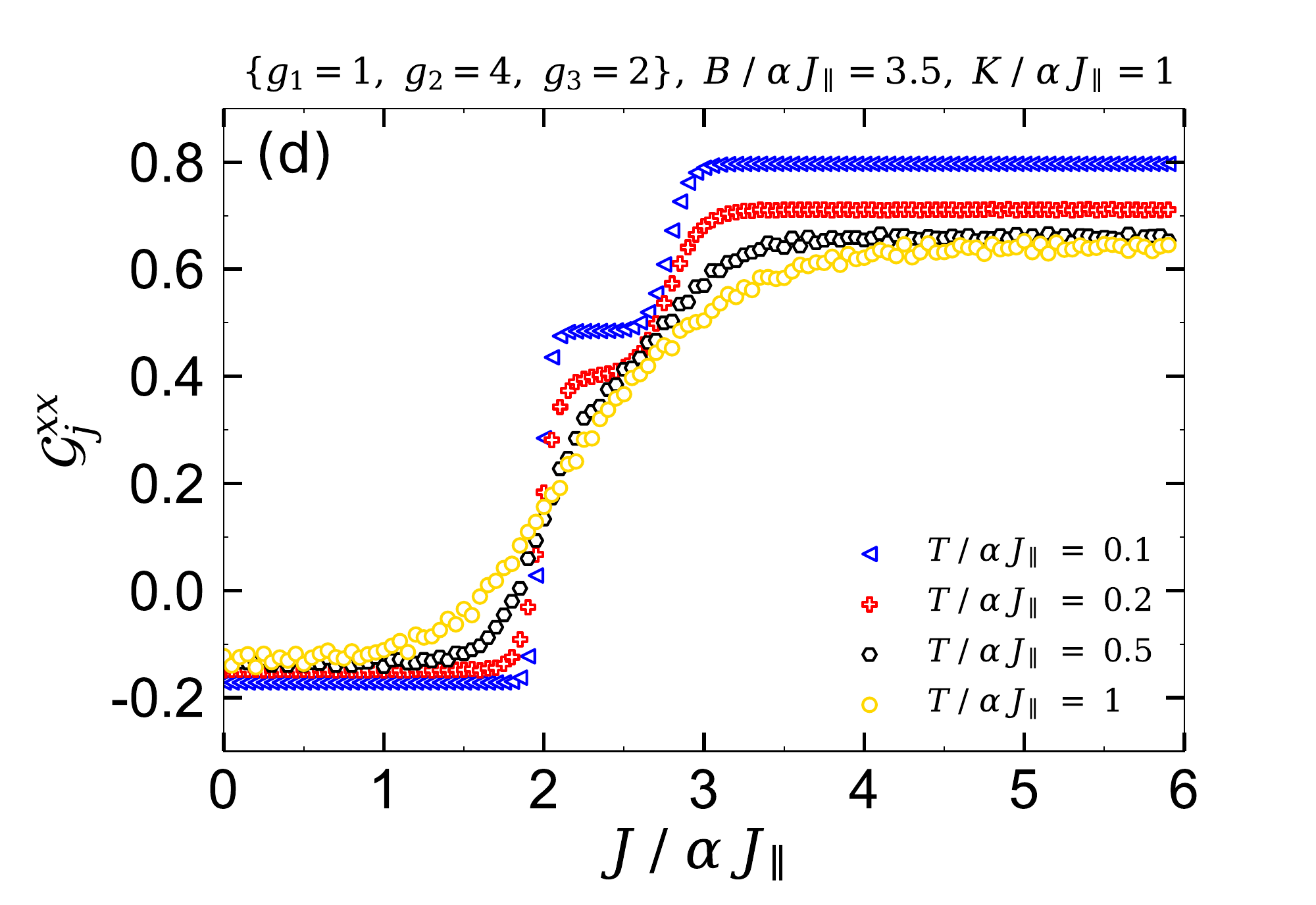} 
}
\caption{Pair correlation function for the Heisenberg dimers under different conditions. (a) $\mathcal{G}_j^{xx}$ with respect to the exchange interaction ratio $J/\alpha J_{\parallel}$ for the set  $\{g_1=1$, $g_2=1$, $g_3=2\}$ and zero value of the parameter
$K/\alpha J_{\parallel}$. The critical magnetic field $B/\alpha J_{\parallel}=5$ has been considered, where other parameters have been assumed as for Fig. \ref{fig:Mag}(a), i.e.,  $\alpha=0.5$, $\Delta= 0.5$, $\gamma=0.5$, and 
${J}_{\perp}/\alpha J_{\parallel}=5$. (b) $\mathcal{G}_j^{xx}$   for the set $\{g_1=1,\; g_2=4,\; g_3=2\}$ and
 $K/\alpha J_{\parallel}=0$, assuming critical magnetic field $B/\alpha J_{\parallel}=2.5$. (c) $\mathcal{G}_j^{xx}$ for the set  
 $\{g_1=1$, $g_2=1$, $g_3=2\}$ and non-zero value $K/\alpha J_{\parallel}=1$ and fixed $B/\alpha J_{\parallel}=7$. (d) $\mathcal{G}_j^{xx}$ for the set  $\{g_1=1$, $g_2=4$, $g_3=2\}$ and fixed values $K/\alpha J_{\parallel}=1$ and $B/\alpha J_{\parallel}=3.5$. }
\label{fig:CorrFun} 
\end{center}
\end{figure*}

We plot in Fig. \ref{fig:CorrFun} the correlation function (\ref{CorrFun}) against the interaction ratio $J/\alpha J_{\parallel}$ for a few selected temperatures under particular conditions assumed in previous plots. In order to compare the correlation function of the Heisenberg dimer-rungs  $\mathcal{G}_j^{xx}$ with the investigated thermodynamic parameters in previous parts, we first consider the set 
$\{g_1=1$, $g_2=1$, $g_3=2\}$ and a zero value for the cyclic four-spin Ising interaction (Fig. \ref{fig:CorrFun}(a)). The magnetic field has been taken as  fixed value $B/\alpha J_{\parallel}=5$ as well, which conveys the field-position of quadruple point marked in Fig. \ref{fig:QPT_BJ}(a).
It is quite surprising that  at low temperature ($T/\alpha J_{\parallel}=0.1$) there are some plateaux and jumps in the pair correlation function curve. By comparing this figure with Fig. \ref{fig:QPT_BJ}(a), one instantly finds out that the correlation function jumps occur in the vicinity of quadruple points. As mentioned before, these intriguing points are intersection of four separated ground states. 
Another elegant remark to pronounce is that, when the temperature increases monotonically, the correlation function plateaux gradually disappear until the correlation function curve becomes smooth at high temperatures. In different situation $\{g_1=1$, $g_2=4$, $g_3=2\}$, by assuming the fixed field $B/\alpha J_{\parallel}=2.5$ the correlation function jumps occur at lower amounts of interaction ratio $J/\alpha J_{\parallel}$ (Fig. \ref{fig:CorrFun}(b)).

Now, the main question that may involve our mind is that whether the cyclic four-spin Ising term $K/\alpha J_{\parallel}$ affects the correlation function $\mathcal{G}_j^{xx}$ or not? To answer this question we illustrate in Fig. \ref{fig:QPT_BJ}(c), the pair correlation function versus the interaction parameter  $J/\alpha J_{\parallel}$ for the set  $\{g_1=1$, $g_2=1$, $g_3=2\}$ and fixed $K/\alpha J_{\parallel}=1$. The field-position of triple point (shown by filled-plus mark in Fig. \ref{fig:QPT_BJ}(c)) is optionally considered. The mentioned triple point is the intersection of three ground states with the magnetization values $M/M_s=3/5$, $M/M_s=4/5$ and $M/M_s=1$. Amazingly, imposing a nonzero cyclic four-spin Ising interaction in each plaquette leads to widen the correlation function plateaux. The correlation function jumps occur at higher interaction ratio 
$J/\alpha J_{\parallel}$.

 Furthermore, we observe that the plateaux appeared in the correlation function curve stay alive at higher temperatures 
(compare black curve marked with honeycombs plotted in both Figs. \ref{fig:CorrFun}(a) and \ref{fig:CorrFun}(c)). Lower right panel \ref{fig:CorrFun}(d) depicts $\mathcal{G}_j^{xx}$ with respect to the interaction parameter $J/\alpha J_{\parallel}$, assuming the set $\{g_1=1$, $g_2=4$, $g_3=2\}$ and fixed magnetic field $B/\alpha J_{\parallel}=3.5$ and $K/\alpha J_{\parallel}=1$. By comparing Figs. \ref{fig:CorrFun}(b) and \ref{fig:CorrFun}(d) with each other, we realize that the correlation function intermediate plateau is broadened by applying a nonzero value of the ratio $K/\alpha J_{\parallel}$.  The width of the intermediate plateau appeared in the correlation function curve is in a good agreement with the width of shaded region between red and blue lines plotted in Fig. \ref{fig:QPT_BJ}.

\section{Conclusions}\label{conclusion}
The present work deals with the magnetic and magnetocaloric properties of the mixed spin-(1/2,1)  two-leg model on an Ising-Heisenberg  ladder, which can be exactly solved by the transfer-matrix technique. Two different sets of Land{\' e} g-factors have been considered for localized spins in the ladder. We performed an extra term so-called cyclic four-spin Ising interaction  in the Hamiltonian of the model, which is important to take in to account when the purpose is investigating the physical properties of spin ladders.
We have also considered  full anisotropic case XYZ for the Heisenberg dimers such that the exchange anisotropy in the $z-$direction has been assumed to be an alternative parameter in each two adjacent plaquettes of a unit block.


We have realized  that there are some quadruple (triple) points in the field-induced ground-state phase diagram of the model  which makes the intersection of four (three) different ground states. The cyclic four-spin Ising term affects  the co-ordinates of these special points.
 It has been demonstrated that at low temperatures, the specific heat curve anomalously behaves nearby the critical magnetic fields at which a magnetization jump occurs.  
Moreover, at high temperatures, exotic vicissitude can be seen in the specific of the model nearby the quadruple points.

  An anomalous magnetocaloric effect  has been observed close to respective magnetization jumps.
More importantly, toning the magnetic field, four-spin Ising interaction, and the exchange coupling parameter slightly above (below) the critical points result in cooling/heating during the adiabatic demagnetization process, where the temperature rapidly falls down and reaches close to the  first-order zero-temperature phase transition.

It is quite surprising that we found an enhanced magnetocaloric effect in the vicinity of discontinuous phase transition points.
One of notable outcomes from our examinations is that the magnetic Gr{\" u}neisen parameter peak created nearby by the quadruple point vanishes at very higher temperatures compared with other peaks. 
Another notable result is that under ultra-cooling the magnetic Gr{\" u}neisen parameter goes to infinity nearby the critical magnetic field at which first-order zero-temperature phase transition occurs.

Finally, it was demonstrated that there are some plateaux in the correlation function curve of each Heisenberg dimer-rung versus the exchange coupling, where a jump between two plateaus mostly occurs nearby the co-ordinates of the described quadruple point.
 We evidenced that by tuning cyclic four-spin Ising term, aforesaid plateaux undergoes a substantial changes. Further, the change in correlation function behavior is in an excellent coincidence with the magnetization variations.

 The most direct application of our considered model probably is as ferromagnetic version of the two-leg ladders including localized particles with different Land{\' e} g-factors in the real world for which an enhanced large magnetocaloric effect can be observed. 

\section*{Acknowledgments}
H. Arian Zad acknowledges the receipt of the grant from the Abdus Salam International Centre for Theoretical Physics (ICTP), Trieste, Italy. The authors are also grateful to Prof.  J. Stre\v{c}ka for insightful discussions. 


\section{Appendix I} \label{App_I}

\begin{figure*}[t!]
\begin{center}

\resizebox{0.48\textwidth}{!}{%
\includegraphics{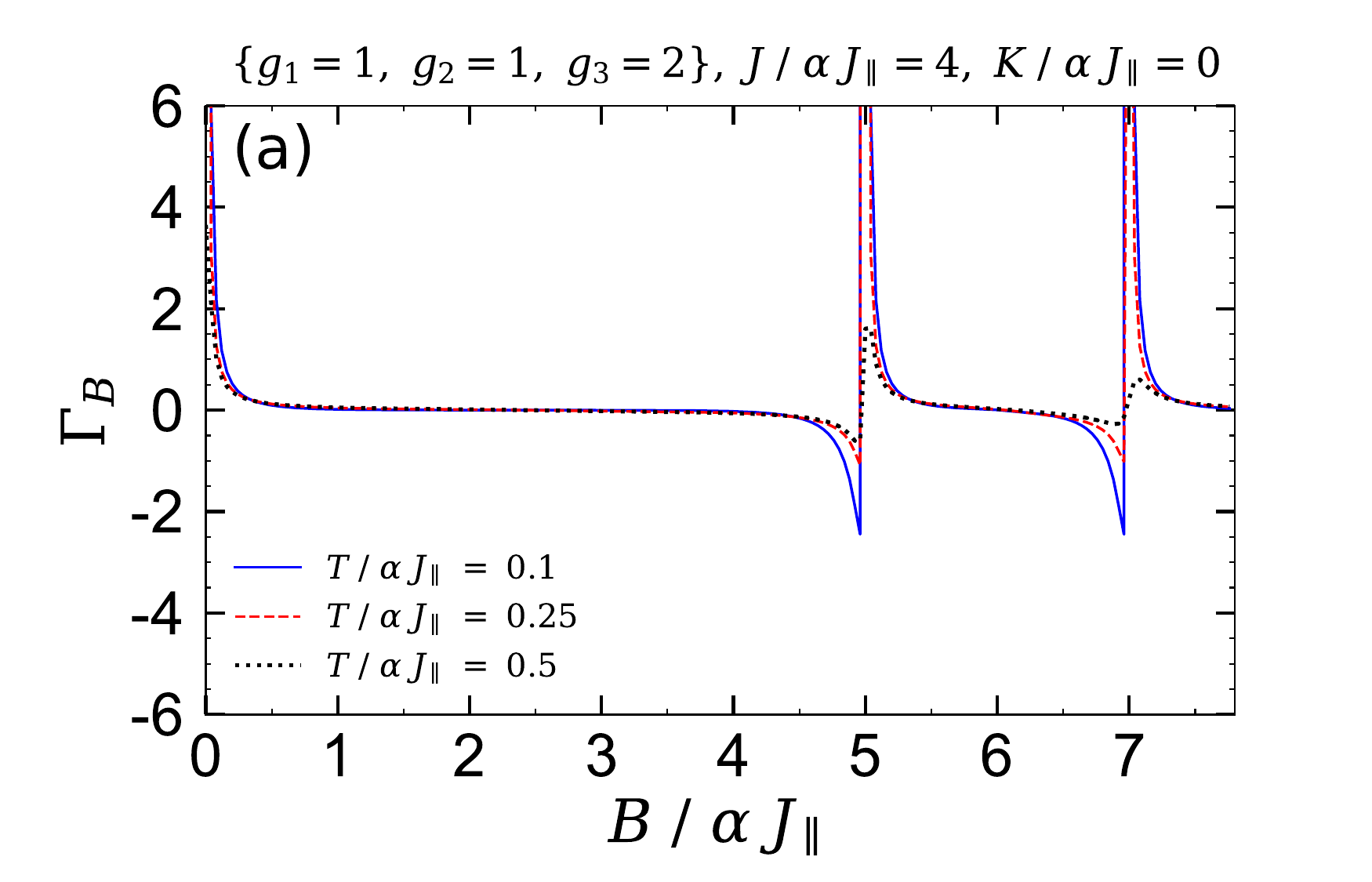}
}
\resizebox{0.48\textwidth}{!}{%
\includegraphics{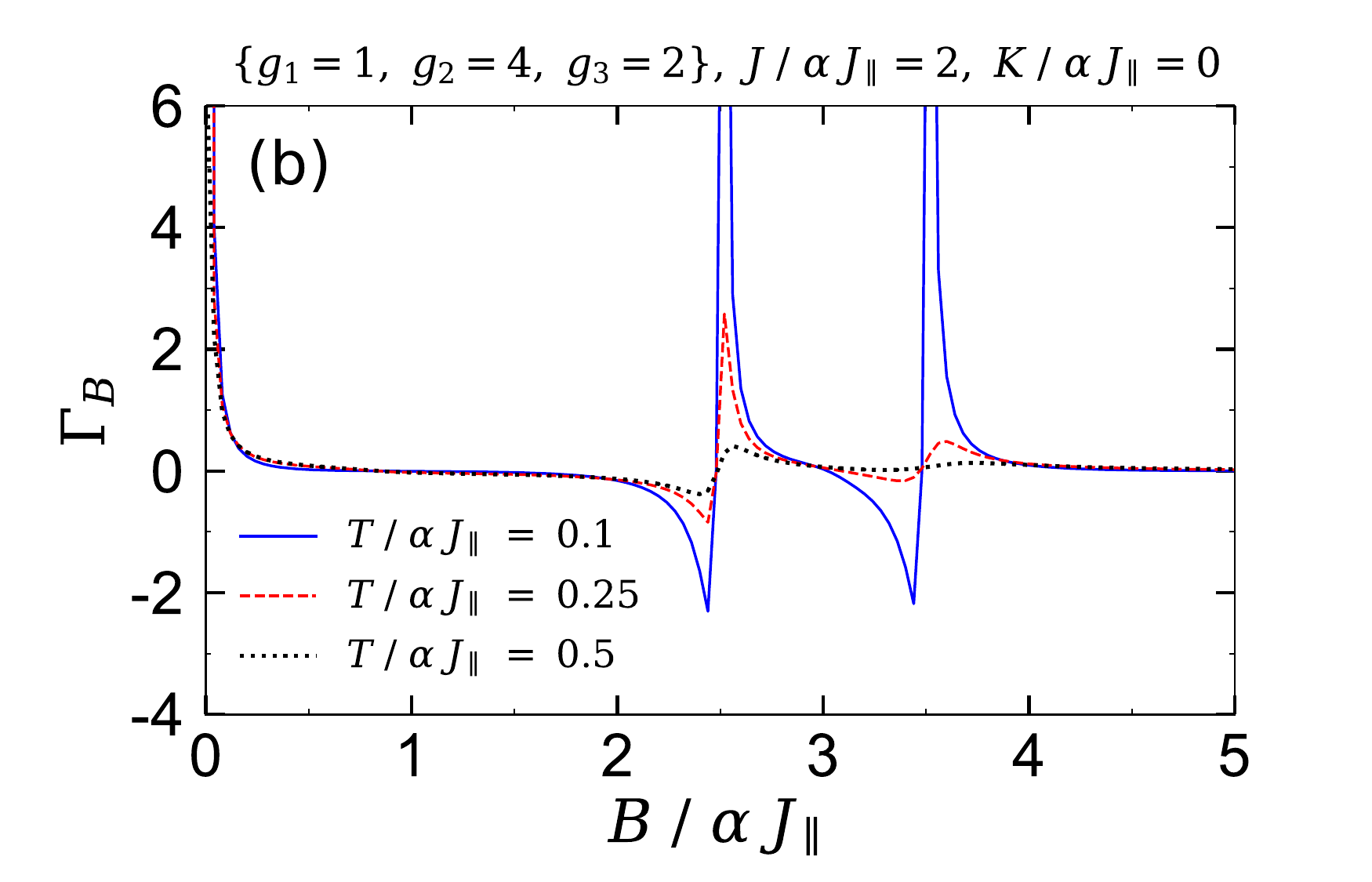} 
}
\resizebox{0.48\textwidth}{!}{%
\includegraphics{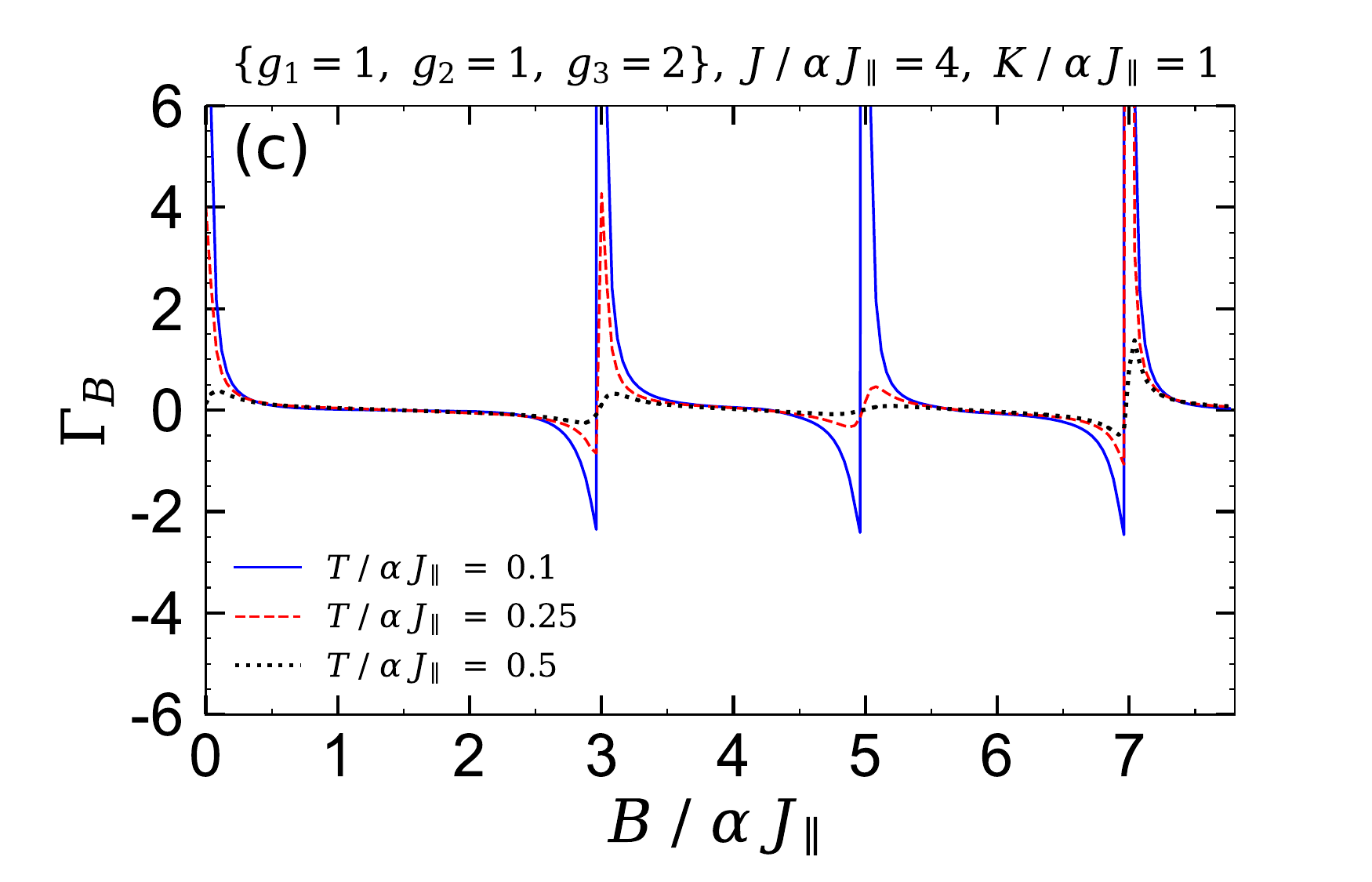} 
}\resizebox{0.48\textwidth}{!}{%
\includegraphics{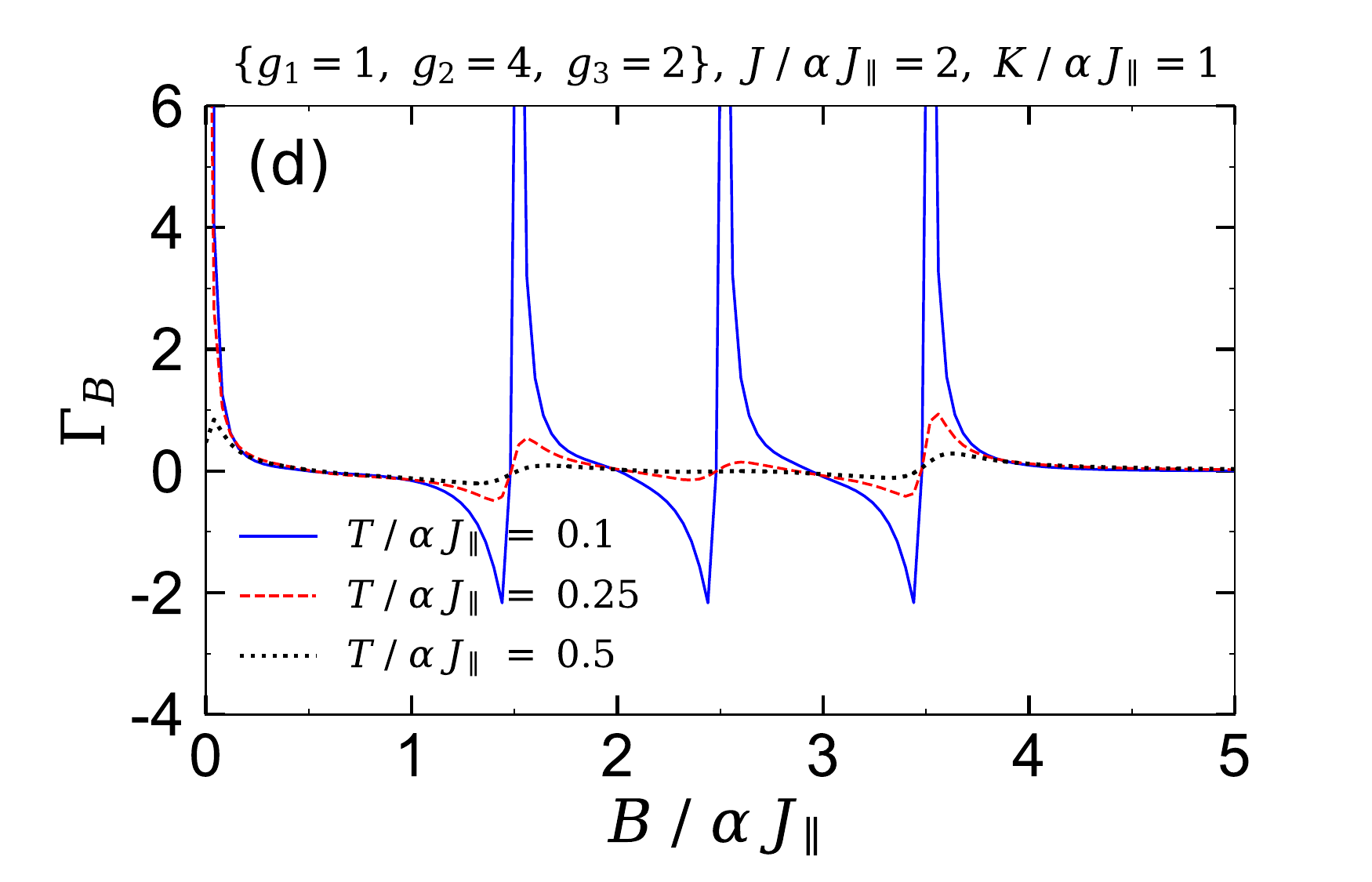} 
}
\caption{$\Gamma_B$ as a function of the ratio $B/\alpha J_{\parallel}$ at three different temperatures for the parameter sets 
(a)  $\{g_1=1$, $g_2=1$, $g_3=2\}$, $J/\alpha J_{\parallel}=4$, $K/\alpha J_{\parallel}=0$.
(b)  $\{g_1=1$, $g_2=4$, $g_3=2\}$,  $J/\alpha J_{\parallel}=2$, $K/\alpha J_{\parallel}=0$.
(c)  $\{g_1=1$, $g_2=1$, $g_3=2\}$, $J/\alpha J_{\parallel}=4$, $K/\alpha J_{\parallel}=1$.
(d)  $\{g_1=1$, $g_2=4$, $g_3=2\}$,  $J/\alpha J_{\parallel}=2$, $K/\alpha J_{\parallel}=1$.
 Other parameters $\alpha$, $\gamma$, $\Delta$ and $J_{\perp}/\alpha J_{\parallel}$ have been taken as Fig. \ref{fig:Mag}.}
\label{fig:Gamma_B}
\end{center}
\end{figure*}

Qualitatively the field and temperature dependence of $\Gamma_B$  around a metamagnetic transition can be understood looking at the two Heisenberg spins only. Here, we would like to investigate this medium by sketching the corresponding figures that are surely of rich pedagogical values for the paper.
We plot in Fig. \ref{fig:Gamma_B} parameter $\Gamma_B$ versus ratio $B/\alpha J_{\parallel}$ at three different selected temperatures.
Four panels of this figure represent $\Gamma_B$ in four different situations. By comparing this figure with Figs. \ref{fig:EntropyK0} and \ref{fig:EntropyK1} one finds that the low-magnetic field behavior of the parameters $\Gamma_B$ and $B\Gamma_B$ is quite different. However, both parameters behaves similar to each other in the vicinity of first-order zero-temperature phase transition points.

\begin{figure*}[t!]
\begin{center}
\resizebox{0.47\textwidth}{!}{%
\includegraphics[trim=20 1 80 1, clip]{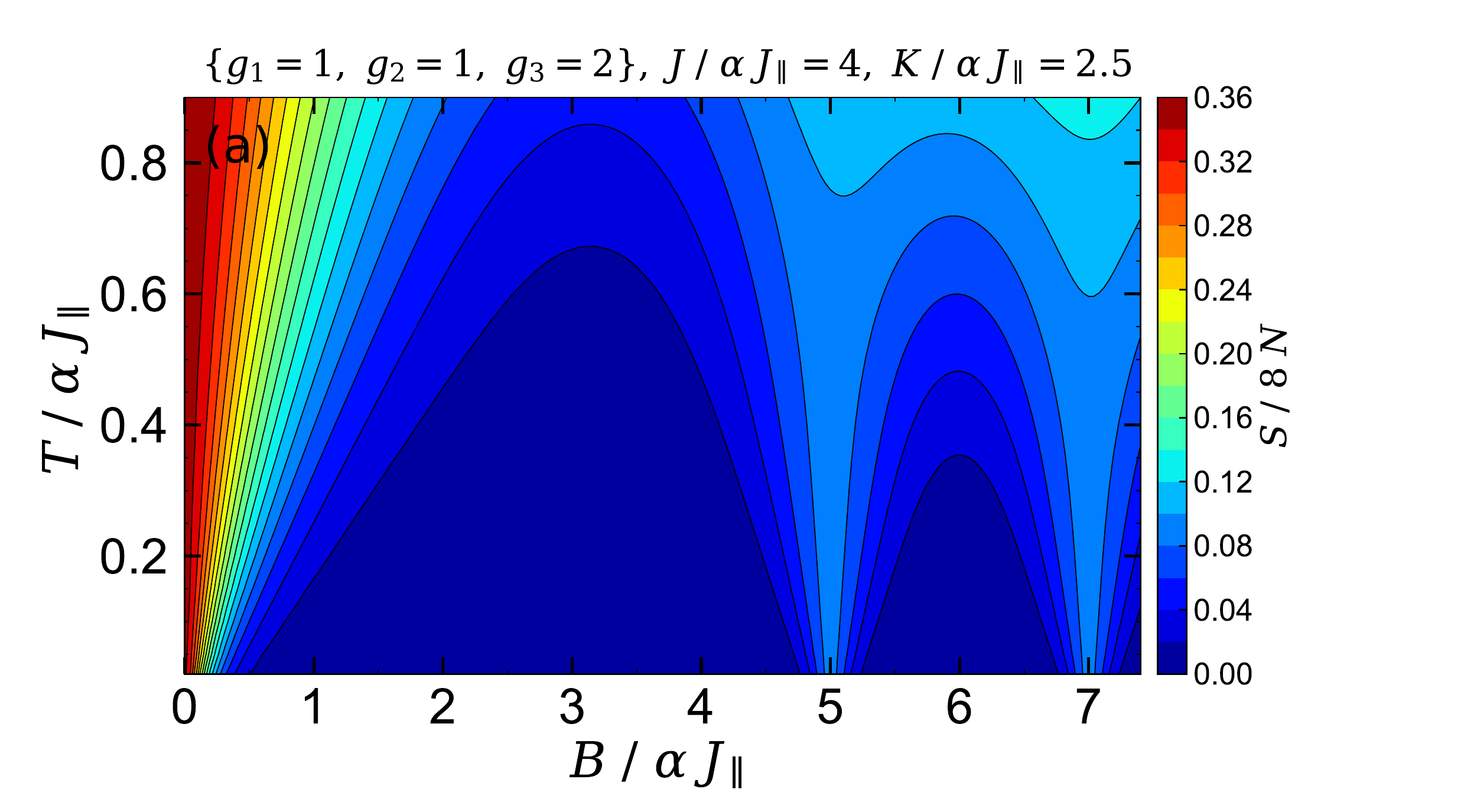}
}
\resizebox{0.47\textwidth}{!}{%
\includegraphics[trim=20 1 80 1, clip]{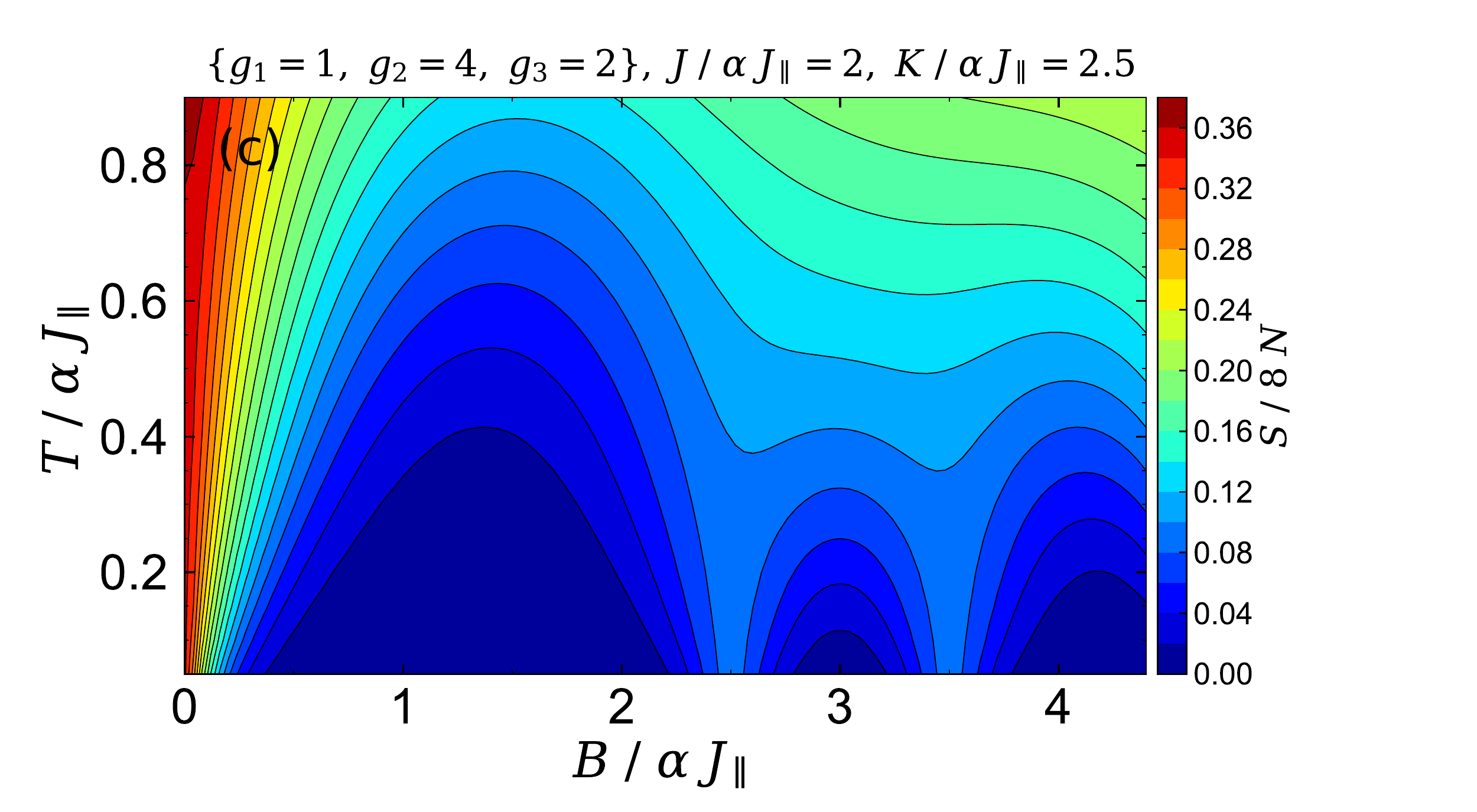}
}
\resizebox{0.47\textwidth}{!}{%
\includegraphics{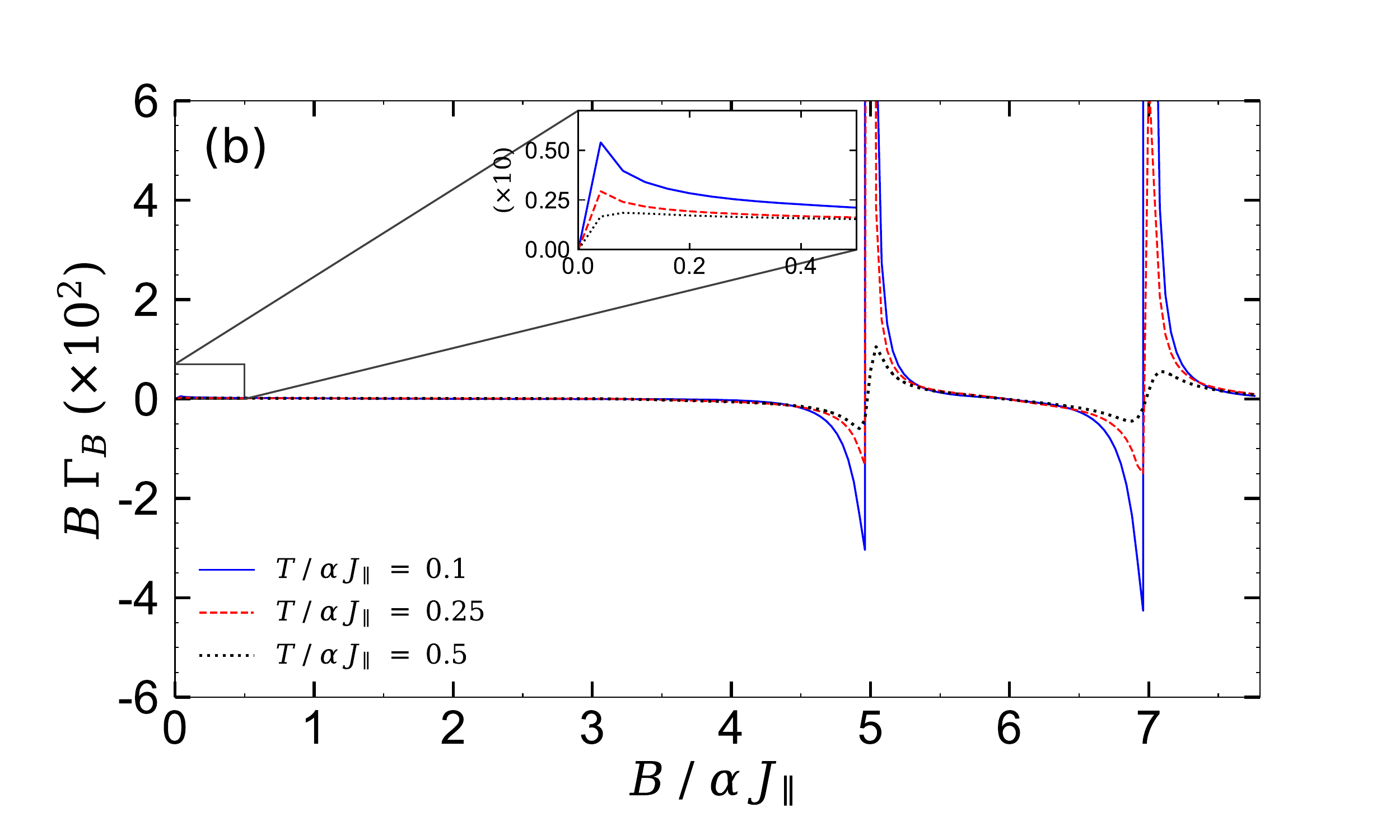}
}
\resizebox{0.47\textwidth}{!}{%
\includegraphics{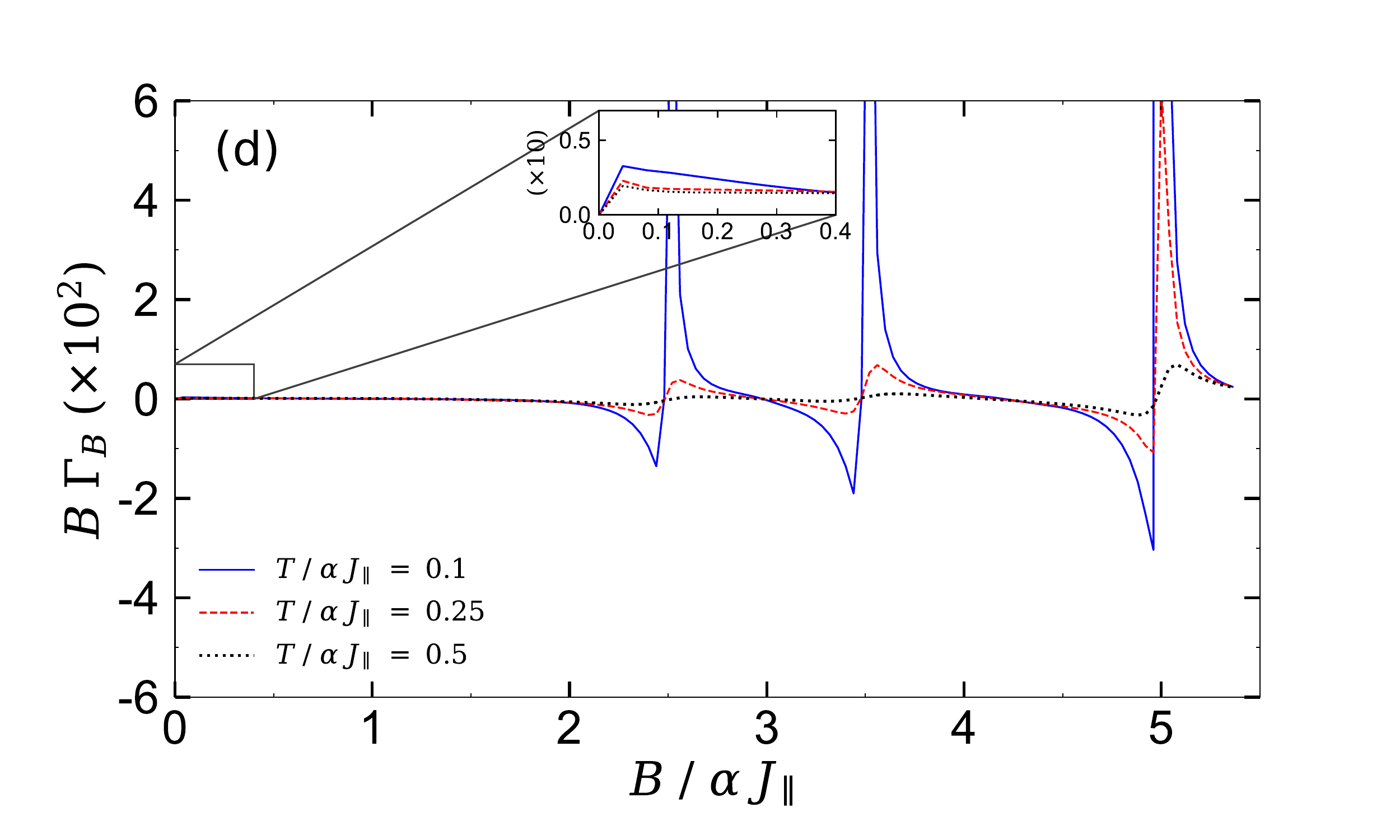} 
} 
\caption{The contour plot of the entropy and some isentropy lines in the field-temperature plane with the corresponding Gr{\" u}neisen parameter as a function the magnetic field for the high amount of the cyclic four-spin Ising interaction $K/\alpha J_{\parallel}=2.5$ under the same conditions considered for the panels \ref{fig:EntropyK0}(a), \ref{fig:EntropyK0}(b), \ref{fig:EntropyK0}(d), \ref{fig:EntropyK0}(e). The reason for choosing value $K/\alpha J_{\parallel}=2.5$ is that, as shown in Fig. \ref{fig:QPT_BK}, there is a critical point in the $(B/\alpha J_{\parallel}-K/\alpha J_{\parallel})$ plane with the co-ordinates
 ($B/\alpha J_{\parallel},\; K/\alpha J_{\parallel})\equiv (0, 2.5)$ at which the ground-state phase boundaries cut each other off and make a fascinating point to consider. 
 }
\label{fig:Entropy_K25}
\end{center}
\end{figure*}

 To better understand the influence of ratio $K/\alpha J_{\parallel}$ on the thermodynamic mechanism of the model, we  illustrate in Figs. \ref{fig:Entropy_K25}(a) and \ref{fig:Entropy_K25}(b), the entropy and magnetic Gr{\" u}neisen parameter in terms of the temperature and the magnetic field for the set $\{g_1=1$, $g_2=1$, $g_3=2\}$ and higher value $K/\alpha J_{\parallel}=2.5$ by assuming fixed $J/\alpha J_{\parallel}=4$ (review Fig. \ref{fig:QPT_BK}(a) in which critical point $(B/\alpha J_{\parallel},\;K/\alpha J_{\parallel})\equiv (0,\;2.5)$ has been marked). 
 While, in Figs. \ref{fig:Entropy_K25}(c) and \ref{fig:Entropy_K25}(d), are plotted the entropy and the Gr{\" u}neisen parameter as functions of the temperature and the magnetic field for the set $\{g_1=1$, $g_2=4$, $g_3=2\}$ and fixed values $K/\alpha J_{\parallel}=2.5$ and $J/\alpha J_{\parallel}=2$ (according to the marked critical point $(B/\alpha J_{\parallel},\;K/\alpha J_{\parallel})\equiv (0,\;2.5)$ in Fig. \ref{fig:QPT_BK}(b)). 
 
Consequently, by imaging different values of the cyclic four-spin Ising interaction parameter $K/\alpha J_{\parallel}$, despite the fact that phase transitions will occur at detected critical points $B_c/\alpha J_{\parallel}=\{0,\;3,\;5\}$ for the case  $\{g_1=1$, $g_2=1$, $g_3=2\}$, and at 
$B_c/\alpha J_{\parallel}=\{0,\;1.5.\;2. 5,\; 3.5,\;5\}$ for the set $\{g_1=1$, $g_2=4$, $g_3=2\}$, the entropy and magnetic Gr{\" u}neisen parameter qualitatively and quantitatively change different from  Fig. \ref{fig:EntropyK0}. 
  

\end{document}